\documentclass[11pt,aps,pre,onecolumn,floatfix,amsmath,amssymb,superscriptaddress,longbibliography]{revtex4-2}
\usepackage[T1]{fontenc}
\usepackage[utf8]{inputenc}

\usepackage{url}
\usepackage{amsmath}
\numberwithin{equation}{section} 
\renewcommand{\thesection}{\Roman{section}} 
\renewcommand{\theequation}{\thesection.\arabic{equation}}
\usepackage{amssymb}
\usepackage{graphicx}
\usepackage{esint}
\usepackage[english]{babel}
\usepackage{xcolor}
\usepackage{setspace}
\usepackage{enumitem}
\usepackage[section]{placeins}
\usepackage{hyperref}
\hypersetup{
    colorlinks=true,    
    linkcolor=blue,     
    citecolor=blue,     
    urlcolor=blue       
}

\usepackage{booktabs}

\makeatletter
\@ifundefined{textcolor}{}
{
 \definecolor{BLACK}{gray}{0}
 \definecolor{WHITE}{gray}{1}
 \definecolor{RED}{rgb}{1,0,0}
 \definecolor{GREEN}{rgb}{0,1,0}
 \definecolor{BLUE}{rgb}{0,0,1}
 \definecolor{CYAN}{cmyk}{1,0,0,0}
 \definecolor{MAGENTA}{cmyk}{0,1,0,0}
 \definecolor{YELLOW}{cmyk}{0,0,1,0}
}
\makeatother

\newcommand{\appendixequations}{
  \renewcommand{\theequation}{\thesection.\arabic{equation}}
  \numberwithin{equation}{section}
}

\begin{document}
\author{Chunyan Li}
\affiliation{Center for Advanced Quantum Studies, School of Physics and Astronomy, Beijing Normal University, Beijing 100875, China}
\affiliation{Key Laboratory of Multiscale Spin Physics, Ministry of Education, Beijing 100875, China}
\author{Zheng Li}
\affiliation{Center for Advanced Quantum Studies, School of Physics and Astronomy, Beijing Normal University, Beijing 100875, China}
\author{Yueyan Li}
\affiliation{School of Science, Tianjin University of Technology and Education, Tianjin 300222, China}
\author{Haiwen Liu}
\email{haiwen.liu@bnu.edu.cn}
\affiliation{Center for Advanced Quantum Studies, School of Physics and Astronomy, Beijing Normal University, Beijing 100875, China}
\affiliation{Key Laboratory of Multiscale Spin Physics, Ministry of Education, Beijing 100875, China}
\affiliation{Interdisciplinary Center for Theoretical Physics and Information Sciences, Fudan University, Shanghai 200433, China}
\author{X. C. Xie}
\affiliation{Interdisciplinary Center for Theoretical Physics and Information Sciences, Fudan University, Shanghai 200433, China}
\affiliation{International Center for Quantum Materials, School of Physics, Peking University, Beijing 100871, China}
\affiliation{Hefei National Laboratory, Hefei 230088, China}

\title{Logarithmic Aging Diffusion from a Multiplicative Event Clock: Rare Event Statistics, Ultraslow Transport, and Ensemble--Time Inequivalence}

\begin{abstract}
    \begingroup
    Logarithmic time dependences occur in many aging materials, but neither a $\ln t$ relaxation law nor a $1/t$ event rate uniquely identifies the underlying stochastic mechanism. We examine a specific log-aging process defined by iterating the age-conditioned forward-recurrence law after every event. This rule makes the event times multiplicative: the logarithmic ratios $U_n=\ln(T_{n+1}/T_n)$ are independent and identically distributed with an explicit non-exponential density. Consequently, both the mean and the variance of the event count grow linearly with $\ln(t/t_0)$, while the density of the $n$th event time has a log-normal central sector and a fixed-$n$ algebraic far tail. These clock statistics generate logarithmic drift and spreading, an Einstein relation under local detailed balance, and ultraslow transit and target-survival laws. They also separate trajectory reproducibility from ensemble--time equivalence: the relative scatter of the time-averaged mean-square displacement decays as $1/\ln(T/t_0)$, although its mean does not converge to the ensemble lag MSD. We distinguish the exact event-level construction from its diffusion-limit generalized Fokker--Planck and random-clock subordination representations, and from a generalized-Langevin closure that can match selected responses and covariances but need not reproduce event counts or rare-duration statistics. The proposed clock is therefore tested not by a single logarithmic curve, but by the joint, no-refitting consistency of multiplier, count, transport, first-passage, and finite-window observables.\\
    
    \par\textbf{Keywords:} logarithmic aging; ultraslow diffusion; multiplicative event clock; continuous-time random walks; event-time and counting statistics; weak ergodicity breaking; first-passage statistics.
    \endgroup
\end{abstract}

\maketitle
\tableofcontents
\clearpage

\section*{Introduction}

Many disordered and glassy systems evolve on time scales that are broad, history dependent, and often inaccessible to direct equilibration~\cite{Berthier_2011,Pollak_2012,Nandkishore_2015,Henkel2010,Cipelletti_2005,Cugliandolo_Course7,doi:10.1142/9789812819437_0006,normal_C4CP03465A,Debenedetti2001,Vincent2007}. Their common signatures include waiting-time-dependent response, intermittent activity, nonstationary correlations, and a mismatch between ensemble and trajectory-level observables. These signatures do not by themselves identify a unique microscopic mechanism. Rather, they motivate an effective stochastic description in which the physical time available for transport or relaxation is itself a random, aging-dependent quantity~\cite{PhysRevLett.134.197102,logaging_PhysRevLett.110.208301,li2025rareeventinducedergodicitybreakinglogarithmic,PhysRevLett.110.020602,PhysRevE.87.032915}.

\begingroup
This Perspective develops such a description for logarithmic aging diffusion. Its central premise is that the relevant operational time grows as $\ln(t/t_0)$, rather than linearly or as a power law in laboratory time. This ultraslow clock produces logarithmic spreading, scale-covariant memory, and genuinely two-time observables. We organize the event-based continuous-time random walk (CTRW), generalized Fokker--Planck equation (GFPE), time-matching or subordination construction, and generalized Langevin equation (GLE) as a \emph{hierarchy of descriptions}, following established connections among these approaches in anomalous transport~\cite{Gil2021,normal_C4CP03465A,CTRW_METZLER20001,disorder_BOUCHAUD1990127,Tauberium_klafter2011,Scher1975,ZASLAVSKY2002461,CTRWandFPE_sandev2018continuous,GLE_PhysRevE.64.051106}. The event process contains the complete clock statistics. The GFPE and subordination descriptions retain the one-point density and selected clock information under their stated limits, whereas deterministic-clock and Gaussian GLE reductions generally retain still less. The framework is not intended as a microscopic model of every aging material. In particular, quenched spatial disorder, collective rearrangements, and strong interactions can generate correlations absent from the independent-multiplier construction; the value of the construction is to isolate falsifiable consequences of an ultraslow aging clock.
\endgroup

\subsection*{Why a Logarithmic One-Time Law Does Not Identify the Clock}

\begingroup
Several inequivalent stochastic mechanisms can generate logarithmic or, more generally, ultraslow one-time observables. The Montroll--Weiss and Scher--Montroll CTRW framework connects independent waiting times to renewal equations and dispersive transport~\cite{CTRW_Montroll_10.1063/1.1704269,Scher1975}. Bouchaud-type trap models and aging  theory explain how a divergent mean waiting time produces forward-recurrence statistics and weak ergodicity breaking~\cite{log_theory_bouchaud1992weak,disorder_BOUCHAUD1990127,Godreche2001Statistics,10.1063/1.1559676,review_Johannes_PhysRevX.4.011028}. Distributed-order fractional kinetics can instead generate powers of $\ln t$ from an exceptionally broad \emph{i.i.d.} waiting-time ensemble~\cite{A.V.Chechkin_2003}. Sinai diffusion produces ultraslow motion through recurrent exploration of a quenched Brownian potential~\cite{Sin83,le1999random}. Record dynamics describes irreversible ``quakes'' whose statistics can become approximately Poissonian in logarithmic time~\cite{P.Sibani_2003,Sibani2007LinearResponse,log_theory_PhysRevE.98.020602,log_theory_Robe_2016}. Finally, a deterministic time change or a nonstationary Gaussian process can reproduce logarithmic spreading without reproducing the same event-count, multiplier, or rare-duration statistics~\cite{clock_PhysRevE.108.034113,beyond_Cherstvy_2021}.

The defining step of the present log-aging process is more specific than the use of a forward waiting time. In an ordinary aged renewal measurement\cite{10.1063/1.1559676}, the forward-recurrence distribution determines the first residual interval after the observation begins, while the microscopic inter-event times remain identically distributed. Here the age-conditioned forward law is iterated as the transition rule at every event time. Consecutive event times therefore obey a multiplicative recursion, and their logarithms have independent increments. This distinction yields an explicit multiplier density and Mellin transform, an inverse-age activity, and the separation between a central log-normal sector and an algebraic duration tail.
\endgroup

\begin{table}[htbp]
\centering
\begingroup\footnotesize\setstretch{1.15}
\setlength{\tabcolsep}{3pt}
\caption{Distinct stochastic mechanisms behind logarithmic or ultraslow one-time laws. The last column lists measurements that retain information discarded by the leading one-time scaling.}
\label{tab:intro_taxonomy}
\begin{tabular}{@{}p{0.16\linewidth}p{0.23\linewidth}p{0.20\linewidth}p{0.31\linewidth}@{}}
\toprule
Model class & Clock or event structure & Leading one-time behavior & Measurement that separates the mechanisms \\
\midrule
Distributed-order or superheavy-tail renewal CTRW~\cite{A.V.Chechkin_2003} &
i.i.d.\ additive waiting times with a slowly varying or distributed-order tail &
$\langle X^2(t)\rangle\propto[\ln(t/t_0)]^\beta$ in standard realizations &
Forward recurrence and count statistics are generated by one renewal survival law; an i.i.d.\ event-time multiplier law is not implied \\
Sinai diffusion~\cite{Sin83,le1999random} &
Quenched Brownian potential with recurrent returns and sample-specific barriers &
$|X(t)|_{\mathrm{typ}}\propto[\ln(t/t_0)]^2$ &
Return correlations, disorder-to-disorder fluctuations, and barrier-resolved trajectories \\
Homogeneous log-Poisson record dynamics~\cite{P.Sibani_2003,log_theory_PhysRevE.98.020602} &
Poisson process in $\ell=\ln(t/t_0)$; exponential i.i.d.\ increments in $\ell$ &
$\langle N(t)\rangle=\operatorname{Var}N(t)\propto\ln(t/t_0)$ and event rate $\propto1/t$ &
Exponential log-time increments and equality of all Poisson count cumulants \\
Deterministic logarithmic time change~\cite{clock_PhysRevE.108.034113} &
Prescribed $s(t)=c\ln(t/t_0)$ with no random event clock &
Gaussian propagator with variance proportional to $\ln(t/t_0)$ &
No multiplier or count fluctuations and no random-clock low-activity sector \\
Present iterated-forward clock~\cite{li2025rareeventinducedergodicitybreakinglogarithmic} &
Non-exponential i.i.d.\ increments $U_n=\ln(T_{n+1}/T_n)$ &
$\langle N\rangle$ and $\operatorname{Var}N\propto\ln(t/t_0)$; central and fixed-$n$ tail sectors &
One multiplier density must predict count cumulants, transport, event-time tails, and finite-window statistics without refitting \\
\bottomrule
\end{tabular}
\endgroup
\end{table}

\subsection*{What Experiments Already Show---and Which Clock Statistics Are Still Missing}

\begingroup
Experiments establish several components of aging, but generally not the complete stochastic law of the underlying clock. Electron glasses exhibit logarithmic conductance relaxation, memory dips, and protocol dependence after gate or field perturbations~\cite{Pollak_2012,Anderson_PhysRevLett.84.3402,Ovadyahu_PhysRevLett.92.066801,electron_glass_PhysRevB.68.184204}. Spin glasses display waiting-time-dependent thermoremanent magnetization, memory, rejuvenation, and fluctuation--response anomalies~\cite{Mydosh1993,PhysRevLett.81.3243,PhysRevB.64.174204,bouchaud1995aging,Cugliandolo_2011}. These observations establish aging, but bulk response alone does not identify the statistics of the elementary rearrangements.

Soft and mechanical matter provide more direct access to events. Dynamic-light-scattering and imaging studies of colloidal gels and glasses reveal age-dependent relaxation, heterogeneous motion, and intermittent cage breaking~\cite{PhysRevLett.84.2275,13prl_glass_doi:10.1126/science.287.5453.627,Cipelletti_2005}. Event-resolved measurements in jammed colloids found an inverse-age rate of irreversible cage-breaking events and logarithmic evolution of both displacement and a mesoscopic length scale, motivating a log-Poisson record interpretation~\cite{log_theory_Robe_2016,log_theory_PhysRevE.98.020602}. Experiments on crumpled sheets, together with simulations of bistable elastic networks, found self-similar instability cascades separated by dwell times proportional to sample age~\cite{log_theory_shohat2023logarithmic,log_aging_nature}. These systems supply unusually concrete microscopic candidates for a multiplicative clock, but they also warn that successive events may be correlated through avalanches and elastic redistribution.

The decisive test is therefore joint and predictive rather than based on a single phenomenological fit. One should identify event times $T_n$, measure the multiplier $Y_n=T_{n+1}/T_n$, test correlations of $\ln Y_n$, and compare the independently inferred clock cumulants with the mean and variance of $N(t)$. The same parameters should then predict transport, first passage, the event-time tail, and both ensemble- and time-averaged observables. This no-refitting sequence is stronger than fitting a logarithm to a single relaxation curve and is the organizing experimental principle of this review.
\endgroup

\subsection*{Central results and organization}

\begingroup
Two results receive structural priority. First, the multiplicative event clock controls a linked family of observables: logarithmic drift and spreading, an aging Einstein relation, transit-time dispersion, target survival, and scale-covariant memory. Second, typical and rare histories play sharply different roles. A central-limit sector in logarithmic time controls ordinary event-count moments, whereas an algebraic low-activity tail controls positive duration moments and characteristic times. Consequently, the conventional EB parameter can vanish as $1/\ln(T/t_0)$ even though the mean time average remains inequivalent to its ensemble counterpart. This is not a contradiction: decreasing relative scatter and ergodicity test different propositions.

The preservation-first modular organization is as follows. Section~I defines the experimental protocol and evidence hierarchy. Section~II develops standard, aged, and log-aging CTRWs and isolates the multiplicative clock. Section~III places event-number statistics, rare durations, time averages, and ergodicity immediately after the clock from which they follow. Section~IV derives transport and first-passage observables. Sections~V--VII present, in decreasing order of retained stochastic information, the GFPE, random-clock subordination, and GLE descriptions. Section~VIII then returns to Sinai landscapes, disordered magnets, electron glasses, spin glasses, jammed colloids, and mechanical instability cascades, now using the theoretical predictions as discriminants. Section~IX separates established results, controlled approximations, microscopic hypotheses, and open experimental tests.
\endgroup

\section{Operational Definition of Aging: From Experimental Protocols to Finite-Window Probability Laws}
\label{sec:operational}

\subsection{Aging Is History Dependence, Not Merely Slow Relaxation}

The first task is to identify what an aging experiment actually measures. Rather than treating every slow relaxation as aging, we distinguish a slow one-time law from a genuine dependence on preparation history and observation window. Experiments therefore enter this Perspective as operational constraints on a stochastic description: the theory must account for waiting-time-dependent responses\cite{PhysRevLett.93.228302}, two-time correlations\cite{13prl_glass_doi:10.1126/science.287.5453.627,PhysRevLett.84.2275}, intermittent activity\cite{10.1063/1.480896}, and the possible separation between ensemble and trajectory-level statistics~\cite{PhysRevLett.90.120601,review_Metlzer_C4CP03465A,RevModPhys.83.587}.

The standard Brownian or Markovian picture provides a useful reference point. In ordinary diffusion or relaxation near equilibrium, microscopic fluctuations have finite correlation times, relaxation rates are characterized by finite time scales, and response functions become time homogeneous after equilibration. In such systems, shifting the origin of the measurement time does not change the statistical outcome. Transport or relaxation can then often be described by a small number of time-independent parameters, such as a diffusion coefficient, a relaxation time or a stationary response function.

Many disordered and glassy systems violate this simple picture. Their relaxation may be so slow that thermodynamic equilibrium is not reached within any reasonable experimental time\cite{log_theory_bouchaud1992weak,Cugliandolo_Course7}. Electron glasses provide a representative example: after a quench, gate-voltage change, infrared excitation, or strong electric-field perturbation, the conductance may relax logarithmically or nearly logarithmically over many decades of time, while memory-dip and two-dip protocols reveal dependence on the previous state~\cite{annurev:/content/journals/10.1146/annurev-conmatphys-062910-140455,Anderson_PhysRevLett.84.3402,Ovadyahu_PhysRevLett.92.066801}. In spin glasses and disordered magnetic systems\cite{Mydosh1993}, analogous aging behavior is observed through waiting-time-dependent correlation and response functions~\cite{PhysRevLett.81.3243,bouchaud1995aging,PhysRevB.64.174204}. In tracer or soft-matter experiments\cite{Cipelletti_2005,Berthier_2011,Waigh_2023}, related physics may appear as heterogeneous displacements, intermittent motion, and non-Gaussian statistics~\cite{PhysRevX.7.021002}. The measured observable differs from system to system, but the common phenomenology is a broad relaxation spectrum, slow approach to equilibrium, and history-dependent dynamics.

A useful way to express this broad spectrum of time scales is through transition rates that depend exponentially on random barriers or random variables\cite{log_theory_bouchaud1992weak,Cecile_Monthus_1996,EfrosShklovskii1984ElectronicProperties}. For example, one may write schematically

\begin{equation}
\gamma=\tau_0^{-1}\exp(-x),
\end{equation}

where $\tau_0$ is a microscopic time scale and $x$ is a broadly distributed positive variable. Even if $\tau_0$ is microscopic, values of $x$ much larger than unity generate relaxation times $\gamma^{-1}$ far beyond experimental time scales. Broad barrier and waiting-time statistics are a standard route to anomalous diffusion and nonergodic dynamics in disordered media~\cite{disorder_BOUCHAUD1990127,CTRW_metzler2000random}. They motivate effective descriptions in terms of trapping events, memory kernels, and random time changes.

\begin{figure}
    \centering
    \includegraphics[]{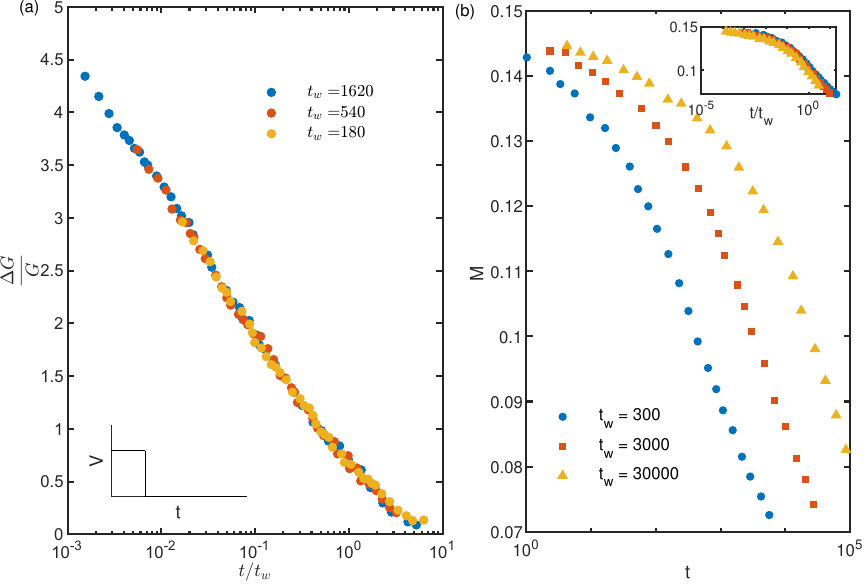}
    \caption{
One-time logarithmic relaxation and protocol-dependent aging are distinct experimental signatures.
(a) Conductance relaxation in an electron glass: the conductance $G$ of thin $\mathrm{In}_2\mathrm{O}_{3-x}$ films relaxes logarithmically after a sudden change in external stress, shown in the inset.
Data are reproduced from Fig.~2 of Ref.~\cite{Ovadyahu_PhysRevLett.92.066801}.
(b) Normalized thermoremanent magnetization $M$ in an $\mathrm{Ag}:\mathrm{Mn}\,2.6\%$ spin glass, showing slow relaxation and aging.
After field cooling and a waiting time $t_w$, the magnetic field was switched off at $t=0$ and the decay of $M$ was recorded.
Data are reproduced from Fig.~1 of arXiv: cond-mat/9607224.
Inset: scaling collapse as a function of $t/t_w$.
}
\label{fig:experiment}
\end{figure}

It is useful to distinguish slow or anomalous relaxation from aging. Slow dynamics refers to the functional form of relaxation or spreading, such as sublinear mean-square displacement in transport problems or logarithmic conductance relaxation in electron glasses. Aging (Fig.~\ref{fig:experiment}), by contrast, refers to the dependence of observables on the age or preparation history of the system. Let the system be prepared at $t=0$ and let a measurement be performed only after a waiting or aging time $t_a$. If, for the same observation time $t$, the measured response, conductance relaxation, displacement distribution, mean-square displacement or two-time correlation function changes with $t_a$, then the system exhibits aging. Thus, the result of an experiment depends not only on how long the system is observed, but also on when and how the observation begins.

From this perspective, aging is a property of the measurement protocol as much as a property of the underlying dynamics. The observation window $[t_a,t_a+t]$ probes a system that may already contain long-lived memory of its preparation and excitation history. In a stationary or equilibrated process, shifting the observation window would not change the statistics. In an aging process, the same shift changes the measured relaxation or transport properties. This explicit dependence on $t_a$ is the experimental origin of the two-time and nonstationary quantities introduced below.

A theory of aging dynamics in disordered systems should therefore answer three basic questions. First, what microscopic or mesoscopic mechanisms generate an extremely broad spectrum of relaxation times? Second, why do measured observables depend on the waiting time or preparation history? Third, why can finite-time measurements fail to represent equilibrium or ensemble behavior even when the macroscopic conditions are nominally fixed? In this work, continuous-time random walks are used as the minimal mechanism that addresses these questions in a transparent way. They are not assumed to be a universal microscopic model for all disordered systems, but they provide a useful first level in a broader stochastic-process framework that also includes Fokker--Planck, memory-kernel, Langevin and subordination descriptions.

\subsection{Waiting-Time Protocols and the Loss of Time-Translation Invariance}

The central role of the observation window makes the definition of time variables essential\cite{doi:10.1142/9789812819437_0006,Cugliandolo_Course7}. A typical aging measurement can be divided into three stages (Fig.~\ref{fig:experiment}). First, the system is prepared or perturbed at an initial time, which we set to be $t=0$. Second, the system evolves for an aging time, or waiting time, denoted by $t_a$. Third, a physical observable is recorded during the observation window $[t_a,t_a+t]$. Here, $t_a$ denotes the age of the system at the beginning of the measurement, while $t$ denotes the duration of the measurement window. The total elapsed time since preparation is therefore $t_a+t$.

The measured observable depends on the physical system under consideration. In an electron glass\cite{Pollak_2012}, it may be the conductance or the amplitude of a memory dip. In a spin glass, it may be a magnetization, a susceptibility, or a two-time correlation and response function. In a transport problem\cite{10.1063/1.1559676,doi:10.1142/S2010194515600071}, it may be a displacement, a mean-square displacement, or a time-averaged observable. The central point is that the same measurement performed over the same duration $t$ may give different results if it starts after different aging times $t_a$.

It is useful to distinguish the experimental aging time $t_a$ from the microscopic waiting time $\tau$ used in continuous-time random walk models. In this work, $t_a$ denotes the time elapsed between preparation and observation, whereas $\tau$ denotes the random time interval between two successive renewal events or jumps in the stochastic process. The former is part of the experimental protocol; the latter is a random variable of the theoretical model. Confusing these two time scales can obscure the distinction between experimentally observed aging and its stochastic-process representation.

If a system has reached equilibrium or a stationary state, the choice of the initial point of the observation window should not affect the statistical outcome. For a stationary process, shifting the observation window from $[0,t]$ to $[t_a,t_a+t]$ does not change the statistics of the measured quantities. In such a case, correlation functions and response functions depend only on time differences. For example, a stationary two-time correlation function satisfies

\begin{equation}
C(t_1,t_2)=C(t_2-t_1).
\end{equation}

This property expresses time-translation invariance: shifting both time arguments by the same amount leaves the correlation function unchanged.

In an aging system, this invariance is broken. Measured quantities depend explicitly on the age of the system at the beginning of observation. For a generic finite-window observable, one may write $\mathcal{O}(t_a,t),$ where the two arguments explicitly denote the age at which observation begins and the duration of the observation window. Aging is signaled by the fact that, for fixed $t$, this quantity changes when $t_a$ is varied. The symbol $\mathcal{O}$ is intentionally general: it may represent a conductance relaxation curve, a memory-dip amplitude, a response function, a correlation function, a displacement statistic, or a time-averaged quantity.

For two-time quantities, the same idea can be written more explicitly. Instead of depending only on the time difference, an aging correlation function must be treated as a genuine two-time object,

\begin{equation}
C=C(t_1,t_2),
\end{equation}
rather than as a function of $t_2-t_1$ alone. In terms of the experimental protocol, one may write

\begin{equation}
C_a(t_a,t)
\equiv
C(t_a,t_a+t).
\end{equation}
If $C_a(t_a,t)$ varies with $t_a$ at fixed $t$, the system exhibits aging. This explicit dependence on both the waiting time and the observation time is the mathematical expression of time-translation symmetry breaking.

Experimentally, this means that repeated measurements performed with the same observation duration but different waiting times need not collapse onto the same curve. In electron glasses, for example, conductance relaxation, memory-dip formation, and two-dip protocols reveal that the response depends on the previous gate-voltage or excitation history. In spin glasses, analogous behavior appears in waiting-time-dependent correlation and response functions. In both cases, the measured response is not determined solely by the elapsed time after the start of measurement; it also reflects the internal state prepared by the preceding history.

It is important to emphasize that aging does not simply mean that a system has evolved for a long time. If a system eventually reaches equilibrium, then its equilibrium statistical properties no longer depend on the preparation time, even though the system may have evolved for a long duration. Such a situation is not usually referred to as aging. Aging refers instead to the persistence of history dependence within the experimentally accessible time window. Thus, the relevant question is not whether the system is old in an absolute sense, but whether its measured observables still depend on its age.

\subsection{From Finite-Window Measurements to Stochastic Observables}

The finite-window notation introduced above provides the starting point for translating aging measurements into a stochastic-process description. In an experiment, one records a response during the interval $[t_a,t_a+t]$; trajectory-level analysis of such finite windows is central to distinguishing competing anomalous-diffusion mechanisms~\cite{review_Vilk_PhysRevResearch.4.033055,review_Metlzer_C4CP03465A}. In a stochastic model, the corresponding quantity is constructed from the segment of a coarse-grained trajectory over the same time window. The purpose of the model is therefore not to reproduce every microscopic detail of the material, but to specify the probability law governing the variables that control the measured finite-time response.

Let $X(t)$ denote such a coarse-grained stochastic variable. Depending on the system, $X(t)$ may represent a particle position, a local configuration, an occupation variable, a trap state, an energy-like variable, or a collective mode. A measured response over the window $[t_a,t_a+t]$ is then represented by a functional of the trajectory segment
\begin{equation}
\{X(s):t_a\leq s\leq t_a+t\}.
\end{equation}
For example, a transport-type measurement naturally involves the increment
\begin{equation}
\Delta X(t_a,t)=X(t_a+t)-X(t_a),
\end{equation}
whereas a two-time measurement may involve a product such as $X(t_a)X(t_a+t)$ or a corresponding correlation function. Other experimental observables can be represented by different functionals of the same trajectory segment.

This trajectory-level viewpoint makes the role of the observation window explicit. The variable $t_a$ determines where the sampling window begins along the stochastic trajectory, while $t$ determines the duration over which the trajectory is sampled. If the process is stationary, changing $t_a$ does not alter the statistics of such finite-window functionals. If the process is aging, the statistics of the same functional may change when the observation window is shifted to a different age.

In CTRW-type descriptions and related hopping or trapping models, the dynamics is often represented as a sequence of random sojourns separated by transition events, such as jumps, hops, or changes of coarse-grained state. A finite-window observable is then controlled not only by the duration of the observation window, but also by the event history accumulated before $t_a$ and by the number and timing of transition events occurring within $[t_a,t_a+t]$. It is useful to introduce the event-counting variable
\begin{equation}
N_a(t_a,t)=N(t_a+t)-N(t_a),
\end{equation}
where $N(t)$ denotes the total number of transition events up to time $t$. Unlike in a strictly renewal process, the statistics of these events need not be independent of the previous history; the waiting times may be age-dependent, time-dependent, or correlated with the evolving state. Consequently, finite-window displacement statistics, response statistics, and time-averaged observables may acquire an explicit dependence on $t_a$.

Thus, the stochastic formulation of aging consists of assigning probability laws to finite-window trajectory functionals without necessarily assuming a renewal structure. For transport-type problems, the next natural step is to describe not only moments of the increment, but its full probability distribution. This leads to the notion of an aging propagator, which provides a probability-level representation of finite-window evolution.

\subsection{The Aging Propagator as a Finite-Window Increment Distribution}

For transport-type problems, the most natural stochastic object is the probability distribution of the change accumulated during the observation window. If $X(t)$ denotes a coarse-grained stochastic variable, one may define the aging propagator as

\begin{equation}
G_a(x,t_a,t)
=
\left\langle
\delta\left[
x-\left(X(t_a+t)-X(t_a)\right)
\right]
\right\rangle .
\end{equation}
Here, $G_a(x,t_a,t)$ is the probability density for observing an increment $x$ during the window $[t_a,t_a+t]$. It is important to emphasize that this quantity is a theoretical probability object. In experiments, one may estimate related distributions from displacement histograms, conductance changes, response amplitudes or other finite-window measurements, but the ideal probability density itself belongs to the stochastic description.

For a process with stationary increments, the corresponding probability distribution depends only on the observation duration,

\begin{equation}
G_a(x,t_a,t)=G_a(x,t).
\end{equation}
In an aging process, by contrast,

\begin{equation}
G_a(x,t_a,t)\neq G_a(x,t),
\end{equation}
in general. This inequality is the probability-level expression of aging: the statistics of changes during the observation window depend on the age of the system at the beginning of that window.

The aging propagator is useful because it separates two questions that are often mixed in experimental discussions. The first question is how much change occurs during the observation time $t$. The second is whether the amount of change depends on when the observation begins. In stationary diffusion or relaxation, only the first question matters. In aging systems, both questions are essential.

Although the notation resembles that of diffusion theory, the same idea can be applied more broadly. If $X(t)$ represents a particle position, $G_a$ describes a displacement distribution. If $X(t)$ represents a coarse-grained configuration, energy, occupation or conductance-related variable, $G_a$ describes the distribution of changes in that quantity over the observation window. Thus, the propagator should be understood as a general probability representation of finite-time evolution, not only as a single-particle displacement density.

Historically, age-dependent probability descriptions of this type appear naturally in trap models, hopping models, and stochastic descriptions of glassy dynamics, where the observation window begins after a finite waiting time~\cite{Tauberium_klafter2011,10.1063/1.1559676,Godreche2001Statistics,PhysRevLett.90.104101}. In the present work, however, we use $G_a(x,t_a,t)$ mainly as a conceptual bridge: it translates experimental waiting-time dependence into a probability-level object that can later be computed within specific stochastic models.

\subsection{Why One-Time Scaling Does Not Identify Aging}

Before introducing specific stochastic mechanisms, it is useful to distinguish slow or anomalous dynamics from aging. Slow dynamics refers to the functional form of relaxation or spreading. For example, in a transport problem one often characterizes the spreading by the mean-square displacement,

\begin{equation}
\langle x^2(t)\rangle\sim t^\alpha .
\end{equation}
The case $\alpha=1$ corresponds to normal diffusion, while $0<\alpha<1$ corresponds to subdiffusion\cite{RevModPhys.95.031003,doi:10.1073/pnas.1016325108,PhysRevLett.106.048103,sub_PhysRevLett.96.098102} and $\alpha>1$ to superdiffusion\cite{PhysRevLett.84.3017,Reverey2015Superdiffusion}. In glassy relaxation\cite{Phillips_1996,PhysRevLett.71.1482}, slow dynamics may instead appear as logarithmic, stretched-exponential or power-law relaxation of a response function. These forms describe how an observable changes with time.

Aging is a different concept. Aging refers to the dependence of observables on the age or preparation history of the system. A system may display slow relaxation without aging if it eventually reaches a stationary regime within the experimental time window. Conversely, a system may show aging even when a single-time observable appears deceptively simple over a limited range of times. Thus, anomalous scaling and aging should not be identified with each other.

This distinction is essential for mechanism identification. The same scaling law may arise from different stochastic processes. Subdiffusive mean-square displacement, for example, can be generated by broad waiting times, correlated Gaussian processes, time-dependent diffusivities, spatial heterogeneity or confinement effects~\cite{FPE_PhysRevLett.103.180602,normal_C2SM25701G,normal_MEROZ20151}. These mechanisms may give similar one-time scaling exponents but differ in their two-time functions, response properties, finite-time averages and dependence on $t_a$.

For this reason, the exponent $\alpha$ alone is not sufficient to identify the underlying dynamics. A robust stochastic interpretation should compare several levels of statistics: age-dependent responses, two-time correlation functions, probability distributions of increments, finite-time averages and, when accessible, fluctuations between repeated realizations. The purpose of the stochastic-process framework is precisely to organize these observables and relate them to different classes of dynamics.

\subsection{Broad Time-Scale Distributions as an Effective Starting Point}

The experimental signatures discussed above suggest that a central ingredient of aging dynamics is the existence of a broad spectrum of time scales. As indicated earlier\cite{https://doi.org/10.1002/bbpc.19780820975,fractioonaldynamics_10.1063/1.1535007}, such broad time scales may arise from transition rates that depend exponentially on barriers, distances or collective rearrangement variables. Even moderate variations in the underlying random variable can generate relaxation times spanning many decades. This mechanism naturally leads to ultraslow relaxation, incomplete equilibration and sensitivity to the waiting history.

A minimal stochastic representation of this idea can be formulated in terms of broadly distributed waiting or trapping times. In this description, the dynamics is viewed as a sequence of intermittent periods of inactivity or slow evolution, interrupted by rearrangements, jumps or configuration changes. The statistical properties of these intervals encode the absence of a single characteristic time scale and provide a convenient route to aging-dependent propagators and correlation functions.

At this stage, this picture should be viewed as a conceptual mechanism rather than a complete microscopic theory. The elementary dynamical episodes may represent particle jumps in transport, escapes from traps in a glassy landscape, rearrangements between metastable configurations, or effective transitions between coarse-grained states. What matters for aging is not the microscopic identity of the episode, but the absence of a finite characteristic time scale when the relevant distribution of time scales is sufficiently broad.

If the relevant waiting-time distribution has a finite mean, the process can often approach an effective stationary long-time regime. In that case, the influence of the preparation time becomes weak. If the mean waiting time diverges, however, no finite characteristic time exists. The statistics within the observation window then depend on how long the system has already evolved before the measurement begins. This is a stochastic origin of aging in broad-time-scale models.

Continuous-time random walks and trap models provide familiar realizations of this idea in transport and glassy relaxation problems. In the present section, we use these models only to motivate the stochastic language. Their detailed aging observables, including aging propagators, forward waiting times, no-event probabilities and explicit propagator formulas, will be discussed later when the CTRW framework is developed systematically.

\section{From Additive Renewal Times to an Iterated Forward-Time Clock}
\label{sec:clock}
\label{sec:ctrw_log_aging}

With the experimental protocol and its physical scope fixed, we can introduce a minimal event-based model. The CTRW is useful precisely because it separates spatial motion from the clock that schedules events. It is not proposed as the microscopic dynamics of every aging material; it is the controlled setting in which the consequences of broad waiting times and an ultraslow clock can be derived explicitly.

The derivation proceeds in three steps. We first recall the stationary renewal CTRW, then show why an observation beginning at age $t_a$ is controlled by a forward rather than a freshly sampled waiting time, and finally impose the scale-covariant statistics that produce a logarithmic clock. This progression isolates the additional assumption responsible for log-aging behavior and prepares the observable predictions of the next section.

\begin{figure}
    \centering
    \includegraphics[width=\linewidth]{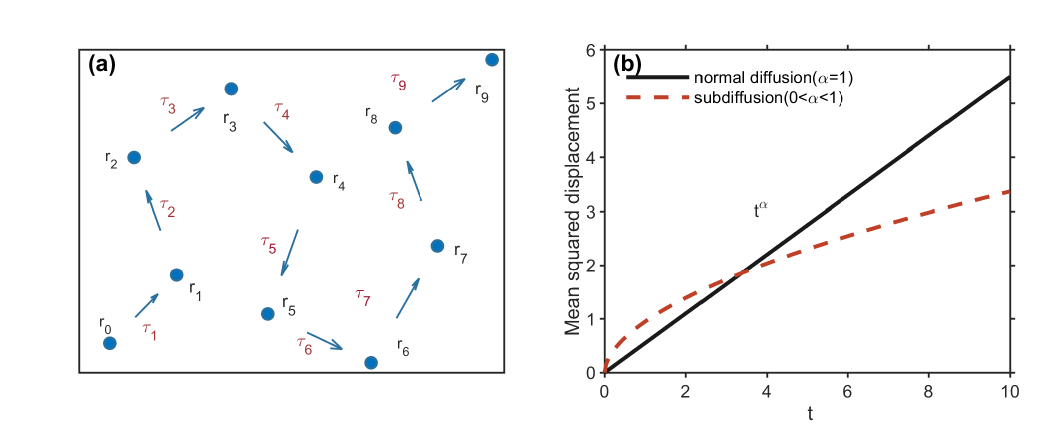}
   \caption{Standard decoupled CTRW: additive renewal times and power-law subdiffusion. 
(a) Random-walk dynamics in a heterogeneous energy landscape.
The particle alternates between mobile episodes and trapping periods induced by defects, traps, or energy barriers.
Escape from a trap occurs after a stochastic waiting time, after which the particle makes a transition to another state.
(b) The distribution of waiting times determines the long-time transport behavior.
When the distribution has a finite characteristic time, the transition rate is effectively constant at long times and the dynamics reduces to normal diffusion, with $\langle x^2(t)\rangle\propto t$.
When the distribution is broad and scale-free, no characteristic waiting time exists; the effective transition rate decreases with time and the dynamics becomes subdiffusive, with $\langle x^2(t)\rangle\propto t^\alpha$ for $\alpha<1$.
}
\label{fig:ctrw}
\end{figure}

\subsection{Standard Decoupled CTRW and Renewal Structure}
\label{sec_CTRW_normal}

The continuous-time random walk was introduced by Montroll and Weiss~\cite{CTRW_Montroll_10.1063/1.1704269} and later applied to anomalous transport in disordered systems by Scher and Montroll~\cite{Scher1975}. Modern CTRW treatments provide a systematic route from broad waiting times to anomalous transport laws~\cite{CTRW_METZLER20001,CTRW_metzler2000random,CTRW_Metzler_2004,CTRW_metzler2004restaurant}. A walker remains inactive for a random time and then performs a spatial jump; repetition of this waiting--jump sequence generates transport in physical time.

In the standard decoupled CTRW [Fig.~\ref{fig:ctrw}(a)], the waiting times $\{\tau_i\}$ are independent and identically distributed random variables drawn from a waiting-time density $\psi(\tau)$, while the jump lengths $\{\xi_i\}$ are drawn from a jump-length distribution $f(\xi)$. The decoupling assumption means that $\tau_i$ and $\xi_i$ are statistically independent. If $N(t)$ denotes the number of jumps up to time $t$, the position of the walker can be written as

\begin{equation}
X(t)=\sum_{i=1}^{N(t)}\xi_i .
\end{equation}
The event times are
\begin{equation}
T_n=\sum_{i=1}^{n}\tau_i ,
\end{equation}
and $N(t)$ is the largest integer $n$ such that $T_n\leq t$. Thus, the stochasticity of the process comes both from the jump lengths and from the random number of events completed within the observation time.

Let $P(x,t)$ be the propagator, namely the probability density of finding the particle at position $x$ at time $t$ after preparation at the origin. The survival probability, or probability that no jump occurs during a time interval $t$, is

\begin{equation}
\phi(t)=1-\int_0^t \psi(\tau)\,d\tau .
\end{equation}
The probability density that the $j$-th event occurs at time $t$ is the $j$-fold convolution of $\psi$, denoted by $\psi_j(t)$. Summing over all possible event numbers gives the standard renewal representation of the propagator. In Fourier--Laplace space, this leads to the Montroll--Weiss equation,

\begin{equation}
P(k,\lambda)
=
\frac{1-\tilde{\psi}(\lambda)}
{\lambda\left[1-\hat{f}(k)\tilde{\psi}(\lambda)\right]},
\end{equation}
where $\tilde{\psi}(\lambda)$ is the Laplace transform of $\psi(t)$ and $\hat{f}(k)$ is the Fourier transform of $f(\xi)$.

The long-time behavior is controlled by the tail of $\psi(\tau)$~\cite{CTRW_metzler2000random,CTRW_Metzler_2004,CTRW_shlesinger1974asymptotic}. If the mean waiting time is finite, $\langle \tau\rangle<\infty$, the number of events grows linearly with time and normal diffusion is recovered when the jump variance is finite. In contrast, if
\begin{equation}
\psi(\tau)\sim
\frac{A}{\tau^{1+\alpha}},
\qquad
0<\alpha<1,
\end{equation}
the mean waiting time diverges. Then the number of jumps grows sublinearly, typically as
\begin{equation}
\langle N(t)\rangle\sim t^\alpha ,
\end{equation}
and, for unbiased jumps with finite variance [Fig.~\ref{fig:ctrw}(b)],
\begin{equation}
\langle x^2(t)\rangle\sim t^\alpha .
\end{equation}
This is the standard CTRW mechanism for subdiffusion. It also produces ergodicity breaking, because a single finite trajectory cannot efficiently sample the ensemble when immobilization times have no characteristic scale.

The standard CTRW is renewal in operational time: after each event, the next waiting time is sampled anew from the same density $\psi(\tau)$. Aging does not require abandoning this renewal structure at the microscopic level. Rather, it enters when the observation starts after a finite age $t_a$. In that case, the first waiting time inside the observation window is not a newly sampled $\psi(\tau)$, but a forward recurrence time conditioned on the history before observation. This distinction provides the entry point to aging CTRW.

\subsection{Aging CTRW and Forward Waiting-Time Statistics}
\label{sec_CTRW_aging_forward}
\begin{figure}
    \centering
    \includegraphics[width=0.5\linewidth]{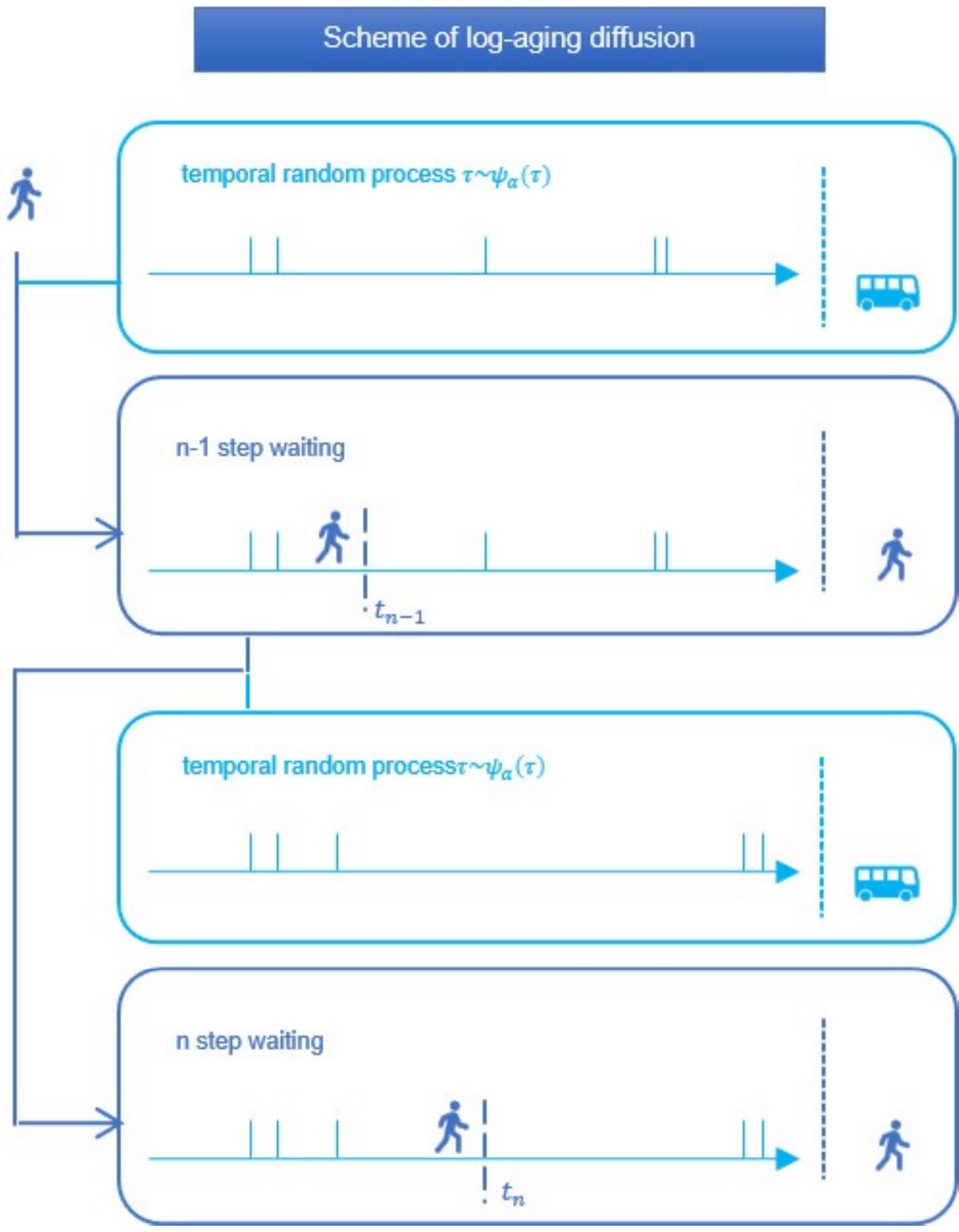}
\caption{
Forward-waiting-time construction and its iteration in the log-aging model. The bus analogy represents an underlying heavy-tailed renewal stream with $\psi_\alpha(\tau)\sim A\tau^{-1-\alpha}$. At an arbitrary observation age, an ordinary aged renewal process uses the residual or forward waiting time only until the next renewal. The log-aging construction makes an additional modeling step: at each event time $T_n$, it samples the next residual interval from the age-conditioned law and thereby generates a multiplier $Y_n=T_{n+1}/T_n$. In the scaling limit these multipliers are i.i.d. with density $g_\alpha(y)$, making the process additive in logarithmic time.}
\label{fig:forward}
\end{figure}

Consider a CTRW prepared at time $0$ but observed only after an aging time $t_a$. The measurement window is therefore $[t_a,t_a+t]$. The central object is the forward waiting time: the time from the beginning of observation to the next renewal event. We denote its probability density by $\psi_1(t,t_a)$. This density should not be confused with the microscopic waiting-time density $\psi(\tau)$ between successive renewal events. It is a residual-time distribution conditioned on the age of the process.

For heavy-tailed renewal processes and trap-like dynamics, the forward waiting-time distribution has the scaling form~\cite{review_Johannes_PhysRevX.4.011028,PhysRevLett.90.104101}

\begin{equation}
\psi_1(t,t_a)
=
\frac{\sin(\pi\alpha)}{\pi}
\frac{t_a^\alpha}{t^\alpha(t+t_a)},
\qquad
0<\alpha<1 .
\end{equation}
This expression encodes the inspection-paradox mechanism of aging: an observer who enters the process at a random time is more likely to find the system inside a long waiting interval than inside a short one. In physical terms, a system that has already aged for a long time is likely to be trapped in a long-lived metastable configuration.

The corresponding no-event probability during the observation time $t$ is

\begin{equation}
\phi_1(t,t_a)
=
1-\int_0^t \psi_1(\tau,t_a)\,d\tau .
\end{equation}
Its asymptotic behavior is

\begin{equation}
\phi_1(t,t_a)
\sim
\begin{cases}
1-\dfrac{\sin(\pi\alpha)}{\pi(1-\alpha)}
\left(\dfrac{t}{t_a}\right)^{1-\alpha},
& t\ll t_a, \\[1.2em]
\dfrac{\sin(\pi\alpha)}{\pi\alpha}
\left(\dfrac{t_a}{t}\right)^\alpha,
& t\gg t_a .
\end{cases}
\end{equation}
These expressions show explicitly that the statistics inside the measurement window depend on both $t_a$ and $t$. They therefore provide a realization of the time-translation symmetry breaking discussed in the previous sections.

The scale dependence on $t/t_a$ is the key feature that motivates the logarithmic-aging construction below. It is essential, however, to distinguish an ordinary aged renewal measurement from the process defined in this work. In the former, the forward law describes the first residual interval inside a chosen observation window; subsequent microscopic waits are again drawn from the same renewal density. In the log-aging process, the age-conditioned forward law is iterated after every event. This iteration is the additional model assumption that creates a multiplicative event clock.

\subsection{The Multiplicative Event Clock and Its Mellin Representation}
\label{sec_CTRW_log}

The logarithmic-aging CTRW [Fig.~\ref{fig:forward}] describes a regime in which the effective number of events grows logarithmically with physical time. Related microscopic and phenomenological routes to logarithmic aging have been discussed in complex disordered systems~\cite{PhysRevLett.134.197102,logaging_PhysRevLett.110.208301,review_chechkin2017ageing}. Let $T_n$ be the physical time of event $n$, let $R_n=T_{n+1}-T_n$ be the forward residual interval sampled at age $T_n$, and define the event-time multiplier

\begin{equation}
Y_n=\frac{T_{n+1}}{T_n}=1+\frac{R_n}{T_n}>1 .
\label{eq:multiplier_definition}
\end{equation}

For the forward law above, the normalized multiplier density is

\begin{equation}
g_\alpha(y)
=
\frac{\sin(\pi\alpha)}{\pi}
\frac{(y-1)^{-\alpha}}{y},
\qquad y>1 .
\label{eq:multiplier_density}
\end{equation}

Thus $T_n=t_0\prod_{j=0}^{n-1}Y_j$. The process is nonstationary in laboratory time but has independent, identically distributed increments $Z_j=\ln Y_j$ in logarithmic time. Since the relevant variable is a ratio of event times rather than a time difference, the Mellin transform becomes the natural analogue of the Laplace transform. For a function $F(z)$, we use

\begin{equation}
\mathcal{M}[F](p)
=
\int_0^\infty z^{p-1}F(z)\,dz .
\end{equation}

The shifted Mellin transform of the multiplier density, normalized at $p=0$, is

\begin{equation}
\widehat G_\alpha(p)
=
\mathcal M[g_\alpha](p+1)
=
\mathbb E[Y^p]
=
\frac{\Gamma(\alpha-p)}
{\Gamma(\alpha)\Gamma(1-p)},
\qquad p<\alpha ,
\label{eq:normalized_multiplier_mellin}
\end{equation}

so that $\widehat G_\alpha(0)=1$. The first two cumulants of $Z=\ln Y$ are

\begin{equation}
\kappa_1=\Psi(1)-\Psi(\alpha),
\qquad
\kappa_2=\Psi_1(\alpha)-\Psi_1(1),
\label{eq:log_clock_cumulants}
\end{equation}

where $\Psi$ and $\Psi_1$ are the digamma and trigamma functions. These positive constants determine both the mean and the variance of the event count at long times.

In this representation, the propagator of the log-aging CTRW takes a Montroll--Weiss-like form in Fourier--Mellin space,

\begin{equation}
\widehat P(k,p)
=
t_0^p
\frac{\widehat{G}(p)-1}
{p\left[1-\hat{f}(k)\widehat{G}(p)\right]},
\end{equation}
where $\widehat{G}(p)\equiv\widehat G_\alpha(p)$ is the normalized Mellin transform in Eq.~\eqref{eq:normalized_multiplier_mellin}, and $t_0$ is an initial microscopic time scale. This expression is structurally analogous to the Montroll--Weiss equation, with the Mellin variable $p$ playing the role of the Laplace variable for scale-covariant aging dynamics.

The analogy is useful but should not be overinterpreted. In the ordinary CTRW, Laplace transforms encode convolution in time and time-translation-invariant renewal statistics. In the log-aging case, Mellin transforms encode scale invariance in time ratios. Thus, the mathematical structure reflects a different physical symmetry: not stationarity in time differences, but aging covariance under rescaling of time.

A central consequence is that the first two event-count cumulants grow as\cite{logaging_PhysRevLett.110.208301}

\begin{equation}
\langle N(t)\rangle
\sim
\frac{\ln(t/t_0)}{\kappa_1},
\qquad
\operatorname{Var}N(t)
\sim
\frac{\kappa_2}{\kappa_1^3}\ln(t/t_0).
\label{eq:count_cumulants_main}
\end{equation}

For unbiased jumps with finite variance, this logarithmic operational time immediately gives
\begin{equation}
\langle x^2(t)\rangle
\propto
\frac{\ln(t/t_0)}{\kappa_1} .
\end{equation}
Thus, logarithmic aging can be viewed as ordinary spatial random walking subordinated to an ultraslow internal clock.

\paragraph{Connection to Sinai-type landscapes.}

Sinai-type diffusion provides an important physical motivation for logarithmic internal clocks~\cite{Sin83,le1999random}. In a one-dimensional random potential with Brownian spatial correlations,

\begin{equation}
\overline{(U(x)-U(y))^2}=2\sigma |x-y|,
\end{equation}
barriers grow with the length scale explored by the particle. Real-space renormalization eliminates small barriers and leaves progressively deeper valleys that dominate the long-time dynamics. This produces ultraslow motion and strong aging effects.

The connection to log-aging CTRW is therefore conceptual rather than exact. Sinai diffusion involves quenched spatial disorder, returns to previously visited valleys and sample-specific correlations, whereas the construction uses effective waiting events. Nevertheless, both descriptions share the idea that the relevant escape time grows with the history of exploration, leading to a logarithmically slow internal clock. This is why Sinai-type landscapes and Griffiths-like regions provide natural physical settings in which log-aging stochastic descriptions become relevant.

\section{Event Statistics, Rare Durations, and Ergodicity Breaking}
\label{sec:rare}
\label{sec_ergodicity}

The preceding section constructed the multiplicative clock at the event level. Before using its mean to derive transport, we must determine how broadly the clock fluctuates and whether its central and far-tail sectors control the same observables. This ordering places the event-count distribution immediately after the transition law that generates it and makes the later transport formulas consequences of a statistically characterized clock.

At leading order, ensemble transport and first-passage laws are largely controlled by the average growth of the operational time,

\begin{equation}
\langle N(t)\rangle
\sim
\frac{\ln(t/t_0)}{\kappa_1} .
\end{equation}
However, aging dynamics cannot be fully characterized by ensemble averages alone. In log-aging CTRW\cite{li2025rareeventinducedergodicitybreakinglogarithmic}, the number of events performed within a finite observation time is itself a broadly distributed random variable. Different trajectories may therefore experience very different dynamical histories even when they are observed over the same physical time interval. This leads to large trajectory-to-trajectory fluctuations, finite-time non-self-averaging and weak ergodicity breaking.

In this section, we analyze these effects through jump-number statistics, time-averaged observables, and the rare-event structure of the process. The main message is that typical trajectories and rare trajectories are controlled by different asymptotic regimes. Typical fluctuations are described by a log-normal central form, whereas rare low-activity histories generate algebraic event-time tails. The central sector controls ordinary count moments; the far tail controls positive duration moments and characteristic times. Rare events therefore do not dominate every statistical average, even though they can dominate the time scales most relevant to relaxation.

\begin{figure}
    \centering
    \includegraphics[]{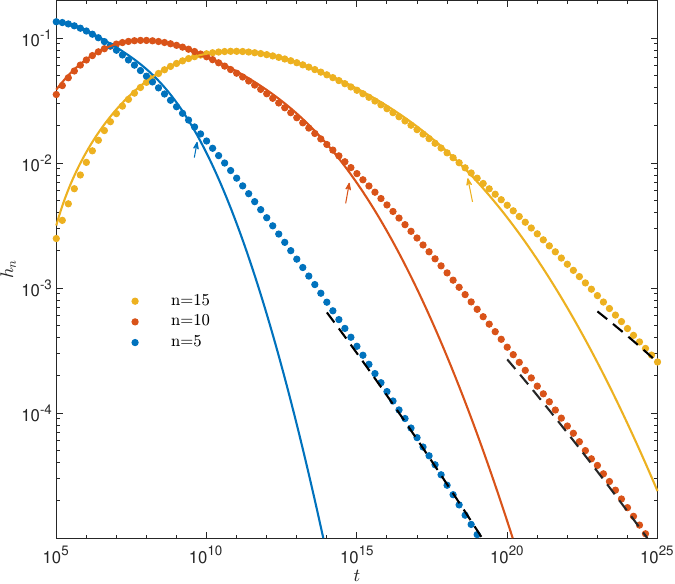}
\caption{
Central and far-tail regimes of the exact-count probability
$h_n(t)=\Pr[N(t)=n]$.
Symbols show numerical results for $n=5$, $10$, and $15$.
The colored solid curves are the central large-$n$ approximation, which is Gaussian in $\ln(t/t_0)$.
The black dashed lines are the fixed-$n$, long-time asymptote
$h_n(t)\propto t^{-\alpha}[\ln(t/t_0)]^n$.
The two fits therefore represent different orders of limits: typical clock fluctuations near the peak and rare low-activity histories in the far-time tail.
}
\label{fig:hn}
\end{figure}

\subsection{Exact-Count Probabilities and Event-Time Densities}
\label{sec_jump_number_statistics}

A natural starting point is the probability that the walker has performed exactly $n$ events by time $t$. We denote this probability by $h_n(t)$. To distinguish it from $h_n(t)$, we denote by $\rho_n(t)$ the probability density for the time of the $n$th event. Together, these quantities contain more dynamical information than the mean number of jumps because they resolve the full distribution of the internal clock.

For the log-aging process, the structure implies the recursion

\begin{equation}
\rho_n(t)
=
\int_0^t
\rho_{n-1}(\tau)\psi_1(t-\tau,\tau)\,d\tau ,
\end{equation}

\begin{equation}
h_n(t)
=
\int_0^t \rho_n(t_1)
\left[
\int_{t-t_1}^\infty
\psi_1(t_2,t_1)\,dt_2
\right]dt_1 .
\end{equation}
where $\psi_1(t_2,t_1)$ is the forward waiting-time density conditioned on the age $t_1$. The first relation gives the density of the $n$th event time, whereas the second gives the probability that exactly $n$ events have occurred by time $t$. To have $n$ events by time $t$, the $n$th event must occur at some earlier time $t_1$, and the following waiting interval must extend beyond $t$. Accordingly, $\int_0^\infty \rho_n(t)\,dt=1$ for each $n$, whereas $\sum_{n=0}^\infty h_n(t)=1$ for each $t$.

The distributions $\rho_n(t)$ and $h_n(t)$ have two important asymptotic regimes [Fig.~\ref{fig:hn}]. For large $n$, the typical fluctuations of the accumulated logarithmic time are governed by a central-limit-type mechanism in logarithmic variables. One obtains the log-normal asymptotic form

\begin{equation}
\rho_n(t)
\simeq
\frac{1}
{t\sqrt{2\pi n\kappa_2}}
\exp\left[
-
\frac{
\left(\ln(t/t_0)-n\kappa_1\right)^2
}
{2n\kappa_2}
\right],
\qquad
n\to\infty ,
\end{equation}

\begin{equation}
h_n(t)\simeq \kappa_1t\rho_n(t)\simeq\frac{\kappa_1}
{\sqrt{2\pi n\kappa_2}}
\exp\left[
-
\frac{
\left(\ln(t/t_0)-n\kappa_1\right)^2
}
{2n\kappa_2}
\right],
\qquad
n\to\infty ,
\end{equation}
where the factor $\kappa_1$ is the mean width of one renewal cell in logarithmic time. Equivalently, normalizing the Gaussian as a density in event number gives the inverse-renewal form with mean $\ln(t/t_0)/\kappa_1$ and variance $\kappa_2\ln(t/t_0)/\kappa_1^3$.

where $\kappa_1$ and $\kappa_2$ are the first two cumulants of the logarithmic waiting-time increment. This expression shows that the typical physical time after $n$ renewal events is multiplicative rather than additive. Equivalently, the logarithm of time is approximately Gaussian for large $n$.

In contrast, for fixed $n$ and long physical time, the same distribution displays a power-law tail\cite{logaging_PhysRevLett.110.208301},

\begin{equation}
\rho_n(t)
\simeq
\frac{
C_\alpha^n\left[\ln(t/t_0)\right]^{n-1}t_0^\alpha
}
{(n-1)!\,t^{1+\alpha}},
\qquad
t\to\infty .
\end{equation}
\begin{equation}
h_n(t)\simeq
\frac{
C_\alpha^{n+1}\left[\ln(t/t_0)\right]^n t_0^\alpha
}
{\alpha n!\,t^{\alpha}},
\qquad
t\to\infty .
\end{equation}
Here $C_\alpha=\sin(\pi\alpha)/\pi$. The extra power of $\ln(t/t_0)$ in $h_n$ arises because ``exactly $n$ events'' requires the $(n+1)$st event time to exceed $t$; it is not obtained by simply multiplying $\rho_n$ by $t$ in the fixed-$n$ tail.

This second asymptotic form describes rare trajectories in which the system becomes trapped for an exceptionally long time after only a small number of events. These trajectories are atypical in probability, but they have very long durations and can strongly influence time-dependent averages and characteristic time scales.

The coexistence of these two regimes is a central feature of log-aging dynamics. The log-normal part describes the typical accumulation of many multiplicative waiting factors, whereas the power-law tail reflects rare but extremely long trapping events. Therefore, the mean behavior of the internal clock does not fully characterize the process. One must also understand the full distribution $\rho_n(t)$ and $h_n(t)$.

\subsection{Inverse-Age Mean Event Rate}
\label{sec_density_renewal_events}

Another useful quantity is the time-dependent density of events,

\begin{equation}
M(t)
=
\sum_{n=1}^{\infty}\rho_n(t).
\end{equation}

The quantity $M(t)dt$ is the mean number of events occurring in the time window $[t,t+dt]$. It therefore plays the role of an event rate, or renewal density.

For log-aging diffusion, the long-time asymptotic behavior takes the form

\begin{equation}
M(t)
\sim
\frac{1}{\kappa_1 t},
\qquad
t\to\infty .
\end{equation}

A notable feature is that the leading decay is independent of the exponent $\alpha$. The parameter $\alpha$ affects the logarithmic clock through the coefficient $\kappa_1$ and the detailed fluctuation structure, but the leading activity decays as the universal scale-free form $1/t$.

This result has a transparent interpretation. In logarithmic aging, the process does not possess a characteristic time scale for activity. The rate of events decreases inversely with the age of the system. Thus, as the system becomes older, events become progressively rarer. This is consistent with the physical picture of a walker exploring deeper and deeper traps or valleys in an effective energy landscape.

The decay $M(t)\sim 1/t$ also distinguishes log-aging transport from standard subdiffusive CTRW. In a power-law CTRW, the renewal rate typically decays as a nontrivial power determined by the waiting-time exponent. In log-aging dynamics, the leading event density is scale-free in logarithmic time, reflecting the marginal nature of the aging clock.

\begin{figure}
    \centering
    \includegraphics[]{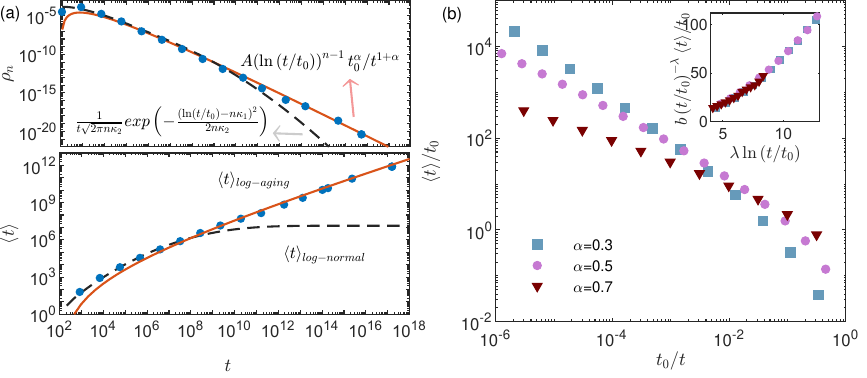}
    \caption{
Central and far-tail sectors of the $n$th-event-time density $\rho_n(t)$. (a) The central large-$n$ sector is asymptotically log-normal in physical time, whereas the fixed-$n$ far tail is $t^{-1-\alpha}[\ln(t/t_0)]^{n-1}$. The truncated first duration moment,
$\langle t\rangle_{\mathrm{tr}}(t_c)=\int_{t_0}^{t_c}t'\rho_n(t')\,dt'$,
exposes the tail because duration weighting amplifies rare, long-lived histories. (b) Scaling of the truncated duration moment for several $\alpha$. The figure diagnoses a duration-weighted rare-event sector; it should not be interpreted as evidence that the same tail dominates ordinary event-count moments or every low-frequency response in an arbitrary microscopic Griffiths system.}
\label{fig:rhon}
\end{figure}

\subsection{Central Log-Normal Event Times and Fixed-$n$ Algebraic Tails}
\label{sec_lognormal_powerlaw}

The two asymptotic forms of $\rho_n(t)$ reveal two complementary statistical mechanisms.

The first mechanism is typical multiplicative accumulation. If the physical time after $n$ events can be represented as a product of random scale factors, then

\begin{equation}
t_n
=
t_0
\prod_{j=1}^{n} q_j ,
\end{equation}
where $q_j$ are positive random variables, then

\begin{equation}
\ln(t_n/t_0)
=
\sum_{j=1}^{n}\ln q_j .
\end{equation}
Under broad conditions, the central limit theorem applied to the logarithmic increments $\ln q_j$ gives a Gaussian distribution for $\ln(t_n/t_0)$. This is the origin of the log-normal form of $\rho_n(t)$ in the typical regime.

The second mechanism is rare-event trapping. Even if most trajectories follow the log-normal scaling, a small fraction of trajectories may remain immobilized for extremely long times. These events generate the power-law tail

\begin{equation}
\rho_n(t)
\propto
t^{-(1+\alpha)}
\left[\ln(t/t_0)\right]^{n-1}.
\end{equation}
The power-law tail is especially important because moments involving physical time may be dominated by the far tail. For example, the characteristic time associated with fixed $n$,

\begin{equation}
\langle{t}\rangle
=
\int_0^\infty t\,\rho_n(t)\,dt ,
\end{equation}
diverges when the tail is sufficiently broad (Fig.~\ref{fig:rhon}). This divergence indicates that a typical time scale is not representative of the full dynamics. The process is then controlled by a competition between typical log-normal fluctuations and rare long trapping events.

This competition provides a microscopic route to weak ergodicity breaking. The system continues to evolve, so it is not permanently trapped in a strict sense. Nevertheless, finite-time trajectories remain dominated by long immobilization periods, and different trajectories sample the available state space very unevenly~\cite{Margolin2006}. As a result, time averages do not converge rapidly to ensemble averages, and their distributions remain broad even at long measurement times.

\begin{figure}
    \centering
    \includegraphics[]{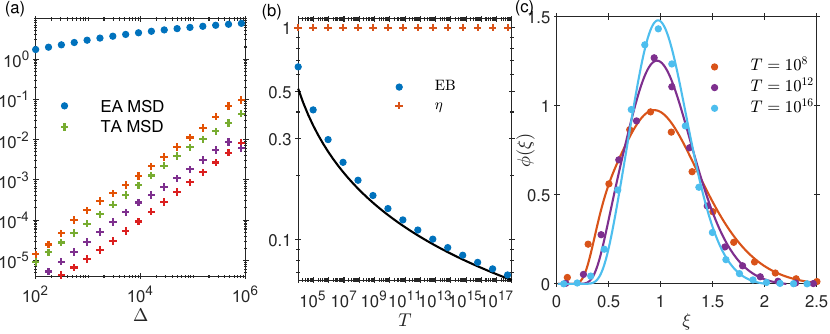}
    \caption{
Narrowing TA-MSD fluctuations do not restore ensemble--time equivalence. (a) The ensemble lag MSD and individual TA-MSD amplitudes correspond to different sampling protocols. (b) The conventional ergodicity-breaking parameter decreases as $1/\ln(T/t_0)$, so the relative trajectory-to-trajectory dispersion narrows ultraslowly, whereas the ensemble--time deviation parameter $\eta$ remains nonzero because the mean TA-MSD does not approach the ensemble lag MSD. (c) The normalized amplitude $\xi=\overline{\delta^2}/\langle\overline{\delta^2}\rangle$ is approximately described by the central clock sector near its peak; the enhanced small-$\xi$ sector records rare, nearly inactive trajectories.}
\label{fig:eb}
\end{figure}

\subsection{Time-Averaged Mean-Square Displacement}
\label{sec_tamsd}

Single-particle tracking experiments often measure time-averaged observables rather than ensemble averages. The most common example is the time-averaged mean-square displacement over a lag time $\Delta$ and total measurement time $T$ (Fig.~\ref{fig:eb}(a)); its trajectory-to-trajectory statistics provide a standard diagnostic of nonergodicity~\cite{PhysRevLett.93.190602,PhysRevLett.101.058101},

\begin{equation}
\overline{\delta^2}(\Delta,T)
=
\frac{1}{T-t_0-\Delta}
\int_{t_0}^{T-\Delta}
\left[
x(t'+\Delta)-x(t')
\right]^2
dt' .
\end{equation}

Writing the measurement interval explicitly from the preparation cutoff $t_0$ removes the otherwise spurious singularity at $t'=0$. For zero-mean independent jumps of variance $\sigma_\xi^2$, the clock gives the exact leading event increment

\begin{equation}
\left\langle N(t+\Delta)-N(t)\right\rangle
\simeq
\frac{1}{\kappa_1}
\ln\!\left(1+\frac{\Delta}{t}\right).
\label{eq:event_increment_tamsd}
\end{equation}
Consequently,
\begin{equation}
\left\langle\overline{\delta^2}(\Delta,T)\right\rangle
\simeq
\frac{\sigma_\xi^2}
{\kappa_1(T-t_0-\Delta)}
\int_{t_0}^{T-\Delta}
\ln\!\left(1+\frac{\Delta}{t}\right)dt .
\label{eq:mean_tamsd_exact_window}
\end{equation}
In the experimentally common sector $t_0\ll\Delta\ll T$, this becomes
\begin{equation}
\left\langle\overline{\delta^2}(\Delta,T)\right\rangle
\sim
\frac{\sigma_\xi^2\Delta}{\kappa_1 T}
\left[
\ln\!\left(\frac{T}{\Delta}\right)+1
\right].
\label{eq:mean_tamsd_asymptotic}
\end{equation}
By contrast, the ensemble displacement accumulated over a lag $\Delta$ immediately after preparation is
\begin{equation}
\left\langle
\left[X(t_0+\Delta)-X(t_0)\right]^2
\right\rangle
\simeq
\frac{\sigma_\xi^2}{\kappa_1}
\ln\!\left(1+\frac{\Delta}{t_0}\right).
\label{eq:ensemble_lag_msd}
\end{equation}

For an ergodic Brownian process, the time-averaged MSD converges to the ensemble-averaged MSD in the long-measurement-time limit. In log-aging dynamics, this convergence fails. The reason is that the number of jumps occurring inside each lag interval remains strongly random and depends on the aging history of the trajectory.

For log-aging diffusion, the ensemble-averaged MSD grows logarithmically,

\begin{equation*}
\langle x^2(t)\rangle
\propto
\frac{\ln(t/t_0)}{\kappa_1} .
\end{equation*}
The time-averaged MSD, by contrast, samples increments along a single finite trajectory. It is therefore controlled by the local activity within the measurement window. Since the event density decays as
\begin{equation}
M(t)\sim \frac{1}{\kappa_1t},
\end{equation}
different portions of the trajectory contribute very unevenly to the time average. Early portions of the trajectory are more active, whereas late portions are increasingly dominated by long waiting times.
Consequently, $\overline{\delta^2}(\Delta,T)$ remains a random variable at finite $T$, while its mean carries an explicit measurement-time dependence through Eq.~\eqref{eq:mean_tamsd_asymptotic}. The comparison of Eqs.~\eqref{eq:mean_tamsd_asymptotic} and \eqref{eq:ensemble_lag_msd}, rather than the scatter alone, provides the direct signature of weak ergodicity breaking.

A useful normalized variable is

\begin{equation}
\xi
=
\frac{
\overline{\delta^2}(\Delta,T)
}
{
\left\langle
\overline{\delta^2}(\Delta,T)
\right\rangle
}.
\end{equation}
The distribution of $\xi$ characterizes the amplitude scatter of time-averaged observables [Fig.~\ref{fig:eb}(c)]. For log-aging dynamics, in the typical regime, the multiplicative structure of the clock leads naturally to a log-normal-like distribution of amplitudes. Deviations from this typical form at small or large $\xi$ reflect rare trajectories with exceptionally few or exceptionally many events.

Thus, the time-averaged MSD reveals information that is invisible in the ensemble MSD. The ensemble MSD measures the average growth of the spatial spread, whereas the time-averaged MSD measures how individual trajectories distribute their activity over a finite measurement window. The discrepancy between the two is one of the clearest manifestations of ergodicity breaking in log-aging transport.

\subsection{Weak Ergodicity Breaking}
\label{sec_weak_ergodicity_breaking}

Weak ergodicity breaking refers to systems in which phase space is not partitioned into permanently disconnected components, but finite-time trajectories nevertheless fail to sample the accessible states in a representative manner~\cite{log_theory_bouchaud1992weak}. In CTRW-type models with broad waiting-time statistics, this phenomenon arises because immobilization periods can be comparable to, or even longer than, the measurement time.

Log-aging diffusion provides a particularly clear realization of this mechanism. As the system ages, the event rate decreases and the number of transition events grows only logarithmically with physical time. Consequently, increasing the observation time by a large multiplicative factor produces only an additive increase in the number of effective jumps. A trajectory observed up to time $T$ may therefore contain only a modest number of transition events, even when $T$ is macroscopically large.

To resolve the statistical nature of this breaking, we analyze fluctuations of the time-averaged MSD. The scatter of individual trajectories in Fig.~\ref{fig:eb}(b) shows that $\overline{\delta^2}$ remains a trajectory-dependent random variable. We quantify the amplitude scatter using the standard ergodicity-breaking parameter~\cite{PhysRevLett.101.058101}. For the normalized variable
\begin{equation}
\xi
=
\frac{
\overline{\delta^2}
}
{
\left\langle \overline{\delta^2} \right\rangle
},
\end{equation}
the EB parameter is
\begin{equation}
\mathrm{EB}
=
\frac{
\left\langle \xi^2 \right\rangle
-
\left\langle \xi \right\rangle^2
}
{
\left\langle \xi \right\rangle^2
}
\sim
\frac{\kappa_2}{\kappa_1\ln(T/t_0)}
\to0 .
\end{equation}
Here, $\kappa_1$ and $\kappa_2$ are model-dependent constants determined by the fluctuations of the event-counting process. As shown in Fig.~\ref{fig:eb}(b), $\mathrm{EB}$ decays logarithmically, proportional to $1/\ln(T/t_0)$. This indicates that the central distribution of time-averaged amplitudes slowly narrows with the measurement window. In this restricted sense, independent long-time measurements become progressively more reproducible. This behavior contrasts with standard scale-free CTRW subdiffusion~\cite{review_Metlzer_C4CP03465A}, where the EB parameter approaches a nonzero plateau.

However, the narrowing of the central distribution should not be interpreted as restored ergodicity. The conventional EB parameter measures the relative trajectory-to-trajectory dispersion of the TA-MSD around its own mean; it does not test whether that mean converges to the ensemble lag MSD. We therefore use $\eta$ only as an ensemble--time deviation parameter. The limits $\mathrm{EB}\to0$ and $\eta\not\to0$ state that trajectories become relatively reproducible while their time average remains inequivalent to the ensemble observable.

To expose this distinction, we introduce the deviation parameter
\begin{equation}
\delta_{\rm ens}^2(\Delta;t_0)
\equiv
\left\langle
\left[X(t_0+\Delta)-X(t_0)\right]^2
\right\rangle
\end{equation}
and define
\begin{equation}
\eta(\Delta,T;t_0)
=
\frac{
\left\langle
\left[
\overline{\delta^2}(\Delta,T)
-
\delta_{\rm ens}^2(\Delta;t_0)
\right]^2
\right\rangle
}
{
\left[\delta_{\rm ens}^2(\Delta;t_0)\right]^2
}.
\end{equation}
For a truly ergodic process, the time-averaged MSD converges to the ensemble-averaged MSD, and hence $\eta\to0$. In log-aging diffusion, by contrast, $\eta$ approaches a finite nonzero value, as shown in Fig.~\ref{fig:eb}(b). Thus, even though time-averaged trajectories become mutually more reproducible, they remain systematically displaced from the ensemble average.

The simultaneous behavior $\mathrm{EB}\to0$ and $\eta\not\to0$ reveals the subtle nature of weak ergodicity breaking in log-aging dynamics. The decrease of EB is a central-sector statement: the relative count fluctuations scale as $\operatorname{Var}N/\langle N\rangle^2\simeq\kappa_2/[\kappa_1\ln(T/t_0)]$. The persistent mismatch is instead caused by the nonstationary allocation of activity over the measurement window, while rare low-activity histories govern the extreme small-amplitude sector and positive duration moments. Thus no single class of rare trajectories should be said to dominate every ensemble observable. The process becomes increasingly reproducible within the time-averaged sector, while its mean time average remains inequivalent to the ensemble lag observable.

This distinction is important for experimental interpretation. A vanishing EB parameter alone may falsely suggest equilibration or restored ergodicity. In log-aging systems, trajectory-to-trajectory fluctuations can narrow while the time average still converges to a value different from the ensemble average. Standard protocols that use decreasing scatter as a proxy for equilibration may therefore underestimate global relaxation time scales, potentially by orders of magnitude.

\subsection{Large-Deviation Interpretation}
\label{sec_large_deviation}

The coexistence of typical log-normal behavior and rare power-law tails can be organized using large-deviation theory\cite{TOUCHETTE20091}. Let

\begin{equation}
z
=
\frac{\ln(t/t_0)}{n}
\end{equation}
be the logarithmic time accumulated per renewal event. For large $n$, the jump-number statistics may be expressed in the form

\begin{equation}
\rho_n(t)
\asymp
\frac{1}{nt}
\exp\left[-n I(z)\right],
\label{eq:ldp_rho_corrected}
\end{equation}
where $I(z)$ is a rate function and $\asymp$ denotes equivalence at exponential order in $n$. The factor $1/(nt)$ is the change-of-variables Jacobian from the density of $z=\ln(t/t_0)/n$ to the density in physical time. Subexponential saddle-point prefactors, which produce the $1/\sqrt{n}$ normalization in the central sector, are suppressed in Eq.~\eqref{eq:ldp_rho_corrected}.

Near its minimum, the rate function is approximately quadratic,

\begin{equation}
I(z)
\simeq
\frac{(z-\kappa_1)^2}{2\kappa_2}.
\end{equation}
This quadratic approximation reproduces the log-normal central regime,

\begin{equation}
\rho_n(t)
\simeq
\frac{1}
{t\sqrt{2\pi n\kappa_2}}
\exp\left[
-
\frac{
\left(\ln(t/t_0)-n\kappa_1\right)^2
}
{2n\kappa_2}
\right].
\end{equation}
However, the far tail of the rate function is not captured by the quadratic approximation. In the large-$z$ regime, rare trajectories correspond to exceptionally large logarithmic waiting increments. If the rate function develops an approximately linear branch, then the probability density acquires an exponential tail in logarithmic time. Since exponential decay in $\ln t$ corresponds to a power law in $t$, this mechanism produces the rare-event tail

\begin{equation}
\rho_n(t)
\propto
t^{-(1+\alpha)}
\left[\ln(t/t_0)\right]^{n-1}.
\end{equation}
This large-deviation viewpoint clarifies why both log-normal and power-law forms appear in the same process. The log-normal form describes fluctuations near the minimum of the rate function and therefore represents typical trajectories. The power-law form comes from the far tail of the rate function and represents rare trajectories. In finite-time experiments and simulations, the number of renewal events may remain small because the clock grows logarithmically. Therefore, the observed distribution can receive significant contributions from both the central and tail regions.

This is an important distinction from ordinary central-limit behavior. In many processes, rare tails become irrelevant for typical observables at long times. In log-aging dynamics, the growth of operational time is so slow that rare events may remain visible and dynamically relevant over experimentally accessible time scales.

\subsection{Why One-Time Probabilities Are Not Enough}
\label{sec_probability_not_enough}

The analysis above illustrates a broader lesson. One-time probability distributions provide only a snapshot of a non-equilibrium process. They describe where the system is at a given time, but they do not fully specify how the system arrived there or how dynamical activity is organized along individual trajectories.

For stationary ergodic processes, this limitation is often not severe. Time averages, ensemble averages and long-time probabilities are tightly connected. For aging and nonstationary processes, however, the temporal organization of events carries essential information. Two processes may have similar one-time probability densities but very different structures, time-averaged observables and rare-event statistics.

Log-aging diffusion is a clear example. The ensemble MSD reflects the average logarithmic growth of the number of jumps. Weak ergodicity breaking, however, requires the full distribution $\rho_n(t)$, the event density $M(t)$, and the temporal allocation of events along finite trajectories. Rare long waiting events control the extreme low-activity sector and duration-weighted quantities, but the central clock sector controls ordinary count moments. Therefore, probability densities at fixed times are not sufficient. One needs the stochastic-process viewpoint, which specifies not only distributions, but also the temporal rules by which trajectories evolve.

This perspective also explains why rare events are not merely corrections. In log-aging dynamics, rare long trapping intervals shape the measurable time averages, broaden trajectory-to-trajectory fluctuations and generate power-law tails in time-related statistics. The process is therefore governed by a hierarchy of dynamical information: ensemble propagators describe average spreading, jump-number distributions describe the internal clock, and time-averaged observables reveal the degree of ergodicity breaking.

The results of this section complete the event-statistical characterization of log-aging CTRW. Section~II constructed the multiplicative clock; the present section has shown how that clock separates central count fluctuations from rare duration tails and produces an enduring mismatch between ensemble and time averages. Section~IV now converts the same clock statistics into transport, response, transit-time, and first-passage predictions.

\section{Observable Consequences of the Multiplicative Clock: Transport, Response, and First Passage}
\label{sec:transport}
\label{sec:observables_log_aging}

The value of the CTRW construction lies in the predictions it makes for measurable quantities. This section translates the logarithmic event clock into displacement, response, transit-time, and target-survival laws. The central physical point is that the spatial dynamics may be ordinary in operational time while every laboratory-time observable is slowed by the same history-dependent clock.

We first derive ensemble transport and first-passage observables, including the generalized Einstein relation. We then use transit-time and target problems to show how boundaries and rare successful trajectories probe the clock differently from bulk spreading. The event-statistics module that precedes this section supplies the fluctuations and rare-duration sector that cannot be inferred from these ensemble laws alone.

\begin{figure}
    \centering
    \includegraphics[width=0.8\linewidth]{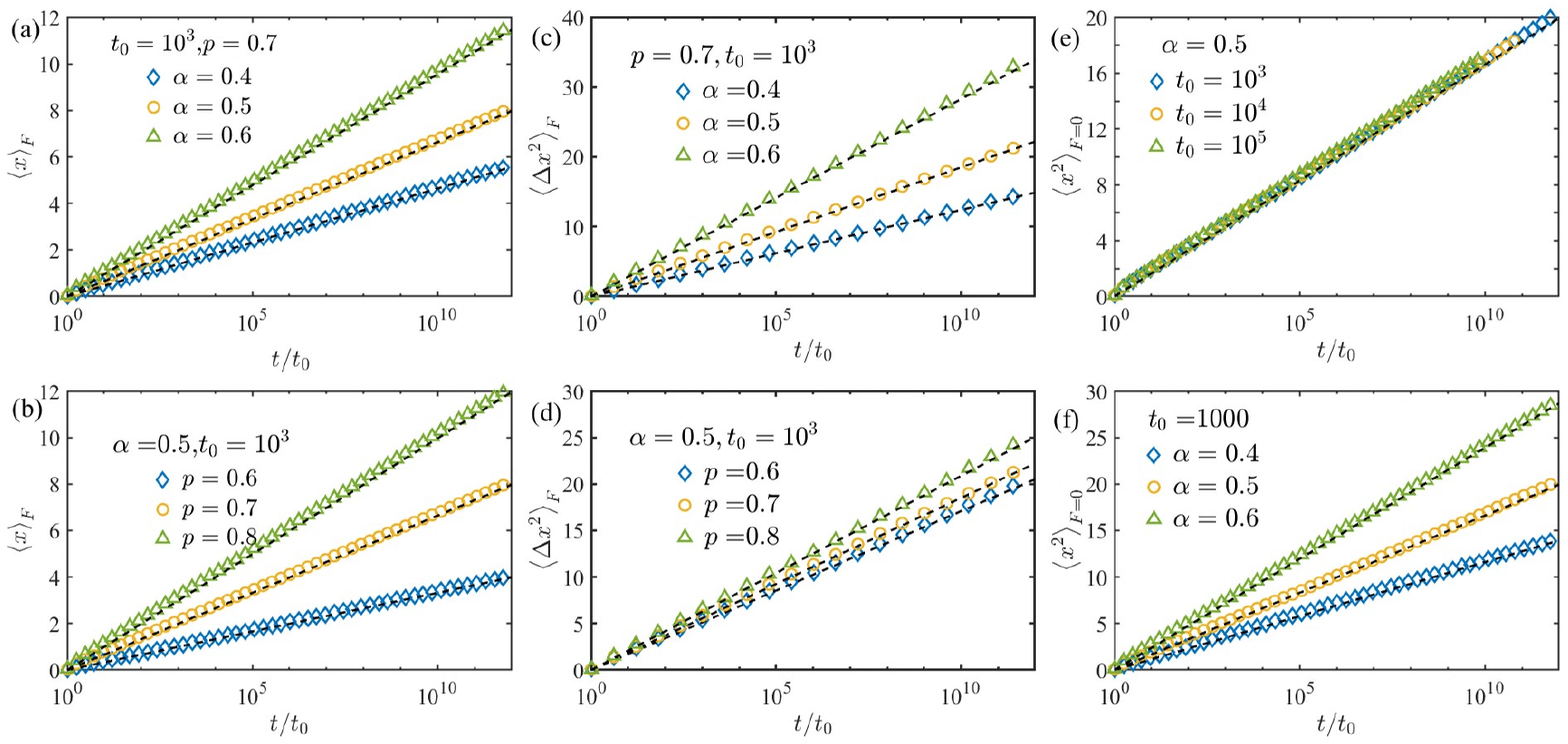}
    \caption{
Logarithmic dynamical scaling in log-aging diffusion. 
Panels (a)--(d) show the mean displacement and displacement variance in the biased case, while panels (e)--(f) show the mean-square displacement in the unbiased case. 
Both cases demonstrate the slow logarithmic growth characteristic of transport governed by log-aging waiting-time statistics.
}
\label{fig:displacement}
\end{figure}

\subsection{Displacement Moments and the Logarithmic Internal Clock}
\label{sec:log_moments}

Let $\widehat P(k,p)$ be the Fourier--Mellin propagator of the log-aging CTRW,

\begin{equation*}
\widehat P(k,p)
=
t_0^p
\frac{\widehat{G}(p)-1}
{p\left[1-\hat{f}(k)\widehat{G}(p)\right]},
\end{equation*}
where $\hat{f}(k)$ is the Fourier transform of the jump-length distribution and $\widehat{G}(p)$ is the Mellin transform of the aging  scaling kernel. The moments of displacement are obtained by differentiating the propagator with respect to $k$,

\begin{equation}
\langle x^n(p)\rangle
=
(-i)^n
\left.
\frac{\partial^n P(k,p)}{\partial k^n}
\right|_{k=0}.
\end{equation}
Assume that the jump-length distribution has finite lower moments and admits the small-$k$ expansion

\begin{equation}
\hat{f}(k)
\simeq
1+i\langle \xi\rangle k
-\frac{1}{2}\langle \xi^2\rangle k^2
+o(k^2).
\end{equation}
Substitution into the Fourier--Mellin propagator gives the first two moments in Mellin space,

\begin{equation}
\langle x(p)\rangle
=
t_0^p
\frac{\tilde{G}(p)}
{p\left[1-\tilde{G}(p)\right]}
\langle \xi\rangle ,
\end{equation}
and

\begin{equation}
\langle x^2(p)\rangle
=
t_0^p
\frac{\tilde{G}(p)}
{p\left[1-\tilde{G}(p)\right]}
\langle \xi^2\rangle
+
2t_0^p
\frac{\tilde{G}^2(p)}
{p\left[1-\tilde{G}(p)\right]^2}
\langle \xi\rangle^2 .
\end{equation}
These expressions separate two contributions. The first term in $\langle x^2(p)\rangle$ comes from the variance accumulated by independent jumps. The second term is present only when the jump distribution is biased and reflects the dispersion associated with the fluctuating number of jumps.

The long-time behavior follows from the small-$p$ expansion of the Mellin kernel. If

\begin{equation}
\tilde{G}(p)
=
1+\kappa_1 p+o(p),
\end{equation}
then the mean number of renewal events grows as

\begin{equation}
\langle N(t)\rangle
\sim
\frac{\ln(t/t_0)}{\kappa_1} .
\end{equation}
For unbiased jumps, $\langle \xi\rangle=0$, and the mean displacement vanishes,

\begin{equation}
\langle x(t)\rangle=0.
\end{equation}
The mean-square displacement is then proportional to the mean number of jumps(Fig.~\ref{fig:displacement}),

\begin{equation}
\langle x^2(t)\rangle
\simeq
\langle \xi^2\rangle \langle N(t)\rangle ,
\end{equation}
so that

\begin{equation}
\langle x^2(t)\rangle
\propto
\frac{\ln(t/t_0)}{\kappa_1}
\qquad
\text{unbiased log-aging CTRW.}
\end{equation}
This logarithmic growth is the simplest transport signature of the log-aging clock. It should be distinguished from standard subdiffusive CTRW, where $\langle x^2(t)\rangle\sim t^\alpha$~\cite{A.V.Chechkin_2003}. In the present case, the spatial random walk in operational time is ordinary, but the conversion from operational time to physical time is logarithmically slow.

\subsection{Biased Transport and the Generalized Einstein Relation}
\label{sec:log_einstein}

We next consider biased transport under a weak constant force $F$ (Fig.~\ref{fig:displacement}). The force changes the jump-length distribution by making jumps along the field slightly more probable than jumps against it. For a lattice model with spacing $a$ in a thermal environment, local detailed balance gives, to linear order in $F$,

\begin{equation}
\langle \xi\rangle_F
\propto
\frac{Fa^2}{2k_B T},
\end{equation}
while the unbiased second moment remains of order

\begin{equation}
\langle \xi^2\rangle_{F=0}\sim a^2 .
\end{equation}
Using the logarithmic growth of the mean number of jumps, the mean displacement under weak bias becomes

\begin{equation}
\langle x(t)\rangle_F
\simeq
\langle \xi\rangle_F \langle N(t)\rangle ,
\end{equation}
or

\begin{equation}
\langle x(t)\rangle_F
\propto
\frac{F a^2}{2k_B T}
\frac{\ln(t/t_0)}{\kappa_1}
\qquad
\text{weakly biased log-aging CTRW.}
\end{equation}
Comparing this expression with the unbiased mean-square displacement,

\begin{equation}
\langle x^2(t)\rangle_{F=0}
\propto
a^2
\frac{\ln(t/t_0)}{\kappa_1},
\end{equation}
one obtains the generalized Einstein relation
\begin{equation}
\langle x(t)\rangle_F
=
\frac{F}{2k_B T}
\langle x^2(t)\rangle_{F=0}
\end{equation}
up to model-dependent conventions for the lattice spacing and jump probabilities.

The persistence of this relation has a clear physical meaning. Aging modifies the temporal pacing of jumps through the logarithmic internal clock, but the local relation between bias and fluctuations at each jump remains thermal. Thus, both drift and spreading are slowed by the same aging clock. The logarithmic time dependence cancels in their ratio, leaving an Einstein-type proportionality between response and fluctuations.
This result provides a useful diagnostic. If both the unbiased spreading and biased drift are logarithmic in time and satisfy the above proportionality, then the dominant aging effect can be interpreted as a slowing down of the operational clock rather than a breakdown of local jump statistics.

\subsection{Transit-Time Dispersion in Log-Aging Transport}
\label{sec:log_transit}

First-passage observables probe aspects of aging dynamics that are not visible in displacement moments alone. A classical example is the transit-time problem in disordered transport~\cite{Scher1975,CTRW_transport_PhysRevB.16.4719}. Consider carriers moving across a sample of length $L$ under a weak bias. The experimentally measured current can be related to the rate of change of the mean displacement,

\begin{equation}
I(t)
\propto
\frac{d}{dt}\langle x(t)\rangle_F .
\end{equation}
Using the logarithmic drift law,

\begin{equation}
\langle x(t)\rangle_F
\propto
\frac{\ln(t/t_0)}{\kappa_1},
\end{equation}
one obtains the early-time current decay
\begin{equation}
I(t)
\propto
\frac{1}{t}
\qquad
t\ll t_{\mathrm{tr}},
\end{equation}
up to prefactors determined by the force, temperature and microscopic jump scale. If the logarithmic variable is written with a regularized microscopic time scale, additional slowly varying factors may appear depending on the precise definition of the current and aging kernel.

At longer times, the finite sample length and absorbing boundary at $L$ become important. The current is then governed by the first-passage probability to the boundary rather than by the bulk drift alone. In log-aging transport, the boundary effect introduces an additional logarithmic suppression. The asymptotic current takes the form

\begin{equation}
I(t)
\propto
\frac{L}{t[\ln(t/t_0)]^2}
\qquad
t\gg t_{\mathrm{tr}} .
\end{equation}

This crossover is qualitatively different from the power-law transit-time dispersion of standard subdiffusive CTRW. In ordinary dispersive transport, the current typically exhibits algebraic regimes controlled by the exponent $\alpha$. In log-aging transport, the dominant decay is close to $1/t$ but modified by logarithmic corrections. Therefore, a crossover from an early logarithmic-clock response to a late-time $t^{-1}[\ln(t/t_0)]^{-2}$ decay provides a direct transport signature of logarithmic aging.

The physical origin of the additional logarithmic factor is the competition between two effects. The walker explores space only through a slowly increasing number of jumps, while the absorbing boundary selects rare trajectories that have reached the sample edge. As time increases, the probability current is therefore controlled not only by the logarithmic operational time but also by the depletion of trajectories that have already been absorbed.

\begin{figure}
    \centering
    \includegraphics[]{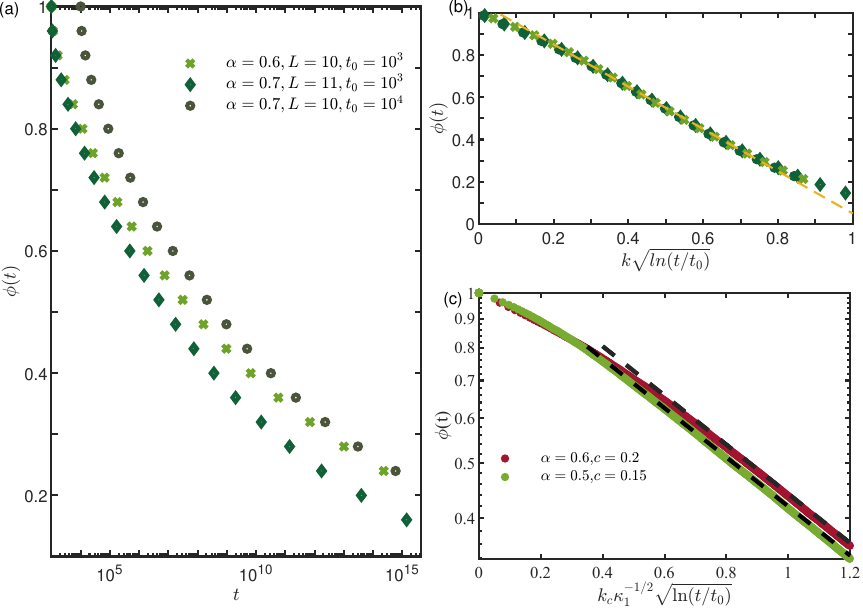}
   \caption{
Survival probability on a one-dimensional lattice. 
(a) Log-aging diffusion exhibits an ultraslow decay of the survival probability for one defect. 
(b) The survival dynamics of log-aging diffusion is in good agreement with the theoretical prediction, indicated by the solid line. 
(c)The survival probability  surrounded by defects of a constant concentration.
}
\label{fig:survival}
\end{figure}

\subsection{Target Problems and First-Passage Statistics}
\label{sec:log_target}

A second class of first-passage observables is provided by target problems, a standard framework for reactive search and survival phenomena~\cite{doi:10.1142/9104,network_nature,RevModPhys.83.81,Guerin2016}. A target problem asks how long it takes for one or more searchers, defects, or reactive particles to find a target. Such observables are sensitive to rare trajectories and spatial exploration, and they therefore provide a stringent test of log-aging dynamics.

Consider first a single target surrounded by one mobile defect in a lattice of volume $V$. The survival probability of the target can be written as

\begin{equation}
\Phi(t)
=
1-\frac{1}{V}
\sum_{r\neq 0}
\int_0^t F(r,\tau)\,d\tau ,
\end{equation}
where $F(r,t)$ is the first-passage density to the target for a defect starting at position $r$. Equivalently, one may write

\begin{equation}
\Phi(t)
=
1-\frac{1}{V}
\int_0^t I(\tau)\,d\tau ,
\end{equation}
where $I(t)$ is the rate at which the defect visits new sites.

The log-aging CTRW framework enters through the statistics of the number of jumps. In Mellin space, the rate of visiting new sites can be represented schematically as

\begin{equation}
\widehat I(p)
=
t_0^{p-1}
\sum_{r\neq 0}
\sum_n
F_n(r)
\left[\widehat{G}(p-1)\right]^n ,
\end{equation}
where $F_n(r)$ denotes the first-passage probability after $n$ operational steps. This expression separates the spatial first-passage problem in operational time from the logarithmic clock that maps operational steps to physical time.

The resulting asymptotic behavior depends on spatial dimension. In three dimensions, random walks are transient, and the number of newly visited sites grows linearly with the number of steps in operational time. Since the operational time grows as $\ln(t/t_0)$, one obtains

\begin{equation}
\int_0^t I(\tau)\,d\tau
\propto
\frac{\ln(t/t_0)}{\kappa_1}
\qquad
\text{in 3D}.
\end{equation}

Therefore, for one defect,

\begin{equation}
\Phi(t)
\simeq
1-
\frac{C_{3D}}{V \kappa_1}
\ln(t/t_0)
\qquad
\text{single defect, 3D}.
\end{equation}

In one dimension, the number of distinct sites visited by a random walk after $n$ steps grows as $n^{1/2}$. Substituting the logarithmic operational time gives

\begin{equation}
\int_0^t I(\tau)\,d\tau
\propto
\sqrt{\frac{\ln(t/t_0)}{\kappa_1}} ,
\end{equation}
and hence

\begin{equation}
\Phi(t)
\simeq
1-
\frac{C_{1D}}{V}
\sqrt{
\frac{\ln(t/t_0)}{\kappa_1}
}
\qquad
\text{single defect, 1D}.
\end{equation}
Here $C_{3D}$ and $C_{1D}$ are constants determined by the lattice geometry and microscopic jump statistics.

For a finite concentration $c$ of independent defects, the survival probability becomes exponential in the explored volume. In the thermodynamic limit, this gives

\begin{equation}
\Phi_c(t)
=
\exp\left[
-c
\int_0^t I(\tau)\,d\tau
\right].
\end{equation}
Thus,

\begin{equation}
\Phi_c(t)
\simeq
\exp\left[
-\frac{cA}{\kappa_1}
\ln(t/t_0)
\right]
\qquad
\text{finite defect concentration, 3D},
\end{equation}
and

\begin{equation}
\Phi_c(t)
\simeq
\exp\left[
-cB
\sqrt{
\frac{\ln(t/t_0)}{\kappa_1}
}
\right]
\qquad
\text{finite defect concentration, 1D}.
\end{equation}
These results show that target survival in log-aging transport is much slower than the stretched-exponential forms commonly encountered in standard subdiffusive or glassy search problems. The logarithmic clock strongly suppresses the rate at which space is explored, thereby increasing target survival over long times.

\subsection{Operational-Time Averaging over the Random Event Count}
\label{sec:observable_subordination}

The preceding results share a common structure. They can all be understood by separating spatial exploration in operational time from the random clock that relates operational time to physical time, a standard subordination viewpoint in anomalous transport~\cite{review_Klfter_10.1063/1.1860472,CTRWandFPE_sandev2018continuous}. If an observable has a known dependence on the number of steps $n$,

\begin{equation}
A=A(n),
\end{equation}
then the corresponding physical-time observable is obtained by averaging over the random event number,

\begin{equation}
A(t)
=
\sum_{n=0}^{\infty}
A(n)h_n(t),
\end{equation}
where $h_n(t)=\Pr[N(t)=n]$ is the probability that exactly $n$ events have occurred by time $t$.

In the leading log-aging regime, the dominant effect of the clock is captured by the replacement

\begin{equation}
n
\rightarrow
\frac{\ln(t/t_0)}{\kappa_1} .
\end{equation}
This replacement explains the logarithmic forms derived above:

\begin{equation}
\langle x^2(n)\rangle\sim n
\quad
\Longrightarrow
\quad
\langle x^2(t)\rangle
\sim
\frac{\ln(t/t_0)}{\kappa_1} ,
\end{equation}

\begin{equation}
S_{3D}(n)\sim n
\quad
\Longrightarrow
\quad
S_{3D}(t)
\sim
\frac{\ln(t/t_0)}{\kappa_1} ,
\end{equation}
and

\begin{equation}
S_{1D}(n)\sim n^{1/2}
\quad
\Longrightarrow
\quad
S_{1D}(t)
\sim
\sqrt{
\frac{\ln(t/t_0)}{\kappa_1}
}.
\end{equation}

Here $S_d(n)$ denotes the number of distinct sites visited after $n$ operational steps in dimension $d$. This subordination structure shows why displacement moments, first-passage quantities and target survival probabilities all acquire logarithmic time dependence.

At the same time, this leading replacement does not capture the full finite-time statistics. The distribution of $N(t)$, its central fluctuations, and the fixed-event rare-duration tail affect different observables in different ways. Section~\ref{sec_ergodicity} developed those clock statistics before the present transport module; the leading mean-clock replacement used here should be read in conjunction with that event-level analysis.

\subsection{Observable Signatures of Logarithmic Aging}
\label{sec:observable_signatures}

The results above identify several experimentally or numerically accessible signatures of log-aging CTRW dynamics.

First, the mean-square displacement grows logarithmically rather than as a power law:

\begin{equation*}
\langle x^2(t)\rangle
\propto
\ln(t/t_0).
\end{equation*}
Second, under weak bias, the mean displacement grows with the same logarithmic clock,

\begin{equation*}
\langle x(t)\rangle_F
\propto
F\ln(t/t_0),
\end{equation*}
and satisfies a generalized Einstein relation with the unbiased fluctuations.

Third, transit-time observables show logarithmic corrections to boundary-controlled current decay. In particular, the long-time current has the asymptotic form

\begin{equation*}
I(t)
\propto
\frac{1}{t[\ln(t/t_0)]^2},
\end{equation*}
up to system-dependent prefactors.

Fourth, target survival probabilities decay extremely slowly. In three dimensions, the finite-concentration survival probability can reduce to a power of time through the logarithmic exponential form,

\begin{equation*}
\Phi_c(t)
\sim
\exp[-A\ln(t/t_0)]
=
\left(\frac{t}{t_0}\right)^{-A},
\end{equation*}
whereas in one dimension it takes the even slower form

\begin{equation*}
\Phi_c(t)
\sim
\exp[-B\sqrt{\ln(t/t_0)}].
\end{equation*}
These signatures differ qualitatively from standard subdiffusive CTRW, where power-law operational clocks lead to power-law or stretched-exponential time dependences. The common feature of the log-aging case is the replacement of operational time by a logarithmic internal clock. This makes transport and first-passage dynamics ultraslow and strongly sensitive to aging history.

The observables discussed in this section characterize ensemble-level transport and first-passage behavior. Their fluctuation and rare-history content cannot be reconstructed from the mean logarithmic clock alone; it is supplied by the event-number and duration distributions developed in Sec.~\ref{sec_ergodicity}. We now turn from observable laws to the continuum memory operator that propagates one-point densities.

\section{Scale-Covariant GFPE in the Spatial Diffusion Limit}
\label{sec:gfpe}
\label{sec_GFPE}

The CTRW resolves individual events, but many applications require a continuous-space equation for probability densities in external fields, confined geometries, or boundary-value problems. The question in this section is therefore: which continuum memory operator encodes the same logarithmic clock? The answer is a nonstationary, scale-covariant kernel fixed by matching the GFPE to the CTRW propagator.

The contrast with ordinary diffusion is structural. Markovian dynamics uses a local time derivative, and stationary anomalous dynamics uses kernels of time differences. Log-aging instead requires a kernel organized by time ratios, because shifting the observation window changes the process. This distinction makes the GFPE a direct continuum expression of broken time-translation invariance.

We derive the temporal kernel by matching the GFPE to the Fourier--Mellin CTRW propagator and recover it independently from gain--loss dynamics. The resulting one-point equation treats eigenmode relaxation, boundaries, and moment equations in the small-jump spatial diffusion limit. Two-time quantities require the joint clock and are derived separately below.

\subsection{Memory Kernel from CTRW--GFPE Matching}
\label{sec_GFPE_kernel}

We consider a generalized Fokker--Planck equation~\cite{FPE_CTRW_PhysRevE.61.132,GFPE_Barkai_PhysRevE.63.046118,GFPE_Sokolov_PhysRevE.66.041101,GFPE_Baule_PhysRevE.71.026101} of the form

\begin{equation}
\int_{t_0}^{t}
K(t,t')
\frac{\partial P(x,t')}{\partial t'}
\,dt'
=
L_{\mathrm{FP}}P(x,t),
\end{equation}
where $P(x,t)$ is the probability density and $L_{\mathrm{FP}}$ is the ordinary Fokker--Planck operator acting on the spatial coordinate. The lower limit $t_0$ denotes the microscopic time scale at which the logarithmic clock is initialized. This is important for log-aging dynamics because the natural time variable is $\ln(t/t_0)$ rather than $t$ itself.

For a constant drift velocity $v$ and diffusion coefficient $D$, we use

\begin{equation}
L_{\mathrm{FP}}P(x,t)
=
-v\frac{\partial P(x,t)}{\partial x}
+
D\frac{\partial^2 P(x,t)}{\partial x^2}.
\end{equation}

For a force field $F(x)$, one may write more generally

\begin{equation}
L_{\mathrm{FP}}P(x,t)
=
-\frac{\partial}{\partial x}
\left[
\mu F(x)P(x,t)
\right]
+
D\frac{\partial^2 P(x,t)}{\partial x^2},
\end{equation}
where $\mu$ is the mobility. The following derivation is simplest for constant drift, but the memory structure is independent of this restriction.

For normal Markovian diffusion, the evolution is local in time and the memory kernel reduces to a delta-like kernel. For log-aging diffusion, the kernel is nonstationary. It cannot be written as $K(t-t')$, because the process is not invariant under translations of the observation time. Instead, the kernel is scale covariant: it is invariant under simultaneous rescaling of both time arguments, up to the scaling implied by the kernel itself.
Since the memory kernel is scale invariant, we write its two-time form as
\begin{equation}
K(t,t')
=
k\left(\frac{t'}{t}\right) .
\end{equation}
Throughout this work, \(\widehat{K}(p)\) denotes the Mellin transform of the associated one-variable ratio kernel \(k(q)\),
\begin{equation}
\widehat{K}(p)
=
\mathcal{M}[k](p)
=
\int_0^\infty q^{p-1} k(q)\,dq,
\qquad
\end{equation}
and should not be interpreted as a two-variable Mellin transform of the full two-time kernel \(K(t,t')\).

Applying a Fourier transform in space and a Mellin transform in time to the generalized Fokker--Planck equation gives

\begin{equation}
\widehat P(k,p)
=
\frac{\widehat{K}(p)t_0^p}
{-p\widehat{K}(p)+ivk+Dk^2}.
\end{equation}
 On the other hand, the log-aging CTRW propagator has the Fourier--Mellin form

\begin{equation*}
\widehat P(k,p)
=
t_0^p
\frac{\tilde{G}(p)-1}
{p\left[1-\hat{f}(k)\tilde{G}(p)\right]},
\end{equation*}
where $\tilde{G}(p)$ is the Mellin transform of the aging kernel and $\hat{f}(k)$ is the Fourier transform of the jump-length distribution.

For small wave number, the jump distribution may be expanded as

\begin{equation}
\hat{f}(k)
\simeq
1-ivk-Dk^2 .
\end{equation}

Matching the generalized Fokker--Planck propagator with the CTRW propagator then gives the memory kernel in Mellin space,

\begin{equation}
\widehat{K}(p)
=
\frac{\widehat{G}(p)-1}
{p\widehat{G}(p)}.
\end{equation}

Thus, the memory kernel is not introduced phenomenologically. It is fixed by the same aging kernel $\widehat{G}(p)$ that controls the CTRW clock.

For the log-aging kernel considered here, inverse Mellin transformation gives

\begin{equation}\label{eq_memorykernel}
K(t,t')
=
\left[
\left(
\frac{t}{t-t'}
\right)^{1-\alpha}
-1
\right]
\Theta(t-t'),
\qquad
0<\alpha<1 .
\end{equation}

Equivalently, the kernel can be viewed as a function of the ratio $t'/t$,

\begin{equation}
K(t,t')
=
\left[
\left(
1-\frac{t'}{t}
\right)^{-(1-\alpha)}
-1
\right]
\Theta(t-t').
\end{equation}

This form makes the aging structure explicit. The kernel is homogeneous under the simultaneous rescaling $t\to at$ and $t'\to at'$, but it is not invariant under time translation. This is the generalized Fokker--Planck manifestation of aging.

For $t\gg t'$, expansion in $t'/t$ yields

\begin{equation}
K(t,t')
\simeq
(1-\alpha)\frac{t'}{t}.
\end{equation}

Thus, the memory of very early times decays algebraically in the ratio $t'/t$. Near $t'\to t$, the kernel has a singular short-lag structure,

\begin{equation}
K(t,t')
\sim
\left(
\frac{t}{t-t'}
\right)^{1-\alpha}.
\end{equation}

The same kernel therefore contains both a long memory of the aging history and a strong sensitivity to recent events.

The limit $\alpha\to1^{-}$ is singular in this parametrization: the multiplier density concentrates near $Y=1$ and $\kappa_1\to0$, so the long-time logarithmic formulas are not uniform in $\alpha$. A separate limiting analysis is required at $\alpha=1$.

\subsection{Generalized Master Equation Derivation}
\label{sec_GFPE_GME}

The same memory kernel can be derived from a generalized master equation\cite{Tauberium_klafter2011}. This derivation clarifies the microscopic origin of the GFPE and shows how the memory term arises from the gain and loss of probability at each site.

Let $j^+(x,t)$ and $j^-(x,t)$ denote the gain and loss fluxes at position $x$ and time $t$. Probability conservation gives

\begin{equation}
\frac{\partial P(x,t)}{\partial t}
=
j^+(x,t)-j^-(x,t).
\end{equation}

The loss flux $j^-(x,t)$ consists of two contributions. The first comes from walkers that were already at position $x$ at the initial time $t_0$ and leave the site at time $t$. The second comes from walkers that arrived at position $x$ at an earlier time $t'$ and leave at time $t$. Therefore,

\begin{equation}
j^-(x,t)
=
\psi_1(t-t_0,t_0)P(x,t_0)
+
\int_{t_0}^{t}
j^+(x,t')
\psi_1(t-t',t')
\,dt' .
\end{equation}

Using probability conservation, $j^+(x,t')=\partial_{t'}P(x,t')+j^-(x,t')$, this can be rewritten as

\begin{equation}
j^-(x,t)
=
\psi_1(t-t_0,t_0)P(x,t_0)
+
\int_{t_0}^{t}
\psi_1(t-t',t')
\left[
\frac{\partial P(x,t')}{\partial t'}
+
j^-(x,t')
\right]
dt' .
\end{equation}

In Mellin space, this relation gives a closed expression for the loss flux in terms of the probability density,

\begin{equation}
\widehat j^-(x,p)
=
-(p-1)
\frac{\widehat{G}(p-1)}
{1-\widehat{G}(p-1)}
\widehat P(x,p-1).
\end{equation}

The gain flux is related to the loss flux by the jump kernel,

\begin{equation}
j^+(x,t)
=
\int f(x,x')j^-(x',t)\,dx',
\end{equation}
where $f(x,x')$ is the probability density for jumping from $x'$ to $x$. Combining the gain--loss relation with probability conservation gives, in Mellin space,

\begin{equation}
\frac{\widehat{G}(p)-1}
{p\widehat{G}(p)}
\left[
-p\widehat P(x,p)-t_0^pP(x,t_0)
\right]
=
\int dx'
\left[
f(x,x')-\delta(x-x')
\right]
\widehat P(x',p).
\end{equation}

Transforming back to the time domain yields the generalized master equation

\begin{equation}
\int_{t_0}^{t}
K(t,t')
\frac{\partial P(x,t')}{\partial t'}
\,dt'
=
\int dx'
\left[
f(x,x')-\delta(x-x')
\right]
P(x',t).
\end{equation}

The continuum generalized Fokker--Planck equation follows by expanding the jump operator on the right-hand side in spatial gradients. For jumps with finite first and second moments, this expansion gives

\begin{equation}
\int_{t_0}^{t}
K(t,t')
\frac{\partial P(x,t')}{\partial t'}
\,dt'
=
L_{\mathrm{FP}}P(x,t).
\end{equation}

This derivation shows explicitly that the nonstationary memory kernel originates from the age-dependent forward waiting-time statistics of the CTRW. The GFPE is therefore a continuum representation of the same logarithmic internal clock.

\subsection{Eigenmode Relaxation and Ultraslow Decay}
\label{sec_GFPE_eigen}

The GFPE is particularly useful when the spatial part of the dynamics can be diagonalized, as in standard spectral treatments of Fokker--Planck operators~\cite{GFP_eigen_risken1996fokker,GFP_eigen_coffey2004langevin}. Suppose that the Fokker--Planck operator has eigenfunctions $F_k(x)$ satisfying

\begin{equation}
L_{\mathrm{FP}}F_k(x)
=
-\lambda_k F_k(x),
\end{equation}
with eigenvalues $\lambda_k\geq0$. We seek separated solutions of the form

\begin{equation}
P(x,t)
=
F_k(x)T_k(t).
\end{equation}

Substitution into the GFPE gives the temporal equation

\begin{equation}
\int_{t_0}^{t}
K(t,t')
\frac{dT_k(t')}{dt'}
\,dt'
=
-\lambda_k T_k(t).
\end{equation}

In Mellin space this becomes

\begin{equation}
\widehat{K}(p)
\left[
-p\widehat T_k(p)-t_0^p\widehat T_k(t_0)
\right]
=
-\lambda_k\widehat T_k(p).
\end{equation}

Taking $T_k(t_0)=1$, one obtains

\begin{equation}
\widehat T_k(p)
=
\frac{t_0^p\widehat{K}(p)}
{\lambda_k-p\widehat{K}(p)}.
\end{equation}

Equivalently, using the logarithmic-clock form, the long-time eigenmode relaxation is controlled by the replacement of ordinary time by the internal time

\begin{equation}
s(t)
=
\frac{\ln(t/t_0)}{\kappa_1}.
\end{equation}
Thus,

\begin{equation}
T_k(t)
\sim
\exp[-\lambda_k s(t)]
=
\left(
\frac{t_0}{t}
\right)^{\lambda_k/\kappa_1}.
\end{equation}

This power-law decay in physical time is the eigenmode signature of logarithmic aging. It is much slower than ordinary exponential relaxation in time, but it is exponential in the logarithmic internal clock. Therefore, the GFPE converts the ultraslow clock into algebraic relaxation of spatial modes.

This result also clarifies how equilibrium or confined correlations decay. If an observable is decomposed into eigenmodes of the Fokker--Planck operator, each mode relaxes as

\begin{equation*}
\left(
\frac{t_0}{t}
\right)^{\lambda_k/\kappa_1}.
\end{equation*}

The slow algebraic decay reflects the long memory of the aging kernel and the logarithmically slow growth of the number of effective renewal events.

\subsection{Free-Space Propagator and Boundary Conditions}
\label{sec_GFPE_solution}

For a free particle without drift,

\begin{equation}
L_{\mathrm{FP}}P(x,t)
=
D\frac{\partial^2 P(x,t)}{\partial x^2}.
\end{equation}

In Fourier--Mellin space, the GFPE gives

\begin{equation}
P(k,p)
=
\frac{t_0^pP(k,t_0)}
{-p+Dk^2/\widehat{K}(p)}.
\end{equation}

In the long-time regime, this expression is equivalent to diffusion in the logarithmic internal time

\begin{equation}
s(t)
=
\frac{\ln(t/t_0)}{\kappa_1}.
\end{equation}

Thus, for an initially localized particle,

\begin{equation}
P(x,t_0)
=
\delta(x),
\end{equation}

the central, diffusion-limit free-space propagator is Gaussian in space with a variance growing logarithmically in physical time,

\begin{equation}
P_0(x,t)
=
\frac{1}
{\sqrt{4\pi D s(t)}}
\exp\left[
-\frac{x^2}{4Ds(t)}
\right].
\end{equation}

Equivalently,

\begin{equation}
P_0(x,t)
=
\sqrt{
\frac{\kappa_1}
{4\pi D\ln(t/t_0)}
}
\exp\left[
-\frac{\kappa_1 x^2}
{4D\ln(t/t_0)}
\right].
\end{equation}

With this convention,

\begin{equation}
\langle x^2(t)\rangle
=
2D s(t)
=
\frac{2D}{\kappa_1}
\ln(t/t_0).
\end{equation}

If the diffusion coefficient is instead defined through the convention $\langle x^2\rangle=D s$, the factors of $2$ in the Gaussian should be adjusted accordingly. The convention-independent result is that the \emph{central sector} has a width proportional to $\ln(t/t_0)$. This Gaussian is not a uniform approximation to the complete event-level distribution: trajectories with anomalously few events retain the algebraic duration tail derived in Sec.~\ref{sec:rare}.

This form is consistent with the deterministic mean-clock limit of the subordination picture. If the full inverse-clock density $h(s,t)$ is retained, the laboratory-time propagator is the Gaussian mixture $\int_0^\infty h(s,t)P_0(x,s)\,ds$, which need not be exactly Gaussian. The continuum expression therefore preserves the Gaussian central form while the event process supplies the non-Gaussian rare sector.

For a half-line geometry with a boundary at $x=0$ and initial position $x_0>0$, the method of images gives a convenient route to reflecting and absorbing boundary conditions. Related boundary-value constructions for nonlocal diffusion equations have been extensively studied~\cite{METZLER2000107,BENCHOHRA20092391}.

\begin{equation}
P(x,t;x_0,t_0)
=
P_0(x-x_0,t)
\pm
P_0(x+x_0,t),
\qquad
x>0.
\end{equation}

The plus sign corresponds to a reflecting boundary,

\begin{equation}
\left.
\frac{\partial P(x,t)}{\partial x}
\right|_{x=0}
=
0,
\end{equation}
whereas the minus sign corresponds to an absorbing boundary,

\begin{equation}
P(0,t)=0.
\end{equation}

The validity of this image construction follows from the fact that the memory kernel acts only on time, while the spatial boundary conditions are imposed through the Fokker--Planck operator. Therefore, standard spatial solution methods can often be retained, with physical time replaced by the logarithmic internal clock.

\subsection{Moment Equations}
\label{sec_GFPE_moments}

The GFPE also provides a convenient route to moment equations. For a constant external force $F$, let the drift velocity be $v=\mu F$. The Fokker--Planck operator is

\begin{equation}
L_{\mathrm{FP}}P
=
-v\frac{\partial P}{\partial x}
+
D\frac{\partial^2P}{\partial x^2}.
\end{equation}

Multiplying the GFPE by $x^n$ and integrating over space gives

\begin{equation}
\int_{t_0}^{t}
K(t,t')
\frac{d}{dt'}
\langle x^n(t')\rangle
\,dt'
=
nv\langle x^{n-1}(t)\rangle
+
n(n-1)D\langle x^{n-2}(t)\rangle .
\end{equation}

This follows from integration by parts,

\begin{equation}
\int_{-\infty}^{\infty}
x^n
\frac{\partial P}{\partial x}
\,dx
=
-n\langle x^{n-1}(t)\rangle ,
\end{equation}
and

\begin{equation}
\int_{-\infty}^{\infty}
x^n
\frac{\partial^2P}{\partial x^2}
\,dx
=
n(n-1)\langle x^{n-2}(t)\rangle .
\end{equation}

For $n=1$,

\begin{equation}
\int_{t_0}^{t}
K(t,t')
\frac{d}{dt'}
\langle x(t')\rangle
\,dt'
=
v.
\end{equation}

Using the logarithmic clock, one obtains

\begin{equation}
\langle x(t)\rangle
=
v s(t)
=
\frac{v}{\kappa_1}
\ln(t/t_0).
\end{equation}

For $n=2$,

\begin{equation}
\int_{t_0}^{t}
K(t,t')
\frac{d}{dt'}
\langle x^2(t')\rangle
\,dt'
=
2v\langle x(t)\rangle
+
2D.
\end{equation}

Thus,

\begin{equation}
\langle x^2(t)\rangle
=
2D s(t)
+
v^2s^2(t),
\end{equation}
or

\begin{equation}
\langle x^2(t)\rangle
=
\frac{2D}{\kappa_1}
\ln(t/t_0)
+
\frac{v^2}{\kappa_1^2}
\left[
\ln(t/t_0)
\right]^2 .
\end{equation}

The expression above is the deterministic or mean-clock closure of the GFPE. The full random clock contains an additional fluctuation contribution under bias. Writing $X(t)=X_0[S(t)]$ and using the law of total variance gives

\begin{equation}
\operatorname{Var}X(t)
=
2D\,\mathbb E[S(t)]
+
v^2\operatorname{Var}S(t).
\label{eq:total_variance_random_clock}
\end{equation}

With the central clock cumulants in Eq.~\eqref{eq:count_cumulants_main}, the corresponding biased second moment, at the leading logarithmic orders, is

\begin{equation}
\langle X^2(t)\rangle
\simeq
\frac{v^2}{\kappa_1^2}L^2
+
\left(
\frac{2D}{\kappa_1}
+
\frac{v^2\kappa_2}{\kappa_1^3}
\right)L,
\qquad
L=\ln(t/t_0).
\label{eq:biased_second_random_clock}
\end{equation}

The term proportional to $\kappa_2$ is absent if the random clock is replaced by its mean. The one-point GFPE reproduces it only when the full inverse-clock distribution, rather than the deterministic time change, is retained.

For unbiased diffusion, $v=0$, this reduces to

\begin{equation}
\langle x^2(t)\rangle
=
\frac{2D}{\kappa_1}
\ln(t/t_0).
\end{equation}

These moment equations are consistent with the CTRW derivation. Their advantage is that they can be generalized more naturally to spatially dependent forces, confinement and boundary conditions. In this sense, the GFPE provides a continuum framework complementary to the event-based CTRW formulation.

\subsection{Two-Time Covariance from the Joint Clock}
\label{sec_GFPE_correlation}

Aging dynamics is intrinsically a two-time phenomenon, but a one-point GFPE does not by itself determine multi-time probability laws. Here the free-particle covariance is obtained from the joint inverse-clock construction associated with the same temporal kernel: the positions at two times share the history accumulated up to the earlier time.

A one-point GFPE does not by itself justify a Markov factorization conditioned only on position. The event representation gives the two-time covariance without that assumption. For zero-mean independent jumps with variance $\sigma_\xi^2$,

\begin{equation}
X(t)=\sum_{j=1}^{N(t)}\xi_j,
\qquad
\mathbb E[\xi_j]=0,
\qquad
\mathbb E[\xi_i\xi_j]=\sigma_\xi^2\delta_{ij}.
\end{equation}

Conditioned on the joint clock, the two positions share exactly the jumps completed before the earlier time. Hence

\begin{equation}
\mathbb E\!\left[
X(t_1)X(t_2)\mid N(t_1),N(t_2)
\right]
=
\sigma_\xi^2 N\!\left(\min\{t_1,t_2\}\right),
\end{equation}

and therefore

\begin{equation}
\langle X(t_1)X(t_2)\rangle
=
\sigma_\xi^2\,
\mathbb E\!\left[
N\!\left(\min\{t_1,t_2\}\right)
\right].
\label{eq:two_time_joint_clock}
\end{equation}

This derivation uses the joint clock rather than an unjustified Chapman--Kolmogorov property for the non-Markovian laboratory-time process. With $\sigma_\xi^2=2D$, it gives the same leading logarithmic covariance as the continuum calculation.

For $t_1,t_2\gg t_0$, the inverse transform yields the asymptotic covariance

\begin{equation}
\langle x(t_1)x(t_2)\rangle
\sim
\frac{2D}{\kappa_1}
\ln
\left(
\frac{\min\{t_1,t_2\}}{t_0}
\right).
\end{equation}

This is the logarithmic-clock analogue of the Brownian covariance

\begin{equation}
\langle B(t_1)B(t_2)\rangle
=
2D\min\{t_1,t_2\}.
\end{equation}

The replacement

\begin{equation}
\min\{t_1,t_2\}
\rightarrow
\frac{1}{\kappa_1}
\ln
\left(
\frac{\min\{t_1,t_2\}}{t_0}
\right)
\end{equation}

expresses the fact that the two positions share the same trajectory history up to the earlier time. The correlation therefore depends on the minimum time, but only through the logarithmic internal clock.

This result also illustrates the difference between one-time and two-time descriptions. The one-time propagator gives the spatial spread at a fixed observation time, whereas the two-time correlation reveals how the trajectory retains memory of its earlier evolution. In log-aging diffusion, this memory is long-lived because the internal clock advances only logarithmically.

\subsection{Scope and Information Retained by the GFPE}
\label{sec_GFPE_summary}

The generalized Fokker--Planck formulation provides a continuum representation of log-aging diffusion. Its central ingredient is a nonstationary memory kernel,

\begin{equation*}
K(t,t')
=
\left[
\left(
\frac{t}{t-t'}
\right)^{1-\alpha}
-1
\right]
\Theta(t-t'),
\end{equation*}
This kernel carries the same time-ratio structure as the event clock, but the displayed Fokker--Planck operator additionally assumes a small-jump spatial diffusion limit. The resulting GFPE is therefore exact at the level of the matched temporal kernel and approximate at the level of the continuum spatial closure.

Within the deterministic or central-clock reduction, the free-space propagator is Gaussian with a variance proportional to $\ln(t/t_0)$, and internal-time exponential modes become algebraic in laboratory time. Retaining the full inverse-clock density instead produces a Gaussian mixture and preserves clock-induced corrections such as the biased variance term in Eq.~\eqref{eq:biased_second_random_clock}. Standard spatial boundary methods remain applicable because the nonlocal operator acts only on time.

The GFPE is consequently a one-point probability-density representation, not a replacement for the complete event process. It is well suited to external fields, confinement, eigenmodes, and boundary-value problems, but fixed-$n$ event-time tails, exact-count probabilities, and TA-MSD amplitude statistics must be supplied by the event law or the full joint clock.

\section{Random-Clock Subordination and Its Deterministic Mean-Clock Reduction}
\label{sec:subordination}
\label{sec_time_matching}

The CTRW specifies event statistics and the GFPE specifies their continuum consequence, but neither yet makes the clock construction transparent. Standard subordination represents anomalous transport as Markovian motion in an internal time observed through a random relation to laboratory time~\cite{MeerschaertSikorskii+2019,GFPE_Baule_PhysRevE.71.026101}. This section adapts that construction to logarithmic aging and thereby identifies the clock, rather than the spatial propagator, as the source of the distinct dynamical class.

The comparison is deliberately sharp: normal diffusion has an effectively deterministic clock, standard subdiffusion has a power-law random clock, and log-aging diffusion has a logarithmic, nonstationary clock. This distinction explains both the different scaling laws and the change from Laplace to Mellin methods. It also provides the bridge from probability-density evolution to a trajectory-level description of memory and noise.

\subsection{Internal Time and Laboratory Time}
\label{sec_internal_lab_time}

\begin{figure}
    \centering
    \includegraphics[width=0.8\linewidth]{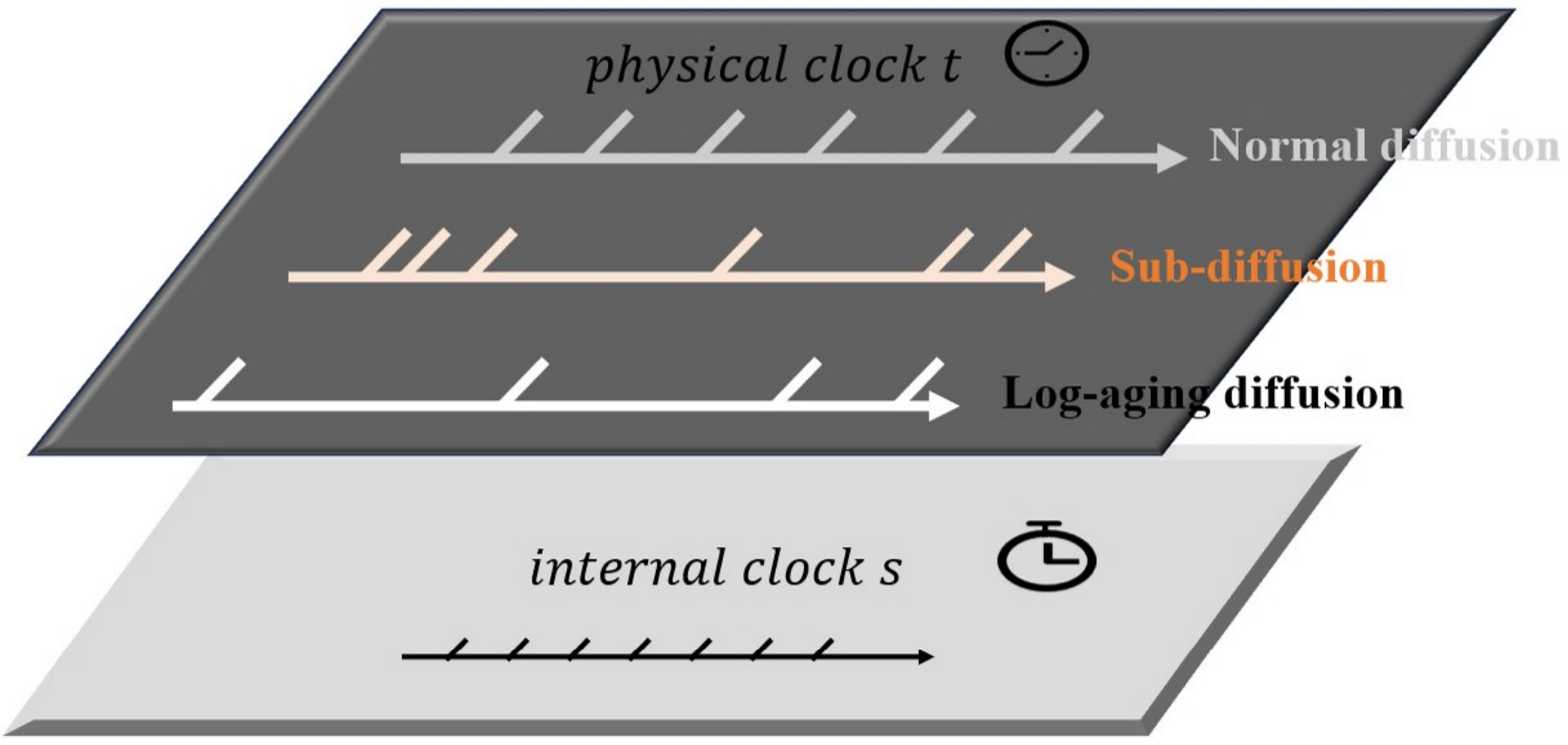}
    \caption{
    Schematic of time matching between internal time $s$ and laboratory time $t$. In internal time the spatial process is Markovian with generator $L_{\mathrm{FP}}$; the inverse-clock density $h(s,t)$ produces the laboratory-time one-point propagator by subordination. Normal diffusion uses a deterministic linear clock,  subdiffusion a broad power-law clock, and the present model a multiplicative clock that is additive in $\ln(t/t_0)$. Replacing $h(s,t)$ by a delta function at its mean is a deterministic-clock approximation and discards clock fluctuations.}
    \label{fig:time_matching}
\end{figure}

A useful representation of generalized diffusion is the integral decomposition (Fig.~\ref{fig:time_matching})

\begin{equation}
P(x,t)
=
\int_0^\infty h(s,t)P_0(x,s)\,ds .
\end{equation}

Here $P_0(x,s)$ is the probability density of a Markovian process evolving in the internal time $s$. It satisfies the ordinary Fokker--Planck equation

\begin{equation}
\frac{\partial P_0(x,s)}{\partial s}
=
L_{\mathrm{FP}}P_0(x,s),
\end{equation}
where $L_{\mathrm{FP}}$ is the same spatial Fokker--Planck operator as in the generalized equation. The function $h(s,t)$ is the probability density that the internal time has value $s$ when the laboratory time is $t$. It is therefore the clock-matching function between the internal and laboratory time descriptions.

The physical interpretation is simple. In the internal time space $(x,s)$, the process is Markovian and memoryless. In the laboratory time space $(x,t)$, the same trajectory appears non-Markovian if the relation between $s$ and $t$ is random or history dependent. The memory kernel in the GFPE is therefore not an independent ingredient; it is generated by the stochastic matching between internal time and laboratory time.

The matching function satisfies

\begin{equation}
\int_{t_0}^{t}
K(t,t')
\frac{\partial h(s,t')}{\partial t'}
\,dt'
=
-\frac{\partial h(s,t)}{\partial s}.
\end{equation}

This equation states that evolution in internal time is driven by the memory-weighted evolution in laboratory time. The form of the kernel $K(t,t')$ determines the type of diffusion.

For stationary renewal processes, such as normal diffusion and standard subdiffusion, the kernel depends only on time differences,

\begin{equation}
K(t,t')=K(t-t').
\end{equation}

The natural transform is then the Laplace transform in laboratory time. One obtains\cite{GFPE_Baule_PhysRevE.71.026101}

\begin{equation}
\tilde h(s,\lambda)
=
\tilde K(\lambda)
\exp[-s\lambda \tilde K(\lambda)].
\end{equation}

For normal diffusion with a characteristic waiting time $\tau$, the clock is effectively deterministic, and one may write

\begin{equation}
\lambda\tilde  K(\lambda)
\simeq
\lambda \tau .
\end{equation}

Thus,

\begin{equation}
\tilde h(s,\lambda)
\simeq
\tau e^{-s\lambda\tau}.
\end{equation}

For standard subdiffusion with exponent $0<\alpha<1$, the memory kernel produces a power-law clock, as obtained from CTRW-to-fractional-kinetic mappings~\cite{FPE_CTRW_PhysRevE.61.132,CTWadFPE_sandev2015diffusion,PhysRevE.61.6308}. In Laplace space,

\begin{equation}
\lambda \tilde K(\lambda)
\sim
A\lambda^\alpha ,
\end{equation}

and hence

\begin{equation}
\tilde h(s,\lambda)
\sim
A\lambda^{\alpha-1}
\exp[-As\lambda^\alpha].
\end{equation}

This is the usual subordination structure of fractional diffusion.

Log-aging diffusion is different. Its memory kernel is not time-translation invariant. Instead, it is scale covariant and organized by ratios of times. The natural transform is therefore the Mellin transform. In Mellin space, the matching function becomes

\begin{equation}
\widehat h(s,p)
=
t_0^p\widehat{K}(p)
\exp[sp\widehat{K}(p)].
\label{eq:inverse_clock_mellin_corrected}
\end{equation}

In the long-time logarithmic regime, the Mellin kernel has the small-$p$ behavior

\begin{equation}
p\widehat{K}(p)
\sim
\kappa_1p .
\end{equation}

Therefore,

\begin{equation}
\widehat h(s,p)
\simeq
t_0^p \kappa_1
\exp[s\kappa_1 p].
\end{equation}

The positive sign is fixed by the Mellin convention used in this work. Indeed, the deterministic logarithmic clock $h(s,t)=\delta[s-\ln(t/t_0)/\kappa_1]$ has Mellin transform $\kappa_1t_0^p\exp(\kappa_1ps)$. A negative sign would instead correspond to a clock running in the wrong direction.

This expression shows that the internal time is matched to laboratory time through the logarithmic clock

\begin{equation}
s(t)
\sim
\frac{\ln(t/t_0)}{\kappa_1}.
\end{equation}

Thus, the three diffusion classes can be summarized as

\begin{equation}
s(t)\sim
\begin{cases}
t, & \text{normal diffusion}, \\[0.4em]
t^\alpha, & \text{subdiffusion}, \\[0.4em]
\ln(t/t_0), & \text{log-aging diffusion}.
\end{cases}
\end{equation}

The spatial process may be the same Markovian diffusion in internal time, but the laboratory-time behavior is determined by the clock-matching rule.

\subsection{Coupled Langevin Equations}
\label{sec_coupled_langevin}

The time-matching construction can also be expressed through coupled Langevin equations, a standard representation of subordinated transport~\cite{subordicate_PhysRevE.82.011117,sub_auto_PhysRevE.78.031112}. Let $X(s)$ denote the spatial process in internal time and $t(s)$ the laboratory time accumulated up to internal time $s$. A general representation is

\begin{equation}
\frac{dX(s)}{ds}
=
F(X(s))+\eta(s),
\end{equation}

and

\begin{equation}
\frac{dt(s)}{ds}
=
\tau(s),
\qquad
\tau(s)>0 .
\end{equation}

Here $\eta(s)$ is a white noise in internal time, and $\tau(s)$ is a positive stochastic process that controls the rate at which laboratory time accumulates. The observed process in laboratory time is obtained by inversion,

\begin{equation}
X(t)=X(s(t)),
\end{equation}

where $s(t)$ is the inverse process of $t(s)$.

The probability density of the direct clock process is

\begin{equation}
p(t,s)
=
\left\langle
\delta(t-t(s))
\right\rangle .
\end{equation}

The density of the inverse clock is

\begin{equation}
h(s,t)
=
\left\langle
\delta(s-s(t))
\right\rangle .
\end{equation}

Equivalently,

\begin{equation}
h(s,t)
=
-\frac{\partial}{\partial s}
\left\langle
\Theta(t-t(s))
\right\rangle .
\end{equation}

The two densities have different normalizations and different physical meanings. The density $p(t,s)$ is normalized with respect to laboratory time $t$ for fixed internal time $s$. It describes how much laboratory time has elapsed after a given internal time. The density $h(s,t)$ is normalized with respect to internal time $s$ for fixed laboratory time $t$. It describes how much internal time has been accumulated by a given laboratory time.

This distinction is important. The propagator in laboratory time uses $h(s,t)$ because the observed position is obtained by averaging the internal-time propagator over the random internal time available at laboratory time $t$:

\begin{equation}
P(x,t)
=
\int_0^\infty h(s,t)P_0(x,s)\,ds .
\end{equation}

In Mellin space, the inverse-clock density for log-aging diffusion is

\begin{equation}
\widehat h(s,p)
=
t_0^p\widehat{K}(p)
\exp[sp\widehat{K}(p)].
\end{equation}

The factor $t_0^p$ explicitly reflects the dependence on the initial microscopic time scale. This is the signature of aging: the clock cannot be defined independently of the time at which the process is initialized.

The Mellin transform also reveals the relation between the statistics of the direct and inverse clocks. Formally,

\begin{equation}
h(s,p)
=
-\frac{\partial}{\partial s}
\left\langle
\int_{0}^{\infty}
t^{p-1}\Theta(t-t(s))\,dt
\right\rangle .
\end{equation}

Within the appropriate Mellin convergence domain, evaluating the survival integral links Mellin moments of the direct process $t(s)$ to the inverse-clock density $h(s,t)$. This is the continuous-time analogue of the relation between the distribution of the time of the $n$th jump and the distribution of the number of jumps completed by time $t$.

\subsection{A Unified View of Diffusion Classes}
\label{sec_unified_diffusion_classes}

The time-matching viewpoint clarifies why normal diffusion, subdiffusion and log-aging diffusion share similar mathematical structures while exhibiting different physical behavior.

In internal time, the spatial process may be the same Markovian diffusion,

\begin{equation}
\frac{\partial P_0(x,s)}{\partial s}
=
L_{\mathrm{FP}}P_0(x,s).
\end{equation}

The distinction between diffusion classes comes from the clock:

\begin{equation}
P(x,t)
=
\int_0^\infty h(s,t)P_0(x,s)\,ds .
\end{equation}

If $h(s,t)$ is sharply peaked around $s=t/\tau$, one recovers normal diffusion. If $h(s,t)$ is broad and generated by an inverse stable subordinator, one obtains subdiffusion with

\begin{equation}
\langle x^2(t)\rangle
\sim
t^\alpha .
\end{equation}

If $h(s,t)$ is generated by a logarithmic aging clock, then

\begin{equation}
\langle x^2(t)\rangle
\sim
\ln(t/t_0).
\end{equation}

Thus, the spatial operator $L_{\mathrm{FP}}$ determines the internal-time motion, while the clock-matching density $h(s,t)$ determines the laboratory-time scaling.

This also explains why the generalized Fokker--Planck equation contains a memory kernel. The kernel is the continuum representation of the random clock. For stationary clocks, the kernel depends on time differences and Laplace transforms diagonalize the dynamics. For aging logarithmic clocks, the kernel depends on time ratios and Mellin transforms diagonalize the dynamics.

Therefore, log-aging diffusion is not merely a limiting case of subdiffusion with a small exponent. It belongs to a different clock class. Standard subdiffusion is governed by power-law operational time,

\begin{equation*}
s(t)\sim t^\alpha ,
\end{equation*}

whereas log-aging diffusion is governed by

\begin{equation*}
s(t)\sim \ln(t/t_0).
\end{equation*}

This difference has direct consequences for aging, memory, ergodicity breaking and rare-event statistics.

\subsection{Two-Time Correlations from Time Matching}
\label{sec_time_matching_correlation}

The time-matching representation is especially useful for two-time observables. For a free particle in internal time, the position correlation is

\begin{equation}
\langle X(s_1)X(s_2)\rangle
=
2D\min\{s_1,s_2\}.
\end{equation}

Equivalently,

\begin{equation}
\langle X(s_1)X(s_2)\rangle
=
2D
\left[
\Theta(s_1-s_2)s_2
+
\Theta(s_2-s_1)s_1
\right].
\end{equation}

In laboratory time, the observed correlation is obtained by averaging over the joint distribution of the inverse clock,

\begin{equation}
\langle X(t_1)X(t_2)\rangle
=
\int_0^\infty ds_1
\int_0^\infty ds_2\,
\langle X(s_1)X(s_2)\rangle
h(s_1,t_1;s_2,t_2),
\end{equation}

where $h(s_1,t_1;s_2,t_2)$ is the two-time density of the inverse clock.

For log-aging diffusion, the double Mellin transform gives

\begin{equation}
\langle \widehat X(p_1)\widehat X(p_2)\rangle
=
\frac{
2D\,t_0^{p_1+p_2}
}
{
p_1p_2
\left(p_1+p_2\right)
\widehat{K}(p_1+p_2)
}.
\end{equation}

Using the small-$p$ behavior of the logarithmic clock,

\begin{equation}
\widehat{K}(p)
\simeq
\kappa_1,
\end{equation}

one obtains, for $t_1,t_2\gg t_0$,

\begin{equation}
\langle X(t_1)X(t_2)\rangle
\simeq
\frac{2D}{\kappa_1}
\ln
\left(
\frac{\min\{t_1,t_2\}}{t_0}
\right).
\end{equation}

Equivalently,

\begin{equation}
\langle X(t_1)X(t_2)\rangle
\simeq
\frac{2D}{\kappa_1}
\left[
\Theta(t_1-t_2)\ln(t_2/t_0)
+
\Theta(t_2-t_1)\ln(t_1/t_0)
\right].
\end{equation}

This is the logarithmic-clock analogue of Brownian covariance. The two positions share the same internal trajectory up to the earlier time, but the accumulated internal time grows only logarithmically with laboratory time. The dependence on $\min\{t_1,t_2\}$ is therefore preserved, while ordinary time is replaced by logarithmic internal time.

\subsection{Multi-Time Correlations and the Limits of a Noise Representation}
\label{sec_noise_memory}

The time-matching viewpoint also clarifies the possible origin of colored noise
in laboratory time. Generalized Langevin descriptions with correlated noise
provide the relevant stationary and fractional background
~\cite{GLE_PhysRevE.72.067701,PhysRevE.81.051118}. In internal time, the
driving noise may be taken to be white,
\begin{equation}
\langle \eta(s_1)\eta(s_2)\rangle
=
2D\delta(s_1-s_2).
\end{equation}

After the stochastic time transformation, the effective noise observed in
laboratory time is no longer white. Strictly speaking, it should be understood
as a generalized noise. In a weak, regularized sense, its covariance may be
represented by averaging the internal-time covariance over the inverse clock,
\begin{equation}
\begin{aligned}
\langle \eta_{\rm eff}(t_1)\eta_{\rm eff}(t_2)\rangle_{\rm reg}
&\simeq
\int_0^\infty ds_1
\int_0^\infty ds_2\,
\langle \eta(s_1)\eta(s_2)\rangle
h(s_1,t_1;s_2,t_2).
\end{aligned}
\end{equation}
The symbol \(\simeq\) emphasizes that the white-noise limit is taken after a
short-time regularization, so that only the finite regular diagonal part of the
two-time inverse-clock density is sampled.

Since the internal noise is delta correlated in \(s\), the laboratory-time
covariance is controlled by the probability that two laboratory times share the
same internal time. Thus the effective noise inherits the multi-time structure
of the inverse clock. As shown heuristically in
Appendix~\ref{app_noise_correlation}, the leading scale-covariant part of the
double Mellin-space covariance can be represented, up to normalization and
regularization-dependent amplitudes, as
\begin{equation}
\langle \widehat\eta_{\rm eff}(p_1)
\widehat\eta_{\rm eff}(p_2)\rangle_{\rm reg}
\sim
2D_{\rm eff}\,
t_0^{p_1+p_2}
\frac{
\widehat{K}(p_1)+\widehat{K}(p_2)
}
{p_1+p_2}.
\end{equation}
Here \(D_{\rm eff}\) denotes the effective noise strength after possible finite
renormalizations associated with the regularization convention.

This expression motivates, but does not by itself prove, a generalized
fluctuation--dissipation relation between effective noise correlations and the
clock kernel. A covariance must be symmetric and positive semidefinite, whereas
the GFPE memory kernel is causal. Accordingly, in the time domain one should
interpret the scale-covariant noise memory in a symmetrized form,
\begin{equation}
\langle
\eta_{\rm eff}(t_1)
\eta_{\rm eff}(t_2)
\rangle_{\rm reg}
\sim
\mathcal{K}_{\rm eff}
\left(
\frac{t_<}{t_>}
\right),
\qquad
t_< = \min(t_1,t_2),
\quad
t_> = \max(t_1,t_2),
\end{equation}
up to prefactors, symmetrization conventions, and possible contact terms.

Thus, the GFPE memory and a colored-noise covariance may originate from the
same clock-matching construction, but they are not the same mathematical
object. In internal time, the reference dynamics is Markovian and driven by
white noise. In laboratory time, a stochastic time transformation generates
aging and multi-time correlations. Establishing a genuine generalized
fluctuation--dissipation theorem additionally requires a microscopic bath or
response model. 

\subsection{Summary of the Time-Matching Picture}
\label{sec_time_matching_summary}

The time-matching picture provides a unified way to understand different diffusion classes. The spatial dynamics is generated in an internal time $s$ by an ordinary Markovian process. The observed laboratory-time dynamics is obtained by matching $s$ to $t$ through a random clock.

Normal diffusion, subdiffusion and log-aging diffusion correspond to different clock classes:

\begin{equation*}
s(t)\sim
\begin{cases}
t, & \text{normal diffusion}, \\[0.4em]
t^\alpha, & \text{subdiffusion}, \\[0.4em]
\ln(t/t_0), & \text{log-aging diffusion}.
\end{cases}
\end{equation*}

This hierarchy explains both the similarities and the differences between the processes. They are similar because they share the same subordination structure. They are different because their clocks have different symmetries. Normal diffusion is Markovian in laboratory time. Standard subdiffusion is non-Markovian but time-translation invariant at the level of the renewal kernel. Log-aging diffusion is nonstationary and breaks time-translation invariance, requiring Mellin rather than Laplace methods.

The clock-matching process therefore provides the conceptual bridge between CTRW, GFPE and Langevin descriptions. It shows that the memory kernel, the logarithmic internal time, the two-time correlations and the colored effective noise all originate from the same stochastic transformation between internal and laboratory time.

\section{Generalized Langevin Representation and Its Information Boundary}
\label{sec:gle}
\label{sec_GLE}

The preceding sections establish the logarithmic clock at the event,
probability-density, and time-matching levels. The remaining question is
dynamical: how does the same time-matching construction lead to an effective
generalized Langevin representation in laboratory time? As shown in
Appendix~\ref{app_time_matching}, the clock-averaged internal-time derivative
is mapped, in Mellin space, to a retarded scale-covariant memory operator. This
provides the origin of the GLE used below.

The generalized-Langevin background for memory friction and correlated noise is
well established~\cite{GFDT_WANG1999341,GLE_PhysRevE.64.051106,PhysRevE.81.051118},
but the resulting Gaussian GLE, even when matched to the response kernel and
second-order statistics, is not generally pathwise equivalent to a subordinated
CTRW. It need not reproduce the discrete event count, the algebraic event-time
tail, or the TA-MSD distribution. We therefore formulate the GLE with an
explicit information boundary: it is derived here as a clock-averaged response
representation and used as a response and covariance model unless equality of
the full finite-dimensional distributions is separately proved.

\subsection{Generalized Fluctuation--Dissipation Relation}
\label{sec_GLE_FDT}

We consider the generalized Langevin equation, following the standard
memory-friction formulation~\cite{GFP_eigen_coffey2004langevin,GLE_PhysRevE.72.067701},
\begin{equation}
m\ddot{x}(t)
=
-\xi
\int_{t_0}^{t}
K_R(t,\tau)\dot{x}(\tau)\,d\tau
+\eta(t)
-\nabla U(x),
\end{equation}
where $m$ is the particle mass, $\xi$ is a friction coefficient, $U(x)$ is an
external potential, and $\eta(t)$ is a stochastic force exerted by the
environment. The retarded kernel $K_R(t,\tau)$ is causal and encodes
history-dependent response. In the present construction, its scale-covariant
form is inherited from the same time-matching kernel that generates the GFPE
memory operator, as shown in Appendix~\ref{app_GLE_derivation}. Nevertheless,
the response kernel and the GFPE memory kernel act on different objects and
should not be regarded as pathwise identical.

For normal equilibrium Langevin dynamics, the fluctuation--dissipation theorem
relates the retarded friction kernel to a symmetric noise spectrum~\cite{RKubo_1966}.
In a nonstationary aging setting, introduce a separate symmetric
positive-semidefinite kernel $K_S(t,t')$. A possible fluctuation--response
relation may be written as
\begin{equation}
\langle \eta(t)\eta(t')\rangle
=
k_B T\,\xi\,K_S(t,t'),
\label{eq:nonstationary_noise_kernel}
\end{equation}
while the connection between $K_S$ and $K_R$ is preparation- and
model-dependent. Aging systems may show fluctuation--dissipation violations,
effective temperatures, or boundary terms~\cite{Cugliandolo_2011,GFDT_WANG1999341}.
Causality alone cannot supply a covariance, because $K_R(t,t')$ is generally
not symmetric. Thus, the time-matching derivation fixes the retarded memory
structure, but it does not by itself fix a unique symmetric covariance kernel.

For log-aging diffusion, the clock-averaged response representation leads to
the retarded scale-covariant kernel, Eq.~\eqref{eq_memorykernel},
\begin{equation*}
K(t,t')
=
\left[
\left(
\frac{t}{t-t'}
\right)^{1-\alpha}
-1
\right]
\Theta(t-t'),
\qquad
0<\alpha<1 .
\end{equation*}
This kernel breaks time-translation invariance and organizes the response
through ratios of times. A corresponding $K_S$ must be constructed and checked
for positive semidefiniteness before the GLE can represent a physical aging
bath. In the remaining formal calculations of this section, $K_R$, or simply
$K$ when no confusion is possible, denotes the chosen retarded kernel; any noise
result additionally assumes an admissible Gaussian closure for $K_S$.

In the overdamped limit and for a free particle, $U=0$, the inertial term can be neglected. The equation becomes

\begin{equation}
\xi
\int_{t_0}^{t}
K(t,\tau)\dot{x}(\tau)\,d\tau
=
\eta(t).
\end{equation}

Taking the Mellin transform gives

\begin{equation}
\widehat x(p)-t_0^p x(t_0)
=
\frac{\widehat\eta(p)}
{\xi p\widehat{K}(p)}.
\end{equation}

Assuming an initial distribution with $\langle x(t_0)\rangle=0$ and that the initial position is uncorrelated with the noise, the two-time position correlation in Mellin space is

\begin{equation}
\langle \widehat x(p)\widehat x(p')\rangle
=
\frac{
\langle \widehat \eta(p)\widehat \eta(p')\rangle
}
{
\xi^2pp'\widehat{K}(p)\widehat{K}(p')
}.
\end{equation}

Using the generalized fluctuation--dissipation relation, the Mellin-space noise correlation takes the form

\begin{equation}
\langle \widehat\eta(p)\widehat\eta(p')\rangle
=
k_B T\,\xi\,
t_0^{p+p'}
\frac{
\widehat{K}(p)\widehat{K}(p')
}
{p+p'}
\left[
\widehat{K}^{-1}(p)
+
\widehat{K}^{-1}(p')
\right],
\end{equation}

which yields

\begin{equation}
\langle \widehat x(p)\widehat x(p')\rangle
=
\frac{k_B T}{\xi}
t_0^{p+p'}
\frac{
\widehat{K}^{-1}(p)
+
\widehat{K}^{-1}(p')
}
{
pp'(p+p')
}.
\end{equation}

In the logarithmic aging regime,

\begin{equation}
\widehat{K}(p)
\simeq
\kappa_1,
\qquad
p\to0.
\end{equation}

Therefore,

\begin{equation}
\langle \widehat x(p)\widehat x(p')\rangle
\simeq
\frac{2k_B T}{\xi \kappa_1}
\frac{
t_0^{p+p'}
}
{
pp'(p+p')
}.
\end{equation}

Inverse Mellin transformation gives the asymptotic two-time correlation

\begin{equation}
\langle x(t)x(t')\rangle
\simeq
\frac{2k_B T}{\xi \kappa_1}
\ln
\left(
\frac{\min\{t,t'\}}{t_0}
\right),
\qquad
t,t'\gg t_0.
\end{equation}

Equivalently, this can be written as

\begin{equation}
\langle x(t)x(t')\rangle
\simeq
\frac{k_B T}{\xi \kappa_1}
\left[
\ln\frac{t}{t_0}
+
\ln\frac{t'}{t_0}
-
\left| \ln\frac{t}{t'} \right|
\right].
\end{equation}

This form is the logarithmic-clock analogue of Brownian covariance. The two positions share the same stochastic history up to the earlier time, while the accumulated internal time grows only logarithmically with laboratory time.

Setting $t=t'$ gives the free-particle variance,

\begin{equation}
\langle x^2(t)\rangle_{F=0}
=
\frac{2k_B T}{\xi \kappa_1}
\ln\left(\frac{t}{t_0}\right).
\end{equation}

\subsection{Constant-Force Response and the Einstein Relation}
\label{sec_GLE_response}

We next consider the response to a constant external force $F$, corresponding to

\begin{equation}
-\nabla U(x)=F.
\end{equation}

In the overdamped limit, the noise average vanishes and the mean displacement satisfies

\begin{equation}
\xi
\int_{t_0}^{t}
K(t,\tau)
\frac{d}{d\tau}
\langle x(\tau)\rangle_F
\,d\tau
=
F.
\end{equation}

In Mellin space,

\begin{equation}
\langle\widehat x(p)\rangle_F
=
\frac{F t_0^p}
{\xi p\widehat{K}(p)}.
\end{equation}

Using $\widehat{K}(p)\simeq \kappa_1$ for small $p$, the long-time mean displacement is

\begin{equation}
\langle x(t)\rangle_F
=
\frac{F}{\xi \kappa_1}
\ln\left(\frac{t}{t_0}\right).
\end{equation}

Comparing this result with the free-particle variance gives

\begin{equation}
\langle x(t)\rangle_F
=
\frac{F}{2k_B T}
\langle x^2(t)\rangle_{F=0}.
\end{equation}

Thus, the generalized Einstein relation is preserved. Aging slows both the spontaneous fluctuations and the linear response by the same logarithmic clock. The ratio between response and fluctuation remains fixed by the thermal energy scale $k_B T$.

This is an important consistency check. It shows that log-aging dynamics can be strongly nonstationary and non-Markovian while still satisfying a generalized fluctuation--dissipation structure, provided the noise and friction are governed by the same memory kernel.

The Mellin-space form of the noise correlation also indicates that the environmental fluctuations are scale-free in laboratory time. In frequency-like language, this corresponds to a broad low-frequency spectrum, consistent with $1/f$-type noise in the logarithmic aging regime. More precisely, the noise is not stationary, so an ordinary power spectral density must be used with care. The robust statement is that the noise correlations are governed by time ratios and by the same memory kernel that controls dissipation.

\subsection{Autocorrelation in a Harmonic Potential}
\label{sec_GLE_Cx}

A particularly useful probe of slow relaxation is the position autocorrelation in a confining potential. We consider a harmonic potential

\begin{equation}
U(x)
=
\frac{1}{2}m\omega^2x^2.
\end{equation}

In the overdamped limit, multiplying the generalized Langevin equation by $x(t_0)$ and averaging gives an equation for the autocorrelation function

\begin{equation}
C_x(t)
=
\langle x(t)x(t_0)\rangle .
\end{equation}

Because the noise at time $t$ is uncorrelated with the initial coordinate in the usual thermal preparation, one obtains

\begin{equation}
m\omega^2 C_x(t)
=
-\zeta
\int_{t_0}^{t}
K(t,\tau)
\frac{dC_x(\tau)}{d\tau}
\,d\tau ,
\end{equation}

where $\zeta$ denotes the friction coefficient in the confined problem. Equivalently,

\begin{equation}
\zeta
\int_{t_0}^{t}
K(t,\tau)
\frac{dC_x(\tau)}{d\tau}
\,d\tau
=
-m\omega^2 C_x(t).
\end{equation}

Applying the Mellin transform gives

\begin{equation}
\widehat C_x(p)
=
-\frac{
C_x(t_0)t_0^p
}
{
p-\dfrac{m\omega^2}{\zeta}
\widehat{K}^{-1}(p)
}.
\end{equation}

In the long-time regime, $\widehat{K}(p)\simeq \kappa_1$, so the inverse Mellin transform yields

\begin{equation}
C_x(t)
=
C_x(t_0)
\left(
\frac{t_0}{t}
\right)^{
m\omega^2/[\zeta \kappa_1]
}.
\end{equation}

Thus, confinement converts the logarithmic internal-time relaxation into a power-law decay in laboratory time. Indeed, if the relaxation in internal time is exponential,

\begin{equation}
C_x(s)
\sim
e^{-\Omega s},
\end{equation}

and the clock is

\begin{equation}
s(t)
=
\frac{\ln(t/t_0)}{\kappa_1},
\end{equation}

then

\begin{equation}
C_x(t)
\sim
\exp
\left[
-\Omega
\frac{\ln(t/t_0)}{\kappa_1}
\right]
=
\left(
\frac{t_0}{t}
\right)^{\Omega/\kappa_1}.
\end{equation}

The power-law form is therefore a direct consequence of logarithmic time matching.

\begin{figure}
    \centering
    \includegraphics[]{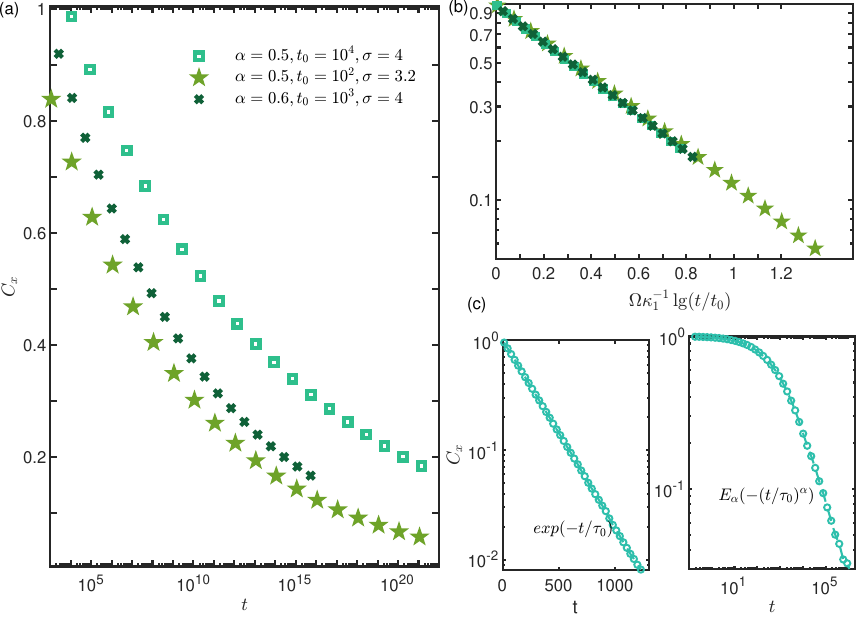}
    \caption{
Position autocorrelation $C_x(t)=\langle x(t)x(t_0)\rangle$, in unit of \(C_x(0)\), for diffusion in a harmonic potential on a lattice. (a) The logarithmic clock converts exponential internal-time relaxation into an extremely slow algebraic laboratory-time decay. (b) Simulation and the response/covariance closure agree over the displayed range. (c) Normal diffusion gives exponential relaxation and standard renewal subdiffusion gives a Mittag--Leffler law. Agreement of this two-point observable supports the clock-based closure but does not by itself establish equality of the full GLE and CTRW path measures.}
    \label{fig:Cx}
\end{figure}

Figure~\ref{fig:Cx} shows numerical results for the normalized autocorrelation function. The simulation\cite{PhysRevLett.134.197102} agrees with the predicted algebraic decay. The exponent is small when the logarithmic clock is slow, producing a very long-lived memory of the initial state.

This slow decay connects log-aging diffusion to glassy relaxation. In trap models of glasses, autocorrelation functions often decay as power laws or aging scaling functions because the system remains trapped in long-lived metastable configurations. The harmonic-potential result above shows how a similar algebraic relaxation emerges from the generalized Langevin description of logarithmic aging.

It also provides a phenomenological connection to Griffiths--McCoy physics. In Griffiths regions, rare slow domains generate broad distributions of relaxation times and may produce anomalously slow, often power-law, correlations. A similar slow law emerges from the present logarithmic clock, but this agreement does not identify a Griffiths mechanism or establish equivalence of the underlying stochastic processes.

\subsection{What the GLE Closure Captures---and What It Does Not}
\label{sec_GLE_interpretation}
In the present work, this GLE is not used as a microscopic pathwise equation
for the event process, but as an effective clock-averaged
response-and-covariance representation generated by the time-matching
construction.
The generalized Langevin formulation highlights the dynamical meaning of the log-aging memory kernel. In aging systems, the relation between fluctuations and dissipation can depart from its stationary equilibrium form~\cite{Cugliandolo_2011,GFDT_WANG1999341,RKubo_1966}. In the GFPE, the kernel appears as a memory term in the probability-density evolution; in the GLE, a causal kernel may appear as a friction memory in the trajectory equation. A noise covariance must, however, be symmetric and positive semidefinite; it cannot be identified with a causal response kernel without a separate microscopic derivation of the appropriate fluctuation--dissipation relation.

The three objects

\begin{equation*}
K(t,t'),
\qquad
\langle \eta(t)\eta(t')\rangle,
\qquad
s(t)\sim \ln(t/t_0)
\end{equation*}

play different mathematical roles. The clock specifies how operational time accumulates, the retarded kernel specifies causal response, and the noise covariance specifies fluctuations. Scale covariance may motivate compatible functional forms for all three, but it does not make them identical or derive one from another. A physical GLE additionally requires a symmetric positive-semidefinite covariance and a preparation-dependent fluctuation--response relation.

Under the Gaussian closure adopted above, the GLE reproduces the logarithmic free-particle covariance, the constant-force response, and the conversion of exponential internal-time relaxation into algebraic laboratory-time relaxation. These are second-order and linear-response statements. They do not establish the discrete event count, the multiplier density, the fixed-$n$ algebraic duration tail, or the full distribution of the TA-MSD.

The free-particle result

\begin{equation*}
\langle x^2(t)\rangle
\propto
\ln(t/t_0)
\end{equation*}
shows the ultraslow growth of fluctuations. The constant-force response

\begin{equation*}
\langle x(t)\rangle_F
\propto
F\ln(t/t_0)
\end{equation*}
shows that linear response is slowed by the same clock. The harmonic autocorrelation

\begin{equation*}
C_x(t)
\propto
\left(
\frac{t_0}{t}
\right)^\gamma
\end{equation*}
shows that exponential relaxation in internal time becomes algebraic relaxation in laboratory time.

The agreement of these observables should therefore be read at the level of the
intended reduced description. It does not by itself imply equality with the
underlying event process, which would require agreement of the full
finite-dimensional path distributions, not only of one- and two-point
observables. Together, these results demonstrate that log-aging diffusion is
not merely diffusion with a small effective diffusion coefficient; it is a
nonstationary dynamical regime whose different reduced descriptions retain
different parts of the clock statistics.

\section{Microscopic Routes, Physical Realizations, and Experimental Discriminants}
\label{sec:physical}

Section I defined the observables that reveal aging. The next question is physical: what can the effective events, waiting times, and clocks represent in real disordered matter? This section supplies that dictionary while keeping a strict distinction between a useful coarse-grained description and a microscopic derivation.

Across the examples below, a waiting interval may represent residence in a trap, pinning of a domain wall, localization in a random-energy valley, persistence of a many-body electronic configuration, or a metastable collective spin state. A jump may represent a particle displacement, a hopping event, a domain rearrangement, or a change in a coarse-grained observable. The shared content is not a literal common event, but the emergence of broad time scales, nonstationarity, and history-dependent observables.

Sinai-type landscapes, disordered magnets, electron glasses, and mean-field spin glasses are used here to mark both the reach and the limits of the framework. In each case, the question is whether an effective clock captures the relevant slow dynamics and which correlations are discarded by a renewal reduction. The section is therefore a guide to interpretation, not a complete review or a claim of microscopic equivalence.

\subsection{Coarse Graining: From Microscopic Dynamics to Effective Stochastic Variables}

Many disordered and glassy systems are characterized by rugged energy landscapes, broadly distributed relaxation times and metastable configurations. The microscopic dynamics may be very different from one system to another. A tracer particle may diffuse in a random potential, a domain wall may move through a disordered magnetic background, an electron may hop between localized states, or a spin configuration may evolve among many metastable states in a high-dimensional free-energy landscape. Nevertheless, at a coarse-grained level these systems often share similar dynamical signatures: long periods of inactivity, intermittent rearrangements, nonexponential relaxation, history dependence and aging.

This motivates an effective stochastic description. In such a description, microscopic details are partially integrated out and replaced by stochastic ingredients such as waiting-time distributions, jump statistics, memory kernels or random time changes. For example, a long residence time in a metastable configuration may be represented as a waiting time\cite{Scher1975}, while a rearrangement or hopping event may be represented as a jump. If the residence-time distribution is broad, the resulting dynamics may display slow relaxation, nonstationarity and aging.

The same stochastic object may therefore have different physical meanings in different systems. A waiting-time distribution in a trap model describes residence times in metastable states. In hopping transport, it may describe the time between activated electronic hops. In a disordered magnet, it may represent the time over which a pinned domain remains immobile before rearranging. In a random landscape, it may describe residence in a valley before escape. This flexibility is not a weakness of the stochastic description; it is precisely what allows different microscopic systems to be compared within a common language.

%The CTRW picture is particularly useful when the slow dynamics can be approximated as a sequence of trapping and transition events. However, many physical systems contain spatial correlations, quenched disorder, interactions, or collective degrees of freedom. In such cases, the CTRW should be regarded as an effective or limiting description rather than as a microscopic model. The broader stochastic framework developed in this work allows renewal aging, memory-kernel dynamics, and trajectory-level descriptions to be compared within a unified language.

\subsection{Sinai Diffusion and Quenched Random Landscapes}

A paradigmatic example of slow dynamics in a disordered environment is Sinai diffusion\cite{Sin83}. In this problem, a particle moves in a one-dimensional quenched random force field, or equivalently in a random potential landscape. Since the typical potential barriers grow with distance, transport becomes extremely slow. Instead of ordinary diffusive scaling, the typical displacement grows only logarithmically with time, commonly expressed as
\begin{equation}
|x(t)|\sim (\ln t)^2,
\end{equation}
up to model-dependent constants and scaling conventions.

Sinai diffusion illustrates how quenched disorder can generate ultraslow motion, strong sample-to-sample fluctuations, and aging~\cite{le1999random,PhysRevE.57.6296}. The particle spends long periods trapped in deep valleys of the random potential and occasionally escapes over large barriers. This phenomenology is closely reminiscent of trapping dynamics and is therefore naturally connected to the physical intuition underlying renewal aging and continuous-time random walk (CTRW) descriptions.

There is, however, an important distinction. In standard CTRW models, the waiting times are usually treated as independent random variables drawn from an annealed distribution. In Sinai diffusion, by contrast, the disorder is quenched in space: the same random landscape is visited repeatedly by the particle. Consequently, spatial correlations, returns to previously visited valleys, and sample-specific effects play an essential role. The process is therefore not a simple renewal CTRW. Nevertheless, the concepts of broad trapping-time distributions, aging propagators, and two-time observables remain highly useful for characterizing its slow nonequilibrium dynamics.
\begin{figure}
    \centering
    \includegraphics[width=\linewidth]{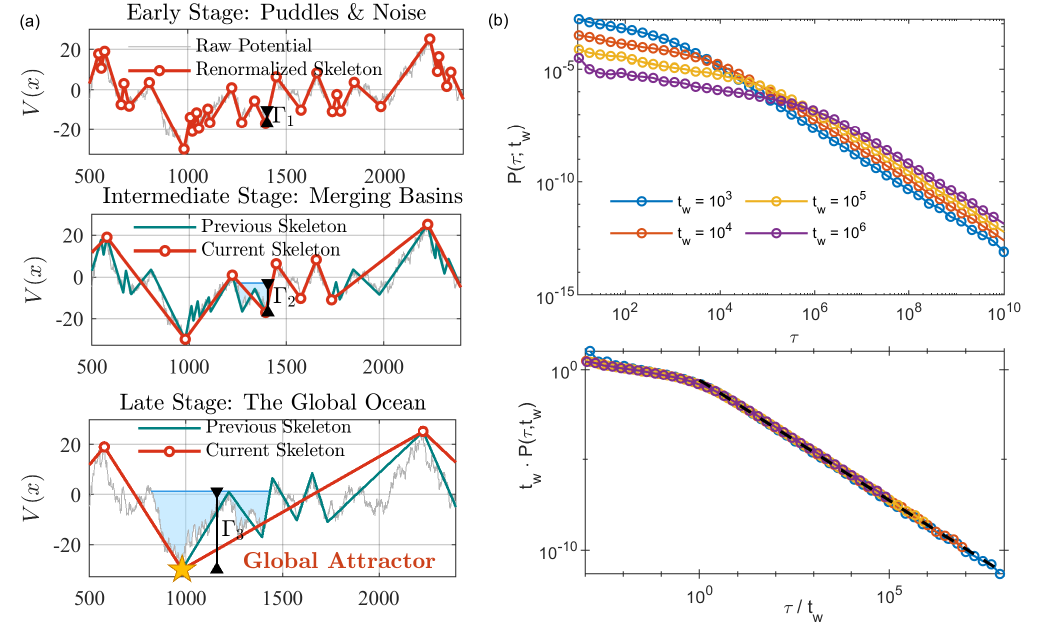}
\caption{
(a) Renormalization-group picture for dynamics in a rigid Sinai-type landscape. 
The particle moves in a quenched random potential with correlations satisfying $\overline{[U(x)-U(y)]^2}\sim \|x-y\|$. 
During the evolution, barriers smaller than the current dynamical scale are overcome and become effectively equilibrated, while the remaining larger barriers control the subsequent slow dynamics. 
Thus, coarse graining eliminates small barriers and leaves an effective landscape dominated by larger barriers. 
(b) Aging dynamics associated with the remaining dominant barriers. 
The statistics of the next trapping or barrier-crossing event depend on the history accumulated before the observation time. 
As a result, the longer the system has evolved, the longer the typical additional waiting time becomes. 
Inset: scaling collapse of the waiting-time distribution as a function of $t/t_w$.
}

    \label{fig:sinai}
\end{figure}

\subsubsection{Biased Sinai dynamics and effective barrier statistics}

We now consider a biased Sinai process, in which a particle moves in a random potential under a constant external bias. Strong-disorder renormalization group analysis shows that the effective energy barriers opposing the motion follow an exponential distribution~\cite{le1999random} (Fig.~\ref{fig:sinai}(a)),
\begin{equation}\label{eq_rhoE}
\rho(E)\sim \exp(-2\delta E),
\end{equation}
where \(\rho(E)\) denotes the density of effective trap barriers and \(\delta\) is a parameter determined by the bias. The emergence of an exponential barrier distribution can be understood as a robust consequence of extreme-value statistics. During the renormalization procedure, fast local equilibration processes are successively eliminated, while the slowest barriers, which dominate the long-time dynamics, are retained. In this sense, the SDRG focuses on the record-like barriers that still remain relevant at the observation time.

As time increases, thermal activation allows the particle to overcome progressively higher barriers\cite{PhysRevE.57.6296}. Barriers lower than the current dynamical scale are effectively equilibrated, while barriers exceeding all previously encountered ones control the subsequent slow relaxation. It is therefore natural to focus on the sequence of record barriers,
\begin{equation}
E_0<E_1<E_2<\cdots<E_n .
\end{equation}
The joint probability density for such an ordered sequence may be written as
\begin{equation}
\rho(E_1,E_2,\ldots,E_n)
=
\frac{\rho(E_1)}{\int_{0}^{\infty}\rho(E)dE}
\frac{\rho(E_2)}{\int_{E_1}^{\infty}\rho(E)dE}
\cdots
\frac{\rho(E_n)}{\int_{E_{n-1}}^{\infty}\rho(E)dE}.
\end{equation}
For an exponential distribution, Eq. (\ref{eq_rhoE}),
this expression factorizes in terms of the barrier increments
\begin{equation}
\Delta E_i=E_{i+1}-E_i .
\end{equation}
Indeed, one obtains
\begin{equation}
\frac{\rho(E_i+\Delta E_i)}
{\int_{E_i}^{\infty}\rho(E)dE}
\sim
\exp(-2\delta \Delta E_i).
\end{equation}
Thus the increments \(\Delta E_i\) are independent and identically distributed exponential random variables,
\begin{equation}
p(\Delta E_i)\sim \exp(-2\delta \Delta E_i).
\end{equation}
This memoryless property is a direct consequence of the exponential form of the effective barrier distribution. It implies that the sequence of record barriers grows as a sum of independent increments,
\begin{equation}
E_n=E_0+\sum_{i=0}^{n-1}\Delta E_i .
\end{equation}
Therefore, the typical barrier height increases linearly with the number of record events,
\begin{equation}
E_n\propto n .
\end{equation}
Equivalently, the record-breaking process in barrier space has a Poissonian structure\cite{P.Sibani_2003}. This observation provides a simple physical basis for the logarithmic aging behavior of the biased Sinai dynamics.

\subsubsection{Arrhenius mapping from barrier increments to power-law event times}

According to the Arrhenius activation law, the time required to overcome a barrier of height \(E\) is
\begin{equation}
t=t_0\exp\left(\frac{E}{k_B T}\right),
\end{equation}
where \(t_0\) is a microscopic time scale and \(T\) is the temperature. Hence, the barrier scale explored up to time \(t\) is
\begin{equation}
E(t)=k_B T\ln\left(\frac{t}{t_0}\right).
\end{equation}
Since the barrier increments are exponentially distributed, the corresponding distribution of activation times becomes a power law. Starting from
\begin{equation*}
p(\Delta E)\sim \exp(-2\delta \Delta E),
\end{equation*}
and using
\begin{equation}
\tau=t_0\exp\left(\frac{\Delta E}{k_B T}\right),
\end{equation}
one obtains
\begin{equation}
p(\tau)d\tau=p(\Delta E)d\Delta E .
\end{equation}
Because
\begin{equation}
\Delta E=k_B T\ln\left(\frac{\tau}{t_0}\right),
\qquad
d\Delta E=k_B T\frac{d\tau}{\tau},
\end{equation}
the waiting-time distribution takes the asymptotic form
\begin{equation}
\psi(\tau)\sim \frac{1}{\tau^{1+\alpha}},
\end{equation}
with
\begin{equation}
\alpha=2\delta k_B T .
\end{equation}
Thus the exponential distribution of barriers is transformed by Arrhenius activation into a heavy-tailed distribution of trapping times. This is the basic mechanism behind aging in the biased Sinai process.

The same result can be expressed in terms of the absolute time of successive record events. Suppose that the previous record event occurred at time \(t_{n-1}\), and that the next record event occurs at time \(t_n>t_{n-1}\). Since the corresponding barrier increment is
\begin{equation}
\Delta E_n
=
k_B T\ln\left(\frac{t_n}{t_{n-1}}\right),
\end{equation}
the exponential distribution of \(\Delta E_n\) gives
\begin{equation}
\psi(t_n|t_{n-1})
\sim
\frac{t_{n-1}^{\alpha}}{t_n^{1+\alpha}},
\qquad
t_n>t_{n-1}.
\end{equation}
This expression shows explicitly that the statistics of the next event depends on the time of the previous event(Fig.~\ref{fig:sinai}(b)). In other words, the longer the system has evolved, the longer one typically has to wait for the next record-breaking event. This dependence on the history is the physical origin of aging.

The probability that no new record event occurs up to time \(t_n\), given that the previous event occurred at \(t_{n-1}\), is
\begin{equation}
\Pi(t_n|t_{n-1})
=
\int_{t_n}^{\infty}
\psi(t'|t_{n-1})dt'
\sim
\left(\frac{t_{n-1}}{t_n}\right)^{\alpha}.
\end{equation}
This form may also be interpreted as a ratio of characteristic length scales, \(L(t)\sim t^\alpha\),
\begin{equation}
\Pi(t_n|t_{n-1})
\sim
\frac{L(t_{n-1})}{L(t_n)}.
\end{equation}
Therefore, the two-time quantities depend on the ratio of the two observation times, rather than on their difference. This is a characteristic feature of aging dynamics.

\subsubsection{Forward Recurrence in an Aged Renewal Process}

It is useful to distinguish the conditional record-time distribution above from the standard forward waiting-time distribution in renewal theory. The former assumes that the previous event occurred exactly at \(t_{n-1}\), while the latter describes an observation made at an arbitrary age \(t_a\), which generally lies inside an ongoing trapping interval. Therefore, the forward waiting-time density requires an additional average over the time of the last renewal event before \(t_a\), as well as over the number of renewals that have already occurred.\cite{Tauberium_klafter2011}

Let \(t_a\) be the age of the process, namely the time elapsed between the preparation of the system and the beginning of the observation. The forward waiting time \(t\) is defined as the time interval between the observation time \(t_a\) and the next renewal event. Suppose that exactly \(n\) renewal events have occurred before \(t_a\). Let \(T_n\) denote the time of the \(n\)-th renewal event, and let \(\psi_n(t')\) be the probability density of \(T_n=t'\). If the \(n\)-th event occurs at time \(t'<t_a\), then the next waiting time must last for a total duration
\begin{equation}
\tau=t_a-t'+t
\end{equation}
in order for the next renewal event to occur at time \(t_a+t\). Therefore, the conditional probability density of the forward waiting time, given that exactly \(n\) renewals have occurred before \(t_a\), is
\begin{equation}
\phi_n(t;t_a)
=
\int_0^{t_a}
\psi_n(t')\,
\psi(t_a-t'+t)\,dt' .
\end{equation}
Here \(\psi(\tau)\) is the waiting-time distribution between two successive renewal events.

The full forward waiting-time density is obtained by summing over all possible values of \(n\):
\begin{equation}
\psi_1(t,t_a)
=
\sum_{n=0}^{\infty}\phi_n(t;t_a)
=
\int_0^{t_a}
\left[
\sum_{n=0}^{\infty}\psi_n(t')
\right]
\psi(t_a-t'+t)\,dt' .
\end{equation}
The quantity
\begin{equation}
m(t')=\sum_{n=0}^{\infty}\psi_n(t')
\end{equation}
is the renewal density. Hence the forward waiting-time distribution can be written compactly as
\begin{equation}
\psi_1(t,t_a)
=
\int_0^{t_a}
m(t')\,
\psi(t_a-t'+t)\,dt' .
\end{equation}
This expression makes clear that the forward waiting-time distribution is not simply the original waiting-time distribution \(\psi(t)\). Instead, it is biased by the age \(t_a\) of the process and by the renewal history before \(t_a\). This is the mathematical origin of aging in renewal processes.

For a broad waiting-time distribution of the form
\begin{equation}
\psi(\tau)
\sim
\frac{A}{\tau^{1+\alpha}},
\qquad 0<\alpha<1,
\end{equation}
the mean waiting time diverges. In this case the renewal process never reaches a stationary regime, and the forward waiting-time distribution depends explicitly on the age \(t_a\). In the scaling limit \(t,t_a\gg t_0\), one obtains the well-known aging form
\begin{equation}
\psi_1(t,t_a)
\simeq
\frac{\sin(\pi\alpha)}{\pi}
\frac{t_a^{\alpha}}{t^{\alpha}(t+t_a)} .
\end{equation}
Equivalently, in terms of the scaled variable
\begin{equation}
u=\frac{t}{t_a},
\end{equation}
this can be written as
\begin{equation}
\psi_1(t,t_a)
\simeq
\frac{1}{t_a}
\frac{\sin(\pi\alpha)}{\pi}
\frac{1}{u^{\alpha}(1+u)} .
\end{equation}
This scaling form shows explicitly that the typical forward waiting time is of the same order as the age \(t_a\). Therefore, an older system remains trapped for a longer residual time. This is the central signature of renewal aging.

The corresponding persistence or survival probability, namely the probability that no renewal event occurs during the time interval \((t_a,t_a+t)\), is
\begin{equation}
\Psi_1(t,t_a)
=
\int_t^{\infty}
\psi_1(t',t_a)dt' .
\end{equation}
For a long-tailed process, this survival probability is a function of the ratio \(t/t_a\), rather than of the time difference alone. This dependence on the ratio between observation time and system age is the characteristic feature of aging systems.

In the biased Sinai problem, the long-tailed waiting-time distribution originates from the exponential distribution of effective barriers. The limit of RG gives the  barrier increments as
\begin{equation*}
p(\Delta E)\sim \exp(-2\delta \Delta E),
\end{equation*}
and the activation time follows the Arrhenius law
\begin{equation}
\tau=t_0\exp\left(\frac{\Delta E}{k_B T}\right),
\end{equation}
then the corresponding waiting-time distribution has the algebraic tail
\begin{equation}
\psi(\tau)\sim \frac{1}{\tau^{1+\alpha}},
\qquad
\alpha=2\delta k_B T .
\end{equation}
Thus the CTRW aging formula for the forward waiting time provides an effective renewal description of the record-barrier dynamics\cite{log_theory_PhysRevLett.71.1482,Sibani2007LinearResponse} in the biased Sinai landscape. The important physical message is that, although the Sinai process is generated by quenched spatial disorder and is not a simple annealed renewal CTRW, its long-time aging behavior can nevertheless be captured by the same forward waiting-time structure. The reason is that the relevant slow events are record barrier crossings, whose statistics become asymptotically renewal-like due to the exponential, memoryless distribution of barrier increments.

\subsubsection{Relation to Griffiths phases and record dynamics}

The broader connection between biased Sinai dynamics and random transverse-field Ising chains is most naturally formulated through their shared strong-disorder and random-walk structures~\cite{PhysRevB.57.11404}. This analogy is physically useful, but it does not establish a literal renewal mapping or a universal identification of the control parameter $\delta$ across the two models.

The recurrence of broad barriers and record-like escapes in systems as different as Sinai landscapes and Griffiths regimes motivates an effective description of aging. Bouchaud's trap model provides a canonical example in which long-time dynamics is dominated by deep states~\cite{log_theory_bouchaud1992weak}. Record-dynamics approaches likewise describe rare, irreversible rearrangements or ``quakes' whose rate decreases with the system age~\cite{log_theory_PhysRevLett.71.1482,log_theory_PhysRevE.98.020602,log_theory_Robe_2016}. These analogies identify a common phenomenology, not a universal microscopic mechanism.

As the system evolves, it explores progressively deeper valleys in the effective energy landscape. Record-like descriptions of aging formalize this progressive exploration through increasingly rare irreversible events~\cite{log_theory_PhysRevE.98.020602,logaging_PhysRevLett.110.208301}. Such mechanisms can generate broad waiting-time statistics and aging two-time observables, but the precise relation between barrier statistics and a logarithmic clock remains model dependent.

\subsection{Disordered Magnets: Activated Dynamics without a Generic Renewal Mapping}

Disordered magnetic systems provide another important class of physical models exhibiting slow relaxation and aging. In the random-field Ising model, spins interact with each other while also experiencing spatially random local fields. The disorder pins domain walls and creates a complex energy landscape for magnetic rearrangements. As a result, relaxation can become slow and history dependent, and the dynamics may involve domain-wall pinning, avalanche-like rearrangements and activated barrier crossing.

In random transverse-field Ising systems, quantum fluctuations compete with disorder. Near infinite-randomness fixed points, the dynamics is often characterized by activated rather than power-law scaling, and the relation to random-walk descriptions provides a useful analytical perspective~\cite{PhysRevB.57.11404}. Griffiths effects\cite{Griffiths1969,DelMaestro_Thesis,PhysRevB.51.6411,IGLOI2005277} can also lead to broad distributions of local relaxation times, because rare regions may remain dynamically slow over very long time scales. These features naturally suggest a connection with stochastic descriptions based on broadly distributed waiting times, slow modes, and nonstationary response functions.

The relation between disordered magnetic systems and aging is therefore indirect but physically meaningful. The elementary stochastic events are not necessarily particle jumps in real space. They may correspond instead to domain-wall motion, spin flips, cluster rearrangements, avalanches or tunneling events. The waiting times represent residence times in metastable magnetic configurations or the time needed to overcome disorder-induced barriers. Thus, CTRW language can serve as a coarse-grained description of certain slow processes, while generalized Fokker--Planck, Langevin or memory-kernel formulations may be more appropriate when collective modes, spatial correlations and history-dependent barriers are important.

\subsection{Electron Glasses and Interacting Hopping Dynamics}

Electron glasses\cite{PhysRevLett.49.758,PhysRevB.57.9736} arise in strongly disordered electronic systems where electrons are localized and interact through long-range Coulomb forces. Transport occurs through hopping between localized states, but the hopping rates are strongly affected by disorder, interactions, and the Coulomb gap. Experimentally, electron glasses exhibit extremely slow relaxation, memory effects, history dependence, and aging phenomena such as waiting-time-dependent conductance relaxation~\cite{electron_glass_PhysRevB.68.184204,Anderson_PhysRevLett.84.3402,PhysRevLett.81.669} .

The connection with the stochastic aging framework is natural. Hopping events between localized sites play a role analogous to jumps, while long residence times in local electronic configurations correspond to trapping times. A broad distribution of hopping rates or activation barriers can generate a broad distribution of effective waiting times, leading to slow relaxation and aging.

At the same time, electron glasses are generally not renewal processes. The Coulomb interaction couples different localized electrons, so a hop by one electron changes the energy landscape seen by others. The waiting-time statistics can therefore depend on the evolving many-body configuration rather than being independent and identically distributed. This suggests that electron-glass dynamics may require non-renewal extensions of CTRW, interacting trap models, memory kernels or generalized Langevin-type descriptions. In this sense, electron glasses provide a physical setting where the limitations of simple renewal aging become clear, while the broader stochastic framework remains useful.

\subsection{Mean-Field Spin Glasses and Trap-Like Aging}

Spin glasses are among the most important physical systems in which aging has been studied. The Sherrington--Kirkpatrick model is a mean-field spin-glass model with infinite-range random couplings. Its equilibrium structure is characterized by a complex free-energy landscape with many metastable states. Dynamically, spin glasses exhibit slow relaxation, history dependence, aging, and violations of equilibrium fluctuation--dissipation relations~\cite{PhysRevLett.43.1754,PhysRevLett.51.911,PhysRevB.38.373,log_theory_PhysRevLett.71.173,Cugliandolo_2011,Vincent2007}.

In spin-glass dynamics, aging is most naturally described through two-time correlation and response functions. A typical spin autocorrelation function can be written as

\begin{equation}
C(t,t_w)=\frac{1}{N}\sum_{i=1}^{N}
\langle S_i(t_w+t)S_i(t_w)\rangle,
\end{equation}
where $t_w$ is the waiting time after preparation. In an aging regime, this function depends on both $t_w$ and $t$, and cannot be reduced to a function of their difference alone. This is directly analogous to the time-translation symmetry breaking discussed earlier.

The connection between spin glasses and renewal-type aging is often made phenomenologically through trap models~\cite{log_theory_bouchaud1992weak,bouchaud1995aging,PhysRevLett.53.958}. In a trap interpretation, the system moves among metastable states separated by broadly distributed barriers. Long residence times in deep traps lead to slow relaxation and aging. The Bouchaud trap model\cite{log_theory_bouchaud1992weak} is a particularly important effective model in this context, because it translates the complex energy-landscape picture of glassy systems into a stochastic process with broad trapping times.

Nevertheless, the Sherrington--Kirkpatrick model is not itself a simple renewal CTRW. Its dynamics involves collective spin degrees of freedom, mean-field correlations and a hierarchical organization of states. Therefore, renewal models and CTRW-type descriptions should be viewed as effective phenomenological reductions of certain aspects of spin-glass aging, rather than complete microscopic descriptions. The more general objects introduced in this work, including aging propagators, two-time functions, response functions and memory kernels, provide a common language for comparing spin-glass aging with other disordered systems.

\subsection{Jammed Colloids and Record-Resolved Aging}
\label{sec:jammed_colloids}

Jammed colloids supply an unusually direct experimental bridge between an abstract clock and resolved rearrangement events. Particle-level measurements separate reversible cage rattling from irreversible cage-breaking ``quakes.'' Robe \emph{et al.} found that the quake rate decreases approximately as the inverse system age and that both the mean-square displacement and a mesoscopic length scale grow logarithmically~\cite{log_theory_Robe_2016}. These observations support the record-dynamics hypothesis that rare irreversible events are approximately stationary in logarithmic time~\cite{P.Sibani_2003,log_theory_PhysRevE.98.020602}.

The agreement at the level of the mean rate, however, does not identify the event law. A homogeneous log-Poisson process has exponentially distributed logarithmic increments $U_n=\ln(T_{n+1}/T_n)$ and therefore satisfies $\operatorname{Var}U_n=\langle U_n\rangle^2$. The present multiplicative clock instead predicts
\begin{equation}
f_\alpha(u)
=
\frac{\sin(\pi\alpha)}{\pi}
\left(e^u-1\right)^{-\alpha},
\qquad u>0,
\end{equation}
or, equivalently, the multiplier density $g_\alpha(y)$ in Sec.~II, with
\begin{equation}
\langle U\rangle=\kappa_1=\Psi(1)-\Psi(\alpha),
\qquad
\operatorname{Var}U=\kappa_2=\Psi_1(\alpha)-\Psi_1(1).
\end{equation}
Thus an event-resolved colloidal data set can distinguish the two mechanisms without fitting a displacement curve: measure the full distribution and serial correlations of $U_n$, not only the average number of quakes per logarithmic decade. The algebra in the first displayed density uses $y=e^u$ and $dy/du=e^u$, so that $f_\alpha(u)=e^u g_\alpha(e^u)=(\sin\pi\alpha/\pi)(e^u-1)^{-\alpha}$.

\subsection{Mechanical Instability Cascades and Crumpled Structures}
\label{sec:instability_cascades}

Mechanical systems provide a second experimentally accessible route to logarithmic aging. In crumpled sheets and disordered mechanical networks, slow relaxation can emerge from cascades of coupled instabilities rather than from a particle repeatedly escaping statistically independent traps. Shohat, Friedman, and Lahini showed that such cascades generate logarithmic aging and reproduce protocol-dependent relaxation over broad time windows~\cite{log_theory_shohat2023logarithmic}. This result is important for the present review because it supplies a concrete counterexample to mechanism-by-curve-fitting: the same logarithmic relaxation may arise from collective deterministic or weakly stochastic instabilities.

The connection to a multiplicative clock must therefore be tested, not assumed. An instability cascade is compatible with the present theory only if suitably defined irreversible transitions satisfy three linked conditions: the distribution of $T_{n+1}/T_n$ collapses across event number and preparation age; correlations between successive logarithmic increments are negligible at the coarse-graining scale; and the measured count cumulants agree with the same $\kappa_1$ and $\kappa_2$ inferred from the multiplier distribution. Systematic violations would be equally informative. Correlated increments would motivate a Markov-additive or long-memory clock, whereas event-size--waiting-time coupling would require a coupled CTRW rather than the decoupled spatial process used here.

\subsection{What Experiments Establish---and What Still Identifies the Clock}
\label{sec:experimental_status}

The existing experimental literature establishes several ingredients separately. Electron-glass protocols demonstrate waiting-time-dependent conductance relaxation and long-lived memory~\cite{electron_glass_PhysRevB.68.184204,Anderson_PhysRevLett.84.3402,PhysRevLett.81.669,Pollak_2012}. Spin-glass measurements establish two-time aging, rejuvenation, memory, and non-equilibrium response~\cite{Vincent2007,Cugliandolo_2011}. Colloidal and soft-glassy measurements reveal intermittency, dynamical heterogeneity, and event rates compatible with $1/t$~\cite{Cipelletti_2005,10.1063/1.480896,log_theory_Robe_2016}. Mechanical instability cascades demonstrate that a collective mechanism can produce logarithmic aging~\cite{log_theory_shohat2023logarithmic}. None of these observations, taken alone, establishes the specific multiplier density, its Mellin transform, or the linked event-time and TA-MSD predictions developed here.

The appropriate claim is therefore quantitative and conditional. The framework can explain an experiment with one parameter set only when event-time, count, transport, and finite-window statistics close consistently. A logarithmic one-time relaxation is necessary for relevance but insufficient for identification. Conversely, failure of the multiplier test does not invalidate the observed aging; it locates the material in a different microscopic class, such as a correlated record process, quenched landscape, interacting trap system, or deterministic instability cascade.

\subsection{A Joint, No-Refitting Test of the Multiplicative Clock}
\label{sec:experimental_protocol}

Table~\ref{tab:experimental_discriminants} converts the theory into an overdetermined experimental test. Events must first be defined independently of the clock fit. The distribution and serial correlations of $U_n=\ln(T_{n+1}/T_n)$ then determine whether an independent-multiplier description is admissible and fix $\alpha$, $\kappa_1$, and $\kappa_2$. Those parameters must subsequently predict the count cumulants, spatial moments after a separate calibration of jump statistics, first-passage behavior, and finite-window observables without refitting. No single logarithmic relaxation curve, inverse-age rate, or EB trend identifies the model; identification requires joint closure across the rows of the table.

\begin{table}[htbp]
\centering
\begingroup\footnotesize\setstretch{1.15}
\setlength{\tabcolsep}{4pt}
\caption{Joint tests of the independent-multiplier clock. Parameters are inferred from event-time ratios and then propagated to count, transport, and finite-window observables; no single row identifies the model.}
\label{tab:experimental_discriminants}
\begin{tabular}{@{}p{0.20\linewidth}p{0.38\linewidth}p{0.32\linewidth}@{}}
\toprule
Measurement & Prediction of the present event law & Interpretation or closest confounder \\
\midrule
Log increment
$U_n=\ln(T_{n+1}/T_n)$ &
i.i.d.\ density
$f_\alpha(u)=(\sin\pi\alpha/\pi)(e^u-1)^{-\alpha}$ for $u>0$ and $0<\alpha<1$,
with mean $\kappa_1$ and variance $\kappa_2$ &
A homogeneous log-Poisson process gives exponential $U_n$; reproducible serial correlations falsify the independent-multiplier assumption \\
Mean and variance of $N(t)$ &
$\langle N\rangle\simeq L/\kappa_1$ and
$\operatorname{Var}N\simeq(\kappa_2/\kappa_1^3)L$,
where $L=\ln(t/t_0)$ &
$M(t)\simeq1/(\kappa_1t)$ alone is not diagnostic; log-Poisson has the same inverse-age rate but Fano factor $1$ rather than $\kappa_2/\kappa_1^2$ \\
$n$th-event-time density $\rho_n(t)$ &
Central log-normal sector and fixed-$n$ tail
$\rho_n(t)\propto t^{-1-\alpha}L^{n-1}$ &
A log-Poisson process can also generate an algebraic fixed-$n$ tail; the exponent and prefactor must close with the independently measured $f_\alpha$ \\
Spatial moments after jump calibration &
$\langle X\rangle\simeq\mu_\xi L/\kappa_1$ and
$\operatorname{Var}X\simeq\sigma_\xi^2L/\kappa_1
\!+\mu_\xi^2\kappa_2L/\kappa_1^3$ &
A deterministic logarithmic time change reproduces the first term but not the random-count variance correction \\
First passage and target survival &
Operational-time laws evaluated at $L/\kappa_1$, including the stated transit-current and dimension-dependent target-survival asymptotes &
Geometry and boundary conditions change prefactors and sometimes asymptotic sectors; the same $\kappa_1$ must be used without refitting \\
TA-MSD, EB, and $\eta$ &
A measurement-time-dependent mean,
$\mathrm{EB}\sim\kappa_2/[\kappa_1\ln(T/t_0)]$, and a nonzero ensemble--time deviation &
Nonstationary Gaussian time changes can also have $\mathrm{EB}\to0$ with ensemble--time mismatch; multiplier/count statistics and the small-$\xi$ sector provide the separation \\
\bottomrule
\end{tabular}
\endgroup
\end{table}

\subsection{Common Structures and the Limits of Renewal Descriptions}

The examples discussed above are microscopically different, but they share several dynamical structures. First, they contain broad distributions of relaxation or trapping times. Second, their dynamics is often controlled by barriers, metastable states or rare events. Third, experimentally or numerically measured observables depend on the waiting time after preparation. Fourth, two-time correlation and response functions generally break time-translation invariance. Finally, finite-time trajectories or finite-window measurements may fail to represent ensemble behavior, producing apparent non-self-averaging or weak ergodicity breaking.

These common features justify the use of stochastic-process concepts such as waiting-time distributions, aging propagators, memory kernels, random time changes, and time-averaged observables. However, it is equally important to recognize the limitations of the simplest renewal picture~\cite{review_Johannes_PhysRevX.4.011028,review_Andrey_C3CP53056F}. Standard CTRW assumes renewal dynamics and statistically independent waiting times and jumps. This assumption is appropriate for some trap-like systems, but it may fail in quenched random landscapes, interacting electronic systems, spin glasses with collective rearrangements, or disordered quantum magnets with strong spatial correlations.

Thus, the stochastic framework developed here should be understood as a hierarchy of descriptions. CTRW provides the simplest event-based picture of aging. Generalized Fokker--Planck equations describe the evolution of probability densities and allow one to incorporate memory kernels, external fields and boundary conditions. Langevin and subordination formulations provide trajectory-level interpretations and can encode random time changes or non-Markovian noise. Together, these descriptions offer a flexible framework for connecting aging phenomenology across different physical systems.

The physical systems discussed in this section motivate a common language but not a universal mechanism. They show that aging can arise from slow exploration of complex landscapes, broadly distributed time scales, interacting rearrangements, or instability cascades. The preceding sections developed the event, GFPE, time-matching, and GLE levels; the discussion below now states precisely which conclusions survive across those levels and which require the specific independent-multiplier clock.

\section{Discussion and Conclusion}
\label{sec:discussion}
\label{sec_conclusion}
\begin{figure}
    \centering
    \includegraphics[width = \linewidth]{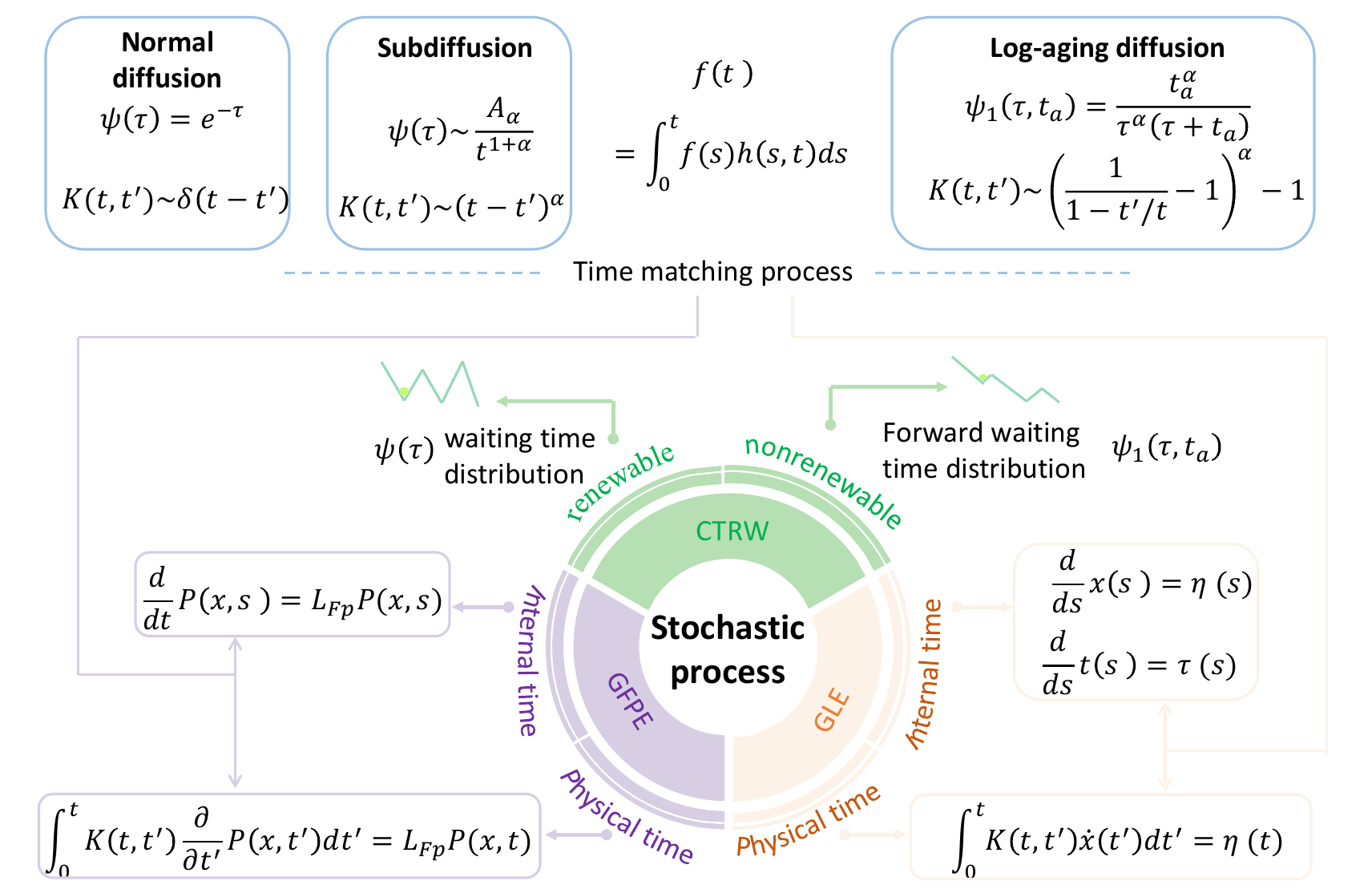}
    \caption{Relations among event-level and reduced descriptions of logarithmic aging. Normal renewal CTRW has i.i.d. waiting times with finite mean; renewal subdiffusion retains i.i.d. waiting times but may have a divergent mean and an age-dependent first residual interval. The present log-aging construction iterates the conditional residual-time law at successive event ages, producing age-dependent physical waiting intervals and independent multipliers in logarithmic time. Its event law generates a random clock whose one-point diffusion limit can be represented by a scale-covariant GFPE. A GLE may reproduce selected response and covariance functions only after an additional closure; the arrows therefore denote controlled mappings or information-reducing representations, not universal pathwise equivalence.}
    \label{fig:overview_legacy}
\end{figure}

\subsection{What the Multiplicative Clock Establishes}

In this work, we have developed a stochastic framework for logarithmic aging diffusion. The central message is that the ultraslow dynamics considered here is not simply diffusion with a small effective diffusion coefficient. Rather, it is governed by a different clock. The motion becomes simple when described in an internal operational time, but appears slow, history dependent and non-Markovian when observed in laboratory time. This separation between internal time and laboratory time provides the organizing principle for the whole theory(Fig.~\ref{fig:overview_legacy}).

The key feature of log-aging diffusion is that the mean operational time grows only logarithmically with the physical observation time. As a result, increasing laboratory time by orders of magnitude produces only an additive increase in the effective number of events. This explains the ultraslow drift and spreading and the persistence of finite-window effects. The statistical conclusion is more specific: central logarithmic increments determine ordinary event-count moments, whereas rare low-activity histories determine fixed-event duration tails and positive duration moments. The logarithmic clock is therefore not a fitted time axis but the common origin of a linked, observable-dependent hierarchy.

A central point of the paper is that log-aging diffusion forms a distinct dynamical class. It should not be viewed merely as an extreme limit of ordinary subdiffusion. In standard subdiffusion, the process is slow because the operational time grows as a power of laboratory time. In log-aging diffusion, the operational time grows only logarithmically. This difference changes the symmetry of the problem. Standard subdiffusive models are often organized by time differences and are naturally treated with Laplace methods. Log-aging dynamics is instead organized by time ratios and by the age of the process, making Mellin methods the natural language.

The CTRW construction provides the event-level stochastic basis for this picture. It shows how iterated age-dependent waiting laws generate an internal clock that advances logarithmically. In this description, transport is controlled by event statistics rather than by physical time directly. The mean displacement and mean-square displacement grow logarithmically, but the full event-time density contains two asymptotic sectors. Typical trajectories follow a Gaussian central limit in $\ln t$; fixed-event, long-duration histories form an algebraic tail. Their influence must be stated observable by observable rather than summarized as universal rare-event dominance.

This structure leads to a particularly instructive form of weak ergodicity breaking. At finite measurement time, individual time averages fluctuate because only $O[\ln(T/t_0)]$ events are sampled. Their relative scatter nevertheless narrows as $\mathrm{EB}\sim1/\ln(T/t_0)$. Ergodicity is not restored because the mean time-averaged MSD remains measurement-time dependent and does not converge to the ensemble lag MSD. Narrowing relative dispersion and restoration of ensemble--time equivalence are therefore distinct limits.

\subsection{A Hierarchy of Exact Results and Reduced Representations}

The generalized Fokker--Planck formulation gives a complementary continuum description. In that representation, the logarithmic clock appears as a nonstationary memory kernel. The kernel is not a function of the elapsed time alone; instead, it reflects the relative position of past times within the whole observation history. This is the continuum expression of aging. It also shows why the process breaks time-translation invariance even when the underlying spatial dynamics is time independent. The GFPE formulation is particularly useful for treating external fields, boundaries and confined geometries.

The time-matching picture provides the conceptual bridge between the CTRW and GFPE descriptions. It makes clear that different diffusion classes can share the same internal Markovian spatial process but differ in the way internal time is matched to laboratory time. Normal diffusion, subdiffusion and log-aging diffusion are therefore distinguished primarily by their clocks. This viewpoint helps clarify why their mathematical structures can look similar while their physical behavior is profoundly different.

The generalized Langevin formulation adds a response-level interpretation, but only within an explicitly stated closure. A causal friction kernel and a symmetric positive-semidefinite noise covariance are different mathematical objects. In a stationary equilibrium bath they are related by fluctuation--dissipation theory; in an aging bath their relation depends on preparation and microscopic dynamics. A Gaussian GLE can reproduce selected two-point functions and linear responses, including the conversion of exponential internal-time relaxation into algebraic laboratory-time relaxation, but it does not thereby reproduce the discrete event count, the full subordination density, or the rare event-time tail. Thermodynamic consistency must be checked at this level rather than inherited from the GFPE by notation. A GLE matched to response and covariance is a separate reduced branch unless equality of finite-dimensional distributions is proved. This hierarchy is the main structural correction to a simple ``four equivalent descriptions'' narrative.

\subsection{Microscopic Scope and Quantitative Explanatory Power}

The physical relevance of log-aging diffusion lies in its ability to describe systems where relaxation is dominated by broad, hierarchical or disorder-induced time scales. Examples include particles moving in complex energy landscapes, trap-like glassy systems, Sinai-type random environments, electron glasses, spin glasses and Griffiths regions in disordered many-body systems. These systems are microscopically different, but they share several phenomenological features: slow relaxation, aging, memory, trajectory heterogeneity and sensitivity to rare events. The framework developed here provides a common stochastic language for these phenomena.

The framework quantitatively explains a data set only when the same clock parameters account for multiple levels at once: the multiplier density, the two count cumulants, logarithmic drift and spreading, the response--fluctuation ratio, the fixed-event duration tail, first-passage scaling, and the ensemble--time-average mismatch. Existing experiments already establish subsets of this phenomenology, especially $1/t$ quake rates in jammed colloids, aging response in glasses, and logarithmic relaxation from instability cascades. The full overdetermined closure proposed in Sec.~\ref{sec:experimental_protocol} remains an experimental objective.

At the same time, the present theory should be viewed as a starting point rather than a complete description of all aging systems. The specific construction assumes independent, identically distributed multipliers in logarithmic time and spatial increments decoupled from the clock. Its physical waiting intervals are age dependent and hence are not an ordinary i.i.d. renewal sequence in laboratory time. Real disordered systems may additionally contain correlations between multipliers, spatially correlated quenched disorder, collective rearrangements, or many-body constraints. In such cases, the next-event statistics may depend on the full evolving state rather than only on the current event time. Extending the theory to correlated clocks and interacting systems is therefore an important direction for future work.

Another open problem concerns the microscopic origin of the logarithmic clock. In this work, the clock is introduced at the stochastic-process level and then shown to generate a consistent hierarchy of descriptions. In specific physical systems, however, one would like to derive it from microscopic ingredients such as barrier distributions, energy landscape structures, rare-region effects or collective rearrangement mechanisms. Establishing such microscopic derivations would strengthen the connection between log-aging diffusion and concrete experimental systems.

The fluctuation--dissipation structure also deserves further investigation. In equilibrium-like settings, noise and friction are related through a generalized fluctuation--dissipation relation. In genuinely aging environments, however, the bath itself may be out of equilibrium. This raises questions about effective temperatures, response functions and the possible violation or modification of fluctuation--dissipation relations. Log-aging diffusion provides a useful testing ground for these questions because its memory kernel has a clear scale-covariant structure.

\subsection{Conclusion}

Logarithmic aging diffusion is best defined not by a logarithmic curve but by a multiplicative stochastic clock and the web of predictions that follows from it. Iterating the conditional residual-time law produces independent increments in logarithmic time, an explicit multiplier density and its Mellin transform, logarithmic event accumulation, and a scale-covariant continuum memory operator in the spatial diffusion limit. The same event law separates typical log-normal timing fluctuations from algebraic rare-duration tails and explains how $\mathrm{EB}\to0$ can coexist with a persistent ensemble--time-average mismatch. CTRW, random-clock subordination, GFPE, and GLE formulations are therefore most useful when ordered by the stochastic information they retain. This ordering turns a formal unification into an experimental program: measure events, close the clock statistics, and require one parameter set to predict transport, response, first passage, and finite-time nonergodicity. Extending that program to correlated, interacting, and driven clocks offers a direct route from a solvable aging process to quantitative theories of glassy, disordered, and mechanically evolving matter.

\appendix
\appendixequations

\section{Mellin Transform Conventions}
\label{app_mellin}

In this Appendix we summarize the Mellin-transform conventions and several technical derivations used in the main text. For a function $f(x)$ defined on $x>0$, we use the Mellin transform

\begin{equation}
   \widehat f(p)
    =
    \mathcal{M}\{f(x)\}(p)
    =
    \int_0^\infty f(x)x^{p-1}\,dx .
\end{equation}

The inverse Mellin transform is

\begin{equation}
    f(x)
    =
    \frac{1}{2\pi i}
    \int_{c-i\infty}^{c+i\infty}
    \widehat f(p)x^{-p}\,dp ,
\end{equation}

where the constant $c$ is chosen inside the fundamental strip of convergence. In the main text, factors of $t_0^p$ appear because the natural dimensionless time variable is $t/t_0$.

For probability densities of positive random variables we also use the shifted, normalized convention
\begin{equation}
\widehat G(p)
\equiv
\mathcal M[g](p+1)
=
\int_0^\infty y^p g(y)\,dy
=
\mathbb E[Y^p],
\end{equation}
so that $\widehat G(0)=1$. This shift is stated explicitly because the standard Mellin transform of $g$ at argument $p$ is $\mathbb E[Y^{p-1}]$.

For two-time functions with scale covariance, the double Mellin transform is useful. Consider a symmetric two-time function of the form

\begin{equation}
    F(t_1,t_2)
    =
    f\left(\frac{t_1}{t_2}\right)\Theta(t_1-t_2)
    +
    f\left(\frac{t_2}{t_1}\right)\Theta(t_2-t_1).
\end{equation}

Its double Mellin transform is

\begin{equation}
\begin{aligned}
  \widehat  F(p_1,p_2)
    &=
    \int_{t_0}^{\infty}
    \int_{t_0}^{\infty}
    F(t_1,t_2)t_1^{p_1-1}t_2^{p_2-1}
    \,dt_1dt_2 .
\end{aligned}
\end{equation}

After separating the two time-ordering sectors and changing variables to the ratio of times, one obtains the identity

\begin{equation}
  \widehat  F(p_1,p_2)
    =
    t_0^{p_1+p_2}
    \frac{\widehat f(p_1)+\widehat f(p_2)}
    {p_1+p_2},
\end{equation}

where $\widehat f(p)$ denotes the corresponding single-ratio Mellin transform. This identity is repeatedly used for two-time correlations and noise kernels in the main text.

\section{Continuous-Time Random Walk Results}
\label{app_CTRW}

\subsection{Second Moment for Biased Log-Aging Diffusion}
\label{app_biased_variance}

Here we derive the asymptotic second moment for biased diffusion. Let the Fourier transform of the jump-length distribution be expanded at small wave number as

\begin{equation}
   \hat f(k)
    =
    1+iak-\frac{b^2}{2}k^2+O(k^3),
\end{equation}

where $a$ is the mean jump length and $b^2$ is the second jump moment. The Fourier--Mellin propagator yields

\begin{equation}
\begin{aligned}
    \langle \widehat x^2(p)\rangle
    &=
    -t_0^p
    \frac{\widehat G(p)}
    {p[\widehat G(p)-1]}
    \left.
    \frac{\partial^2\hat f(k)}{\partial k^2}
    \right|_{k=0}
    \\
    &\quad
    +
    2t_0^p p
    \left[
    \frac{\widehat G(p)}
    {p[1-\widehat G(p)]}
    \left.
    \frac{\partial\hat f(k)}{\partial k}
    \right|_{k=0}
    \right]^2 .
\end{aligned}
\end{equation}

Using

\begin{equation}
    \left.
    \frac{\partial\hat f(k)}{\partial k}
    \right|_{k=0}
    =
    ia,
    \qquad
    \left.
    \frac{\partial^2 \hat f(k)}{\partial k^2}
    \right|_{k=0}
    =
    -b^2,
\end{equation}

we obtain

\begin{equation}
    \langle\widehat x^2(p)\rangle
    =
    t_0^p b^2
    \frac{\widehat G(p)}
    {p[\widehat G(p)-1]}
    +
    2a^2t_0^p p
    \left[
    \frac{\widehat G(p)}
    {p[1-\widehat G(p)]}
    \right]^2 .
\end{equation}

For $p\to0$,

\begin{equation}
    \frac{\widehat G(p)}
    {p[\widehat G(p)-1]}
    \sim
    \frac{1}{\kappa_1p^2}.
\end{equation}

The second term has the expansion

\begin{equation}
    p
    \left[
    \frac{\widehat G(p)}
    {p[1-\widehat G(p)]}
    \right]^2
    \sim
    \frac{1}{\kappa_1^2p^3}
    +
    \frac{
    -6\kappa_1^2-\pi^2
    +6\partial_\alpha^2\ln\Gamma(\alpha)
    }
    {6\kappa_1^3p^2}.
\end{equation}

Therefore, inverse Mellin transformation gives the long-time asymptotic behavior

\begin{equation}
\begin{aligned}
    \langle x^2(t)\rangle
    &=
    \frac{a^2}
    {\kappa_1^2}
    \left[
    \ln\left(\frac{t}{t_0}\right)
    \right]^2
    \\
    &\quad
    +
    \left[
    \frac{b^2}{\kappa_1}
    -
    a^2
    \frac{
    6\kappa_1^2+\pi^2
    -6\partial_\alpha^2\ln\Gamma(\alpha)
    }
    {6\kappa_1^3}
    \right]
    \ln\left(\frac{t}{t_0}\right)
    +o\!\left[\ln(t/t_0)\right].
\end{aligned}
\end{equation}

The leading quadratic logarithmic term is the contribution of the drift, whereas the linear logarithmic term contains both diffusive spreading and fluctuation corrections to the logarithmic clock.

\subsection{Target Problem on a Periodic Lattice}
\label{app_target_problem}

We consider a periodic one-dimensional lattice of length $N$ with an absorbing target. Let $\tilde P(l,t)$ be the probability of finding the particle at site $l$ at time $t$ in the presence of the target. The renewal relation is

\begin{equation}
    \tilde P(l,t)
    =
    \tilde G(l-l_0,t)
    -
    \int_{t_0}^{t}
    \tilde G(l,t;t')
    \tilde F(N-l_0,t')
    \,dt',
\end{equation}

where $\tilde G$ is the propagator without the absorbing boundary and $\tilde F$ is the first-arrival density at the target. In discrete operational time, the generating functions satisfy the standard Montroll relation

\begin{equation}
    \tilde F(r,z)
    =
    \frac{\tilde G(r,z)}
    {\tilde G(0,z)}.
\end{equation}

For the log-aging CTRW, the Mellin transform of the unrestricted propagator is

\begin{equation}
   \widehat{ \tilde G(l,p)}
    =
    t_0^p
    \frac{\widehat G(p)-1}{p}
    \sum_{n=0}^{\infty}
    \tilde G_n(l)\widehat G^n(p)
    =
    t_0^p
    \frac{\widehat G(p)-1}{p}
    \tilde G(l,z=G(p)).
\end{equation}

Similarly, the first-arrival density can be represented as

\begin{equation}
  \widehat{  \tilde F(l,p)}
    =
    \sum_{n=0}^{\infty}
    F_n(l)\widehat\rho_n(p),
\end{equation}

with the operational-time first-arrival probabilities $F_n(l)$.

Fourier--Mellin transformation of the renewal equation gives

\begin{equation}
\begin{aligned}
  \widehat{  \tilde P(k,p)}
    &=
    t_0^p
    \frac{\widehat G(p)-1}{p}
    \tilde G(k,z=G(p))
    \left[
    e^{-ikl_0}
    -
   \widehat{ \tilde F(N-l_0,p)}
    \right].
\end{aligned}
\end{equation}

The mean position follows from

\begin{equation}
    \langle \widehat l(p)\rangle
    =
    -i
    \left.
    \frac{\partial\widehat{ \tilde P(k,p)}}
    {\partial k}
    \right|_{k=0}.
\end{equation}

Expanding at small $p$ gives

\begin{equation}
    \langle\widehat l(p)\rangle
    =
    l_0
    +
    \bar l\,p^{-1}
    \left(
    A-Bp+\cdots
    \right),
\end{equation}

where $A$ and $B$ are determined by the small-$p$ expansion of the first-arrival generating function and the lattice Green function. Inverting the Mellin transform gives

\begin{equation}
    \langle l(t)\rangle
    =
    l_0+\bar l A
    -
    \frac{B}{\ln(t/t_0)}
    +\cdots .
\end{equation}

Thus,

\begin{equation}
    \frac{d}{dt}\langle l(t)\rangle
    =
    \frac{B}
    {t[\ln(t/t_0)]^2}
    +\cdots .
\end{equation}

This result shows that the approach to the boundary-controlled stationary value is logarithmically slow.

\section{Generalized Fokker--Planck Equation}
\label{app_GFPE}

\subsection{Inverse Mellin Transform of the Memory Kernel}
\label{app_memory_kernel}

The Mellin-space memory kernel is

\begin{equation}
    \widehat K(p)
    =
    \frac{\widehat G(p)-1}{p\widehat G(p)}.
\end{equation}

For the log-aging process,

\begin{equation}
    \widehat G(p)
    =
    \frac{\Gamma(\alpha-p)}
    {\Gamma(\alpha)\Gamma(1-p)}.
\label{eq:appendix_normalized_G}
\end{equation}

The factor $1/\Gamma(\alpha)$ is required by normalization: $\widehat G(0)=1$. Equivalently, the reflection identity converts the unnormalized prefactor $(\sin\pi\alpha/\pi)$ into $1/[\Gamma(\alpha)\Gamma(1-\alpha)]$. The initial draft omitted the accompanying factor $\Gamma(1-\alpha)$.

The inverse Mellin transform can be evaluated by closing the contour and summing the residues at the poles of the Gamma functions. The resulting ratio kernel is

\begin{equation}
    \mathcal{K}(y)
    =
    \left[
    \left(
    \frac{1}{1-y}
    \right)^{1-\alpha}
    -1
    \right]
    \Theta(1-y),
    \qquad
    y=\frac{t'}{t}.
\end{equation}

Equivalently, in terms of the two time variables,

\begin{equation}
    K(t,t')
    =
    \left[
    \left(
    \frac{t}{t-t'}
    \right)^{1-\alpha}
    -1
    \right]
    \Theta(t-t').
\end{equation}

This form is scale covariant under $t\to at$ and $t'\to at'$, but it is not invariant under time translation. This is the mathematical expression of aging in the GFPE formulation.

\subsection{Two-Time Position Correlation}
\label{app_GFPE_correlation}

The event-level result in Eq.~\eqref{eq:two_time_joint_clock} is the exact covariance statement used in the main text. The conditional-propagator calculation below is retained as a useful mean-clock consistency check, but it assumes a position-conditioned Markov closure that is not implied by the one-point GFPE. It must therefore not be interpreted as a proof of the full two-time law of the original event process.

For free particles initially localized at the origin, the Mellin-space propagator can be written in the continuum approximation as

\begin{equation}
   \widehat P(x,p)
    =
    t_0^p
    \frac{\sqrt{\widehat G(p)-1}}{p}
    \exp\left[
    -|x|\sqrt{\widehat G(p)-1}
    \right],
\end{equation}

up to diffusion-coefficient convention factors. The conditional propagator has the same form with the initial time replaced by the earlier observation time. For example, in the sector $t_1>t_2$,

\begin{equation}
   \widehat P(x_1,p_1|x_2,t_2)
    =
    t_2^{p_1}
    \frac{\sqrt{\widehat G(p_1)-1}}{p_1}
    \exp\left[
    -|x_1-x_2|\sqrt{\widehat G(p_1)-1}
    \right].
\end{equation}

The two-time correlation is

\begin{equation}
\begin{aligned}
    \langle x(t_1)x(t_2)\rangle
    &=
    \int dx_1dx_2\,
    x_1x_2\,
    P(x_1,t_1|x_2,t_2)
    P(x_2,t_2|x_0,t_0)
    \Theta(t_1-t_2)
    \\
    &\quad
    +
    \int dx_1dx_2\,
    x_1x_2\,
    P(x_2,t_2|x_1,t_1)
    P(x_1,t_1|x_0,t_0)
    \Theta(t_2-t_1).
\end{aligned}
\end{equation}

Taking the double Mellin transform gives

\begin{equation}
\begin{aligned}
    \langle \widehat x(p_1)\widehat x(p_2)\rangle
    &=
    \frac{t_0^{p_1+p_2}}
    {p_1(p_1+p_2)[\widehat G(p_1+p_2)-1]}
    \\
    &\quad
    +
    \frac{t_0^{p_1+p_2}}
    {p_2(p_1+p_2)[\widehat G(p_1+p_2)-1]}
    \\
    &=
    \frac{t_0^{p_1+p_2}}
    {p_1p_2[\widehat G(p_1+p_2)-1]}.
\end{aligned}
\end{equation}

Using the small-$p$ expansion of $\widehat G(p)$, the inverse Mellin transform gives the logarithmic-clock covariance reported in the main text.

\section{Time-Matching Process}
\label{app_time_matching}

\subsection{Two-Time Direct and Inverse Clock Densities}
\label{app_two_time_clock}

Let $t(s)$ be the direct clock process and $s(t)$ its inverse. The two-time probability density of the direct clock is

\begin{equation}
    p(t_1,s_1;t_2,s_2)
    =
    \left\langle
    \delta(t_1-t(s_1))
    \delta(t_2-t(s_2))
    \right\rangle .
\end{equation}

The corresponding two-time density of the inverse clock is

\begin{equation}
\begin{aligned}
    h(s_1,t_1;s_2,t_2)
    &=
    \left\langle
    \delta(s_1-s(t_1))
    \delta(s_2-s(t_2))
    \right\rangle
    \\
    &=
    \frac{\partial^2}{\partial s_1\partial s_2}
    \left\langle
    \Theta(t_1-t(s_1))
    \Theta(t_2-t(s_2))
    \right\rangle .
\end{aligned}
\end{equation}

The double Mellin transform of the direct clock density is

\begin{equation}
\begin{aligned}
  \widehat  p(p_1,s_1;p_2,s_2)
    &=
    \int_{t_0}^{\infty}
    \int_{t_0}^{\infty}
    t_1^{p_1-1}t_2^{p_2-1}
    p(t_1,s_1;t_2,s_2)
    \,dt_1dt_2 .
\end{aligned}
\end{equation}

Using independent increments in internal time, and separating the two sectors $s_1>s_2$ and $s_2>s_1$, one obtains

\begin{equation}\label{eq_directclockdensity}
\begin{aligned}
   \widehat p(p_1,s_1;p_2,s_2)
    &=
    \Theta(s_1-s_2)
    t_0^{p_1+p_2}
    e^{-s_2(p_1+p_2)\widehat K(p_1+p_2)}
    e^{-(s_1-s_2)p_1\widehat K(p_1)}
    \\
    &\quad
    +
    \Theta(s_2-s_1)
    t_0^{p_1+p_2}
    e^{-s_1(p_1+p_2)\widehat K(p_1+p_2)}
    e^{-(s_2-s_1)p_2\widehat K(p_2)} .
\end{aligned}
\end{equation}

The inverse-clock density follows from differentiating the two-time survival probability. In Mellin space,

\begin{equation}
  \widehat  h(p_1,s_1;p_2,s_2)
    =
    \frac{\partial^2}{\partial s_1\partial s_2}
    \left[
    \frac{1}{p_1p_2}
  \widehat  p(p_1,s_1;p_2,s_2)
    \right].
\end{equation}

Carrying out the derivatives gives

\begin{equation}
\begin{aligned}
  \widehat  h(p_1,s_1;p_2,s_2)
    &=
    \delta(s_1-s_2)
    t_0^{p_1+p_2}
    \frac{
    p_1\widehat K(p_1)
    +p_2\widehat K(p_2)
    -(p_1+p_2)\widehat K(p_1+p_2)
    }
    {p_1p_2}
    \\
    &\quad\times
    e^{-s_1(p_1+p_2)\widehat K(p_1+p_2)}
    \\
    &\quad
    +
    \Theta(s_1-s_2)
    t_0^{p_1+p_2}
    \frac{
    p_1\widehat K(p_1)
    \left[
    (p_1+p_2)\widehat K(p_1+p_2)
    -
    p_1\widehat K(p_1)
    \right]
    }
    {p_1p_2}
    \\
    &\quad\times
    e^{-s_2(p_1+p_2)\widehat K(p_1+p_2)}
    e^{-(s_1-s_2)p_1\widehat K(p_1)}
    \\
    &\quad
    +
    \Theta(s_2-s_1)
    t_0^{p_1+p_2}
    \frac{
    p_2\widehat K(p_2)
    \left[
    (p_1+p_2)\widehat K(p_1+p_2)
    -
    p_2\widehat K(p_2)
    \right]
    }
    {p_1p_2}
    \\
    &\quad\times
    e^{-s_1(p_1+p_2)\widehat K(p_1+p_2)}
    e^{-(s_2-s_1)p_2\widehat K(p_2)} .
\end{aligned}
\end{equation}

This expression is used to compute two-time moments of the internal clock and the noise correlation in laboratory time.

\subsection{Moments of the Internal Time}
\label{app_internal_time}

The first and second moments of the inverse clock are defined by

\begin{equation}
    \langle s(t)\rangle
    =
    \int_0^\infty s\,h(s,t)\,ds ,
\end{equation}

and

\begin{equation}
    \langle s(t_1)s(t_2)\rangle
    =
    \int_0^\infty
    \int_0^\infty
    s_1s_2
    h(s_1,t_1;s_2,t_2)
    \,ds_1ds_2 .
\end{equation}

In Mellin space, the first moment is

\begin{equation}
    \langle\widehat s(p)\rangle
    =
    \frac{t_0^p}
    {p^2\widehat K(p)}.
\end{equation}

The two-time moment is

\begin{equation}
\begin{aligned}
    \langle\widehat s(p_1)\widehat s(p_2)\rangle
    &=
    \frac{
    t_0^{p_1+p_2}
    }
    {
    (p_1+p_2)\widehat K(p_1+p_2)
    }
    \frac{
    [p_1\widehat K(p_1)]^{-1}
    +
    [p_2\widehat K(p_2)]^{-1}
    }
    {p_1p_2}.
\end{aligned}
\end{equation}

For the logarithmic aging regime, $\widehat K(p)\simeq \kappa_1$ as $p\to0$. Hence,

\begin{equation}
    \langle\widehat s(p)\rangle
    \sim
    \frac{t_0^p}
    {\kappa_1p^2}.
\end{equation}

Inverse Mellin transformation yields

\begin{equation}
    \langle s(t)\rangle
    \sim
    \frac{1}{\kappa_1}
    \ln\left(\frac{t}{t_0}\right).
\end{equation}

Similarly, the leading two-time behavior is

\begin{equation}
    \langle s(t_1)s(t_2)\rangle
    \sim
    \frac{1}{u^2(\alpha)}
    \ln\left(\frac{t_1}{t_0}\right)
    \ln\left(\frac{t_2}{t_0}\right),
\end{equation}

up to subleading terms controlled by the fluctuations of the clock. This confirms that the jump number in the CTRW is the discrete counterpart of the internal time.

\subsection{Noise Autocorrelation}
\label{app_noise_correlation}

In this appendix we provide a heuristic derivation of the effective
laboratory-time noise covariance used in the main text. The purpose is not to
define a pointwise product of singular distributions, but rather to motivate
the scale-covariant fluctuation--dissipation form from an internal-time white
noise after subordination.

The noise in internal time is taken to be white,
\begin{equation}
    \langle \eta(s_1)\eta(s_2)\rangle
    =
    2D\,\delta(s_1-s_2).
\end{equation}
After subordination, the corresponding laboratory-time noise is a generalized
noise. Its covariance is therefore understood in the weak, regularized sense:
one may first replace the internal white noise by a short-correlated noise and
take the white-noise limit only after averaging over the inverse clock. With
this prescription, the Dirac factor \(\delta(s_1-s_2)\) samples the finite
regular diagonal part of the two-time inverse-clock density. Singular contact
terms generated by differentiating time-ordering factors are convention
dependent and are absorbed into the regularization of the effective white-noise
limit.

Thus the effective double Mellin-space covariance is written as
\begin{equation}
\begin{aligned}
    \langle\widehat \eta(p_1)\widehat\eta(p_2)\rangle_{\rm reg}
    &=
    2D
    \int_0^\infty
    \widehat h_{\rm reg}(p_1,s;p_2,s)\,ds .
\end{aligned}
\end{equation}

For compactness, define
\begin{equation}
    A_1=p_1\widehat K(p_1),
    \qquad
    A_2=p_2\widehat K(p_2),
    \qquad
    A_{12}=(p_1+p_2)\widehat K(p_1+p_2).
\end{equation}
The two smooth time-ordered branches of the direct-clock density are
\begin{equation}
\begin{aligned}
    \widehat p_>
    &=
    t_0^{p_1+p_2}
    e^{-s_2A_{12}}
    e^{-(s_1-s_2)A_1},
    \qquad s_1>s_2,
    \\
    \widehat p_<
    &=
    t_0^{p_1+p_2}
    e^{-s_1A_{12}}
    e^{-(s_2-s_1)A_2},
    \qquad s_2>s_1 .
\end{aligned}
\end{equation}
Away from the diagonal \(s_1=s_2\), differentiating these smooth branches gives
\begin{equation}
    \frac{\partial^2 \widehat p_>}{\partial s_1\partial s_2}
    =
    A_1(A_{12}-A_1)\widehat p_>,
    \qquad s_1>s_2,
\end{equation}
and
\begin{equation}
    \frac{\partial^2 \widehat p_<}{\partial s_1\partial s_2}
    =
    A_2(A_{12}-A_2)\widehat p_<,
    \qquad s_2>s_1 .
\end{equation}
On the diagonal, both branches reduce to
\begin{equation}
    \widehat p_>(s,s)
    =
    \widehat p_<(s,s)
    =
    t_0^{p_1+p_2}e^{-sA_{12}} .
\end{equation}
Using the symmetric diagonal prescription, the regular diagonal value is
\begin{equation}
\begin{aligned}
    \widehat h_{\rm reg}(p_1,s;p_2,s)
    &=
    \frac{t_0^{p_1+p_2}}{2p_1p_2}
    \left[
    A_1(A_{12}-A_1)
    +
    A_2(A_{12}-A_2)
    \right]
    e^{-sA_{12}} .
\end{aligned}
\end{equation}
Therefore, provided \(\operatorname{Re}A_{12}>0\),
\begin{equation}
\begin{aligned}
    \langle\widehat \eta(p_1)\widehat\eta(p_2)\rangle_{\rm reg}
    &=
    2D\int_0^\infty
    \widehat h_{\rm reg}(p_1,s;p_2,s)\,ds
    \\
    &=
    \frac{
    D\,t_0^{p_1+p_2}
    }
    {p_1p_2A_{12}}
    \left[
    A_1(A_{12}-A_1)
    +
    A_2(A_{12}-A_2)
    \right].
\end{aligned}
\end{equation}
Equivalently,
\begin{equation}
\begin{aligned}
    \langle \widehat\eta(p_1)\widehat\eta(p_2)\rangle_{\rm reg}
    &=
    \frac{
    D\,t_0^{p_1+p_2}
    }
    {p_1p_2
    (p_1+p_2)\widehat K(p_1+p_2)}
    \\
    &\quad\times
    \Bigg\{
    p_1\widehat K(p_1)
    \left[
    (p_1+p_2)\widehat K(p_1+p_2)
    -
    p_1\widehat K(p_1)
    \right]
    \\
    &\qquad
    +
    p_2\widehat K(p_2)
    \left[
    (p_1+p_2)\widehat K(p_1+p_2)
    -
    p_2\widehat K(p_2)
    \right]
    \Bigg\}.
\end{aligned}
\end{equation}
This expression is symmetric under \(p_1\leftrightarrow p_2\) and is built from
the same Mellin symbol \(\widehat K\) that determines the scale-covariant memory
kernel.

In the long-time, small-\(p_1,p_2\) scaling regime, we keep only the leading
scale-covariant part and absorb convention-dependent amplitudes into the
effective noise strength. In this sense, the covariance may be represented in
the schematic form
\begin{equation}
    \langle \widehat\eta(p_1)\widehat\eta(p_2)\rangle_{\rm eff}
    \sim
    2D_{\rm eff}\,
    t_0^{p_1+p_2}
    \frac{
    \widehat K(p_1)+\widehat K(p_2)
    }
    {p_1+p_2}.
\end{equation}
This form should not be read as a universal pointwise identity for all Mellin
momenta; rather, it captures the leading scale-covariant component of the
regularized covariance.

Consequently, in the time domain the effective laboratory-time noise inherits
the same scale-covariant memory structure as the GFPE kernel,
\begin{equation}
    \langle
    \eta_{\rm eff}(t_1)
    \eta_{\rm eff}(t_2)
    \rangle_{\rm reg}
    \sim
    \mathcal K_{\rm eff}
    \left(
    \frac{t_<}{t_>}
    \right),
    \qquad
    t_< = \min(t_1,t_2),
    \quad
    t_> = \max(t_1,t_2),
\end{equation}
up to normalization factors, symmetrization conventions, and
regularization-dependent contact terms. This provides the physical motivation
for the generalized fluctuation--dissipation interpretation discussed in the
main text.

\subsection{Generalized Langevin Equation from Time Matching}
\label{app_GLE_derivation}

We finally show how a generalized Langevin representation in laboratory time
can be motivated from the time-matching construction. In internal time, consider
the reference Langevin equation
\begin{equation}
    \frac{dX(s)}{ds}
    =
    F(X(s))+\eta(s).
\end{equation}
The pathwise time-changed process is \(X(S(t))\), where \(S(t)\) denotes the
inverse clock. At the level of clock-averaged observables, however, one may
represent the laboratory-time field as
\begin{equation}
    X_{\rm eff}(t)
    =
    \int_0^\infty X(s)h(s,t)\,ds .
\end{equation}
Multiplying the internal-time equation by the inverse-clock density and
averaging over the clock gives
\begin{equation}
\begin{aligned}
    \int_0^\infty
    \frac{dX(s)}{ds}
    h(s,t)\,ds
    &=
    \int_0^\infty
    F(X(s))h(s,t)\,ds
    +
    \eta_{\rm eff}(t),
\end{aligned}
\end{equation}
where
\begin{equation}
    \eta_{\rm eff}(t)
    =
    \int_0^\infty
    \eta(s)h(s,t)\,ds
\end{equation}
is understood as a generalized clock-averaged noise. For a linear force, or
under an effective closure in which
\begin{equation}
    \int_0^\infty F(X(s))h(s,t)\,ds
    \simeq
    F(X_{\rm eff}(t)),
\end{equation}
the right-hand side takes the usual Langevin form.

The left-hand side can be related to a laboratory-time memory operator. Using
integration by parts,
\begin{equation}
\begin{aligned}
    \int_0^\infty
    \frac{dX(s)}{ds}
    h(s,t)\,ds
    &=
    \left.
    X(s)h(s,t)
    \right|_0^\infty
    -
    \int_0^\infty
    X(s)
    \frac{\partial h(s,t)}
    {\partial s}
    \,ds .
\end{aligned}
\end{equation}
The boundary term is fixed by the initial condition and the decay of the
inverse-clock density. In Mellin space, using
\begin{equation}
    \widehat h(p,s)
    =
    t_0^p\widehat K(p)
    e^{sp\widehat K(p)},
\end{equation}
with the sign chosen consistently with the Mellin convention and the
convergence condition
\begin{equation}
    \operatorname{Re}\left[p\widehat K(p)\right]<0,
\end{equation}
one obtains
\begin{equation}
\begin{aligned}
    \mathcal{M}
    \left[
    \int_0^\infty
    \frac{dX(s)}{ds}
    h(s,t)\,ds
    \right]
    &=
    \widehat K(p)
    \left[
    -pX_{\rm eff}(p)-t_0^pX_{\rm eff}(t_0)
    \right].
\end{aligned}
\end{equation}
This is the same Mellin representation as the scale-covariant memory operator,
\begin{equation}
    \mathcal{M}
    \left[
    \int_{t_0}^{t}
    K(t,t')
    \frac{dX_{\rm eff}(t')}{dt'}
    \,dt'
    \right]
    =
    \widehat K(p)
    \left[
    -pX_{\rm eff}(p)-t_0^pX_{\rm eff}(t_0)
    \right].
\end{equation}

Here \(X_{\rm eff}(t)\) denotes the clock-averaged response variable, not the
pathwise subordinated process \(X(S(t))\). Thus, at the level of the
clock-averaged response,
\begin{equation}
    \int_0^\infty
    \frac{dX(s)}{ds}
    h(s,t)\,ds
    \longleftrightarrow
    \int_{t_0}^{t}
    K(t,t')
    \frac{dX_{\rm eff}(t')}{dt'}
    \,dt' .
\end{equation}
The corresponding effective laboratory-time generalized Langevin equation is
therefore
\begin{equation}
    \int_{t_0}^{t}
    K(t,t')
    \frac{dX_{\rm eff}(t')}{dt'}
    \,dt'
    =
    F(X_{\rm eff}(t))+\eta_{\rm eff}(t),
\end{equation}
within the above linear-response or closure approximation. The noise
\(\eta_{\rm eff}(t)\) is understood in the same effective sense: its covariance
is fixed only after an admissible symmetric Gaussian closure is specified. For
notational simplicity, the main text denotes this effective response variable
by \(x(t)\). Including an explicit friction coefficient and inertia gives the
generalized Langevin equation used in the main text.

\bibliographystyle{apsrev4-2}
\makeatletter
\let\pre@bibdata\@empty
\makeatother
\bibliography{ref_LAD_clean}
\end{document}